\documentclass[longauth, referee]{aa}
\usepackage{graphicx}
\usepackage{amsmath}
\usepackage{csquotes}
\usepackage{setspace}
\usepackage[section]{placeins}
\usepackage{pdflscape}
\usepackage{footnote}
\usepackage{geometry}
\usepackage{amssymb}
\usepackage{grffile}
\usepackage{siunitx}
\usepackage{wasysym}
\usepackage{microtype}
\usepackage{wrapfig}
\usepackage{blindtext}
\usepackage[noabbrev, capitalise]{cleveref}
\usepackage{makecell}
\usepackage{subfiles}
\usepackage{booktabs}
\usepackage{longtable}

\newcommand{\solmass}{$M_{\odot}$}
\newcommand{\simsym}{\mathord\sim}

\bibpunct{(}{)}{;}{a}{}{,} % to follow the A&A style

\begin{document}
\title{Multiple Star Systems in the Orion Nebula}

\author{GRAVITY collaboration\thanks{GRAVITY is developed in a collaboration by the Max Planck Insti-
tute for extraterrestrial Physics, LESIA of Paris Observatory / CNRS / UPMC / Univ. Paris Diderot and IPAG of Université Grenoble Alpes / CNRS, the Max Planck Institute for Astronomy, the University of Cologne, the Centro de Astrof\'{\i}sica e Gravita\c{c}\~{a}o,
and the European Southern Observatory.}: Martina Karl\thanks{\email{martina.karl@tum.de}}\inst{\ref{inst1}} \and Oliver Pfuhl\thanks{\email{pfuhl@mpe.mpg.de}}\inst{\ref{inst1}} \and Frank Eisenhauer\inst{\ref{inst1}} \and Reinhard Genzel\inst{\ref{inst1},\ref{inst16}} \and Rebekka Grellmann\inst{\ref{inst4}} \and Maryam Habibi\inst{\ref{inst1}} \and Roberto Abuter\inst{\ref{inst8}} \and
Matteo Accardo\inst{\ref{inst8}} \and 
Ant\'{o}nio Amorim\inst{\ref{inst6}} \and 
Narsireddy Anugu\inst{\ref{inst7}} \and 
Gerardo \'{A}vila\inst{\ref{inst8}} \and 
Myriam Benisty\inst{\ref{inst5}} \and 
Jean-Philippe Berger\inst{\ref{inst5}} \and 
Nicolas Blind\inst{\ref{inst10}} \and 
Henri Bonnet\inst{\ref{inst8}} \and 
Pierre Bourget\inst{\ref{inst9}} \and 
Wolfgang Brandner\inst{\ref{inst3}} \and 
Roland Brast\inst{\ref{inst8}} \and 
Alexander Buron\inst{\ref{inst1}} \and 
Alessio Caratti o Garatti\inst{\ref{inst3}, \ref{inst18}} \and
Fr\'{e}d\'{e}ric Chapron\inst{\ref{inst2}} \and 
Yann Cl\'{e}net\inst{\ref{inst2}} \and 
Claude Collin\inst{\ref{inst2}} \and 
Vincent Coud\'{e} du Foresto\inst{\ref{inst2}} \and 
Willem-Jan de Wit\inst{\ref{inst9}} \and 
Tim de Zeeuw\inst{\ref{inst1},\ref{inst15}} \and 
Casey Deen\inst{\ref{inst1}} \and 
Françoise Delplancke-Str\"{o}bele\inst{\ref{inst8}} \and 
Roderick Dembet\inst{\ref{inst8}} \and 
Fr\'{e}d\'{e}ric Derie\inst{\ref{inst8}} \and 
Jason Dexter\inst{\ref{inst1}} \and 
Gilles Duvert\inst{\ref{inst5}} \and 
Monica Ebert\inst{\ref{inst3}} \and 
Andreas Eckart\inst{\ref{inst4},\ref{inst14}} \and 
Michael Esselborn\inst{\ref{inst8}} \and 
Pierre F\'{e}dou\inst{\ref{inst2}} \and 
Gert Finger\inst{\ref{inst8}} \and 
Paulo Garcia\inst{\ref{inst7}, \ref{inst9}} \and 
Cesar Enrique Garcia Dabo\inst{\ref{inst8}} \and 
Rebeca Garcia Lopez\inst{\ref{inst3}, \ref{inst18}} \and 
Feng Gao\inst{\ref{inst1}} \and 
\'{E}ric Gendron\inst{\ref{inst2}} \and 
Stefan Gillessen\inst{\ref{inst1}} \and 
Fr\'{e}d\'{e}ric Gont\'{e}\inst{\ref{inst8}} \and 
Paulo Gordo\inst{\ref{inst6}} \and 
Ulrich Gr\"{o}zinger\inst{\ref{inst3}} \and 
Patricia Guajardo\inst{\ref{inst9}} \and
Sylvain Guieu\inst{\ref{inst5}} \and 
Pierre Haguenauer\inst{\ref{inst8}} \and 
Oliver Hans\inst{\ref{inst1}} \and 
Xavier Haubois\inst{\ref{inst9}} \and 
Marcus Haug\inst{\ref{inst1},\ref{inst8}} \and 
Frank Haußmann\inst{\ref{inst1}} \and 
Thomas Henning\inst{\ref{inst3}} \and 
Stefan Hippler\inst{\ref{inst3}} \and 
Matthew Horrobin\inst{\ref{inst4}} \and 
Armin Huber\inst{\ref{inst3}} \and 
Zoltan Hubert\inst{\ref{inst2}} \and 
Norbert Hubin\inst{\ref{inst8}} \and 
Christian A. Hummel\inst{\ref{inst8}} \and 
Gerd Jakob\inst{\ref{inst8}} \and 
Lieselotte Jochum\inst{\ref{inst8}} \and 
Laurent Jocou\inst{\ref{inst5}} \and 
Andreas Kaufer\inst{\ref{inst9}} \and 
Stefan Kellner\inst{\ref{inst1},\ref{inst14}} \and 
Sarah Kendrew\inst{\ref{inst3},\ref{inst12}} \and 
Lothar Kern\inst{\ref{inst8}} \and 
Pierre Kervella\inst{\ref{inst2}} \and 
Mario Kiekebusch\inst{\ref{inst8}} \and 
Ralf Klein\inst{\ref{inst3}} \and 
Rainer Köhler\inst{\ref{inst3}, \ref{inst19}} \and
Johan Kolb\inst{\ref{inst9}} \and 
Martin Kulas\inst{\ref{inst3}} \and 
Sylvestre Lacour\inst{\ref{inst2}} \and 
Vincent Lapeyr\`{e}re\inst{\ref{inst2}} \and 
Bernard Lazareff\inst{\ref{inst5}} \and 
Jean-Baptiste Le Bouquin\inst{\ref{inst5}} \and 
Pierre L\'{e}na\inst{\ref{inst2}} \and 
Rainer Lenzen\inst{\ref{inst3}} \and 
Samuel L\'{e}v\^{e}que\inst{\ref{inst8}} \and 
Chien-Cheng Lin\inst{\ref{inst3}} \and
Magdalena Lippa\inst{\ref{inst1}} \and 
Yves Magnard\inst{\ref{inst5}} \and 
Leander Mehrgan\inst{\ref{inst8}} \and 
%Marcus Mellein\inst{\ref{inst3}} \and 
Antoine M\'{e}rand\inst{\ref{inst8}} \and 
%Javier Moreno-Ventas\inst{\ref{inst3}} \and 
Thibaut Moulin\inst{\ref{inst5}} \and 
Eric M\"{u}ller\inst{\ref{inst8}} \and 
Friedrich M\"{u}ller\inst{\ref{inst3}} \and
Udo Neumann\inst{\ref{inst3}} \and
Sylvain Oberti\inst{\ref{inst8}} \and
Thomas Ott\inst{\ref{inst1}} \and
Laurent Pallanca\inst{\ref{inst9}} \and
Johana Panduro\inst{\ref{inst3}} \and
Luca Pasquini\inst{\ref{inst8}} \and
Thibaut Paumard\inst{\ref{inst2}} \and
Isabelle Percheron\inst{\ref{inst8}} \and
Karine Perraut\inst{\ref{inst5}} \and
Guy Perrin\inst{\ref{inst2}} \and
%Pierre-Olivier Petrucci\inst{\ref{inst5}} \and
Andreas Pfl\"{u}ger\inst{\ref{inst1}} \and
Thanh Phan Duc\inst{\ref{inst8}} \and
Philipp M. Plewa\inst{\ref{inst1}} \and
Dan Popovic\inst{\ref{inst8}} \and
Sebastian Rabien\inst{\ref{inst1}} \and
Andr\'{e}s Ram\'{\i}rez\inst{\ref{inst9}} \and
Jose Ramos\inst{\ref{inst3}} \and
Christian Rau\inst{\ref{inst1}} \and
Miguel Riquelme\inst{\ref{inst9}} \and
Gustavo Rodr\'{\i}guez-Coira\inst{\ref{inst2}} \and
Ralf-Rainer Rohloff\inst{\ref{inst3}} \and
Alejandra Rosales\inst{\ref{inst1}} \and
G\'{e}rard Rousset\inst{\ref{inst2}} \and
Joel Sanchez-Bermudez\inst{\ref{inst9}, \ref{inst3}} \and
Silvia Scheithauer\inst{\ref{inst3}} \and
Markus Sch\"{o}ller\inst{\ref{inst8}} \and
Nicolas Schuhler\inst{\ref{inst9}} \and
Jason Spyromilio\inst{\ref{inst8}} \and
Odele Straub\inst{\ref{inst2}} \and
Christian Straubmeier\inst{\ref{inst4}} \and
Eckhard Sturm\inst{\ref{inst1}} \and
Marcos Suarez\inst{\ref{inst8}} \and
Konrad R.W. Tristram\inst{\ref{inst9}} \and
Noel Ventura\inst{\ref{inst5}} \and
Fr\'{e}d\'{e}ric Vincent\inst{\ref{inst2}} \and
Idel Waisberg\inst{\ref{inst1}} \and
Imke Wank\inst{\ref{inst4}} \and
Felix Widmann\inst{\ref{inst1}} \and
Ekkehard Wieprecht\inst{\ref{inst1}} \and
Michael Wiest\inst{\ref{inst4}} \and
Erich Wiezorrek\inst{\ref{inst1}} \and
Markus Wittkowski\inst{\ref{inst8}} \and
Julien Woillez\inst{\ref{inst8}} \and
Burkhard Wolff\inst{\ref{inst8}} \and
Senol Yazici\inst{\ref{inst1},\ref{inst4}} \and
Denis Ziegler\inst{\ref{inst2}} \and
G\'{e}rard Zins\inst{\ref{inst9}}
} 
\institute{MPE - Max Planck Institute for extraterrestrial Physics, Giessenbachstr., 85741 Garching, Germany\label{inst1} \and 
Department of Physics, Le Conte Hall, University of California, Berkeley, CA 94720, USA\label{inst16} \and 
1. Physikalisches Institut, Universität zu Köln, Zülpicher Str. 77, 50937 Köln, Germany\label{inst4} \and 
European Southern Observatory, Karl-Schwarzschild-Str. 2, 85748 Garching, Germany\label{inst8} \and 
CENTRA -- Centro de Astrof\'{\i}sica e Gravita\c{c}\~{a}o, IST, Universidade de Lisboa, P-1049-001 Lisboa, Portugal; Universidade de Lisboa - Faculdade de Ci\^encias, Campo Grande, 1749-016 Lisboa, Portugal
\label{inst6} \and 
CENTRA -- Centro de Astrof\'{\i}sica e Gravita\c{c}\~{a}o, IST, Universidade de Lisboa, P-1049-001 Lisboa, Portugal; Faculdade de Engenharia, Universidade do Porto, Rua Dr. Roberto Frias, 4200-465 Porto, Portugal\label{inst7} \and 
Univ. Grenoble Alpes, CNRS, IPAG, 38000 Grenoble, France\label{inst5} \and
Observatoire de Gen\`{e}ve, Universit\'{e} de Gen\`{e}ve, 51 Ch. des Maillettes, 1290 Versoix, Switzerland\label{inst10} \and 
European Southern Observatory, Casilla 19001, Santiago 19, Chile\label{inst9} \and 
MPIA - Max-Planck-Institut für Astronomie, K\"{o}nigstuhl 17, 69117, Heidelberg, Germany\label{inst3} 
%\and Chatillon - ONERA - The French Aerospace Lab\label{inst11} 
\and LESIA - Observatoire de Paris, Universit\'{e} PSL, CNRS, Sorbonne Universit\'{e}, Univ. Paris Diderot, Sorbonne Paris Cit\'{e}, 5 place Jules Janssen, 92195 Meudon, France\label{inst2} \and 
Sterrewacht Leiden, Leiden University, Postbus 513, 2300 RA Leiden, The Netherlands\label{inst15} \and 
Max-Planck-Institute for Radio Astronomy, Auf dem Hügel 69, 53121 Bonn, Germany\label{inst14} \and 
ESA - European Space Agency, Space Telescope Science Institute, 3700 San Martin Drive, Baltimore MD 21218, USA\label{inst12}
%\and USACH - Universidad de Santiage de Chile [Santiago]\label{inst13} 
\and Dublin Institute for Advanced Studies, Astronomy \& Astrophysics Section, 31 Fitzwilliam Place, D02 XF86, Dublin, Ireland\label{inst18} \and
University of Vienna, Universitätsring 1, 1010 Wien, Austria\label{inst19}
}

\abstract{
This work presents an interferometric study of the massive-binary fraction in the Orion Trapezium Cluster with the recently comissioned GRAVITY instrument. %It is the most comprehensive and most sensitive 
%interferometric survey of the Trapezium Cluster to date. The resolution and sensitivity of GRAVITY allows resolving companions on scales of 2\textendash 200 milliarcseconds with a magnitude as faint as 10.6 mag (K-band), which corresponds to $\simsym 1$\textendash 100 AU and stellar masses as low as 1.6 \solmass\ at the distance of Orion. 
We observe a total of 16 stars of mainly OB spectral type. We find 
three previously unknown companions for $\theta ^1$ Ori B, $\theta ^2$ Ori B, and $\theta ^2$ Ori C. We determine a separation for the previously suspected companion of NU Ori. We confirm four companions for $\theta ^1$ Ori A, $\theta ^1$ Ori C, $\theta ^1$ Ori D, and $\theta ^2$ Ori A, all with substantially improved astrometry and photometric mass estimates. We refine the orbit of the eccentric high-mass binary $\theta ^1$ Ori C and we are able to derive a new orbit for $\theta ^1$ Ori D. We find a system mass of 21.7~\solmass\ and a period of $53$~days. Together with other previously detected companions seen in spectroscopy or direct imaging, eleven of the 16 high-mass stars are multiple systems. We obtain a total number of 22 companions with separations up to 600~AU. The companion fraction of the early B and O stars in our sample is about 2, significantly higher than in earlier studies of mostly OB associations. The separation distribution hints towards a bimodality. Such a bimodality has been previously found in A stars, but rarely in OB binaries, which up to this point have been assumed to be mostly compact with a tail of wider companions. We also do not find a 
% * <pfuhl@mpe.mpg.de> 2018-05-02T11:58:03.546Z:
%
% > ignificantly higher than in earlier studies of mostly OB associations.
%
% ^.
substantial population of equal-mass binaries. The observed distribution of mass ratios declines steeply with mass, and like the direct star counts, indicates that our companions follow a standard power law initial mass function. Again, this is in contrast to earlier findings of flat mass ratio distributions in OB associations. We exclude collision as a dominant formation mechanism but find no clear preference for core accretion or competitive accretion.
}
\authorrunning{GRAVITY collaboration et al.}
\maketitle
\section{Introduction}
Massive stars, defined as those with masses higher than $8~M_{\odot}$, have an intense impact on the evolution of galaxies. The winds, UV radiation, massive outflows, and the heavy elements produced by high-mass stars influence the formation of stars and planets \citep[see e.g.][]{Bally2005} as well as the structure of galaxies \citep[e.g.][]{Kennicutt1998}. 
Despite their important role, the formation of massive stars is not well understood. High-mass stars have short lifetimes %TODO Beispiel
and spend a significant part of their life hidden within their parental dust and gas clouds. During this embedded phase, some fundamental evolutionary processes are difficult to observe. For a detailed review of high-mass star formation, see e.g. \citet{1987ARA&A..25...23S,Zinnecker2007,Tan2014,Motte2018}

There are several indications that high-mass star formation is not just a scaled-up version of low-mass star formation. 
One indication is that massive stars tend to appear more often in multiple systems than lower mass stars \cite[e.g.][]{Chini2011,Sana2012}. \citet{Zinnecker2007} found that the number of companions per star increases with stellar mass. For example, \citet{Duchene2013} found $0.22 \pm 0.06$ companions for stars with masses $\lesssim 0.1$~\solmass\ and $1.3 \pm 0.2$ companions for primary stars with masses $\gtrsim 16$~\solmass . They also found more multiple systems for stars with higher mass. %\cref{fig:duchene_plot} displays the respective companion frequency (average number of companions per star) and frequency of multiple systems (average number of multiple system per total objects) as a function of mass, as determined by \citet{Duchene2013}. 
At least 60\% of stars with 8\textendash 16~\solmass\ are part of a multiple system. For stars $\gtrsim 16$~\solmass , at least 80\% are found in multiple systems \citep{Duchene2013, 2014ApJS..215...15S}. %A system of two gravitationally bound stars is called a binary, binary system or binary star. We will also use either of these terms to refer to a companion system in general.
The higher number of companions and multiple systems is most likely a result of their formation process. Massive stars are short-lived and thus it is unlikely that they assemble all of their companions by random interactions in reasonable timescales. \citet{Duchene2013} provides a review about stellar multiplicity. \citet{2017ApJS..230...15M} present a detailed study of the distribution and properties of early-type binaries.

In the case of massive star formation, two different scenarios try to explain the birth of a protostellar object from molecular gas. %compression of molecular gas to a protostellar object. %starless core with $\simsym 100~ M_{\odot}$ or a starless clump with $\simsym 1000~ M_{\odot}$. 
\citet{2002Natur.416...59M} proposed that molecular condensations in turbulent gas form a single massive protostar or several gravitationally bound protostars. For this ``monolithic collapse'', the mass of the final product is directly associated with the mass needed for star formation. %When the surrounding gas moves, the core moves with it. 
Thus, the final material for the resulting star is already gathered before the beginning of star formation. 
Monolithic collapse or core accretion, assumes that the initial conditions are similar to low-mass star formation. An isolated core collapses and accretes mass with a disk \citep{2002ApJ...569..846Y}. %\citet{1992ApJ...396..631M} concluded that the thermal motion of these massive cores is dominated by internal turbulence. 
\citet{2002Natur.416...59M} proposed the turbulent core model, assuming mainly non-thermal internal pressure. \citet{Tan2014} pointed out that the accretion rate of the turbulent core model is higher than for competitive accretion. In the past it was believed that radiation pressure of young massive stars could halt accretion \citep[see e.g.][]{Zinnecker2007,Krumholz2015}. This has been solved by introducing non-spherical accretion \citep[see e.g.][]{Krumholz2009}. %The infalling material builds clumps or columns and the radiation escapes through empty pockets or fingers.

\citet{1997MNRAS.285..201B, 2001MNRAS.323..785B} described an alternative scenario in which the core or resulting protostar moves within the cloud, independent of the movement of the surrounding gas. Thus, the material can come from different parts of the parent cloud, as well as from material infalling onto the cloud. %This stellar material is not defined at the beginning of the collapse, but gathered during the formation of stars. 
Each of the forming protostars competes for the material; this mechanism is therefore called ``competitive accretion''. Competitive accretion starts with many low-mass seeds in a parent cloud, which start to accrete mass, e.g. \citet[pp. 213]{Krumholz2016}. Two factors influence the amount of growth for stars with competitive accretion \citep[see e.g.][]{1997MNRAS.285..201B, 2001MNRAS.323..785B}. One is the accretion radius or the accretion domain, meaning the range where gas is gravitationally attracted to the star. %The radius is directly influenced by the stellar mass, implying that the accretion radius grows with stellar mass. 
The second factor is the gas density of the accretion domain. %A star in a denser region, e.g. a center of a protostellar cluster, can attract more mass than a star in a less dense region.% with the same attraction radius. 
Because gas flows down to the center of clusters, the central position in a cluster is beneficial for mass growth. %Furthermore, the potential accretion domain of a protostar centered in a cloud comprises the whole cloud, whereas the potential accretion domain of a star in an off-center position is limited because it has to compete with the gravitational potential of the inner part of the cloud. 
When the accretion volumes start to overlap, the stars are competing for the available material. This might explains why massive stars are rare and mainly form in the most favorable, i.e. densest, conditions. However, there are also a few examples of O-type stars born in isolation \citep{deWit2004, deWit2005, 2013MNRAS.436.3357O}.

% Cores, formed in the previous compression phase, now start collapsing to a protostellar embryo. In the collapse phase, gravity overcomes pressure, turbulence, rotation and magnetic fields. 
% Centrifugal or rotational forces increase during collapse due to conservation of angular momentum, resulting in flattened accretion disks (e.g. \citet{1975ApJ...199..619B, 1999ApJ...525..330Y}).
 
% When the core collapses and becomes optically thick, the core becomes quasi-hydrostatic and accretes more mass. %At its formation, it has approximately a Jupiter mass and a few solar radii \citep{2000ApJ...531..350M}. 
% While accreting, the core contracts on a Kelvin-Helmholtz timescale and, at a certain point, reaches hydrogen-burning temperatures and densities \citep{1987ARA&A..25...23S, Zinnecker2007}. The inner regions of the core are oblivious and not influenced by the accretion process, except for growing mass and angular momentum. Thus, the evolution of the central core is independent of the accretion mechanism.

%\citet{Motte2018} summarized that the accretion rate is highest at the beginning and slows down with progressing stellar evolution. This explains the decreasing outflow of protostars with increasing age. It is important to note that the luminosity of young massive stars is mainly caused by hydrogen burning as soon as they reach the zero-age main sequence phase (see e.g. \citet{Motte2018}). %For low mass stars, the main luminosity comes from the accretion disk (see e.g. \citet{Motte2018}).

The accretion rate for competitive accretion is lower than for monolithic collapse \citep[see e.g.][]{Tan2014, Krumholz2016}. The angular momentum of gas in both cases is large enough to form an accretion disk.

Companion stars can be formed by various mechanisms. In the monolithic collapse scenario, a massive core can fragment into several smaller cores and form a binary or multiple system \citep[see e.g.][and references therein]{Krumholz2016, Tan2014}. %Mechanisms are needed to prevent a premature collapse of the core so that massive stars still can be formed. \citet{2013ApJ...766...97M} concluded that magnetic fields and radiative feedback suppress core fragmentation. Depending on the strength of radiation and magnetic flux, the core tends to form a star cluster or a single star. \citet[p. 295]{Krumholz2016} concluded that binaries with separations $\gtrsim 1000$~AU have to be formed by direct core fragmentation, because they are too distant to be formed in the protostellar disk. %Thus, there are two regimes of companions: close companions formed through disk fragmentation, and wide companions formed through earlier core fragmentations. 
Disk fragmentation can also produce companion stars, see e.g. \citet{doi:10.1146/annurev-astro-081915-023307}. They concluded that for a star with 8~\solmass , the disk cools down sufficiently to undergo disk fragmentation for separations $\geq 50$~astronomical units (AU). \citet[p. 296]{Krumholz2016} stated that the typical accretion rates on a stellar core lead to a high surface density of the disk and eventually result in disk fragmentation. 
%\citet{Zinnecker2007} explained that disk fragmentation results in initially wide binaries with cold rotating Keplerian disks producing circular orbits and and sub-Keplerian disks producing highly eccentric orbits. %A Keplerian disk is a disk whose material velocity is according to Keplerian motion. The material of a sub-Keplerian disk has a smaller velocity. 
%They concluded that $\theta ^1$ Ori C$_2$ (see \cref{ssec:t1C}) could be an example of a companion with a highly eccentric orbit produced by a sub-Keplerian disk. 

% \citet{2002ASPC..267..179M} found that the separations of close binaries are below the Jeans length. They proposed the idea that the initial separation of a wide binary pair shrinks if both components accrete mass. %\citet{Zinnecker2007} concluded that the binary separation $r$ is indirectly proportional to the mass. For a turbulent medium the relation follows $r \propto M ^{-2}$, and an initially low mass wide binary (e.g. 3~\solmass\ and 100~AU) will become a 30~\solmass\ system with a separation of 1~AU \citep{Zinnecker2007}.

Another possible scenario for the formation of companions is the failed merging of two stars. % When two stars nearly collide but go astray, they can form close and eccentric binaries \citep{Zinnecker2007}. 
This process requires a high stellar density, which is much higher than the typical observed density in our Galaxy. \citet{Zinnecker2007} concluded that the cross section is small and that the collision impact parameter requires fine tuning. If there is a disk, the capture of a companion star becomes more likely. %\citet{2006MNRAS.370.2038D} simulated two stars, one with 3~\solmass\ and one with 10~\solmass . The encounter of these two stars disrupted the 3~\solmass\ star, forming a disk around the 10~\solmass\ star. This disk increased the cross section for possible further encounters. The mechanism is called the ``shred and add process''. The so called ``disk-assisted capture'' also works for the accretion disk of young stars. This mechanism is not effective for solar mass stars (e.g. \citet{1998MNRAS.300.1189B}), but has proven effective for massive stars with $\simsym 20$~\solmass\ and large accretion disks of $\simsym 500$~AU \citep{2006ApJ...653..437M}. \citet{2006ApJ...653..437M} also concluded that the result of many encounters is a binary with 100~AU separation. Disk-assisted capture can therefore result in wide binaries with unequal mass. 

Additionally, a binary system can capture a third, massive companion --- a mechanism called ``three body capture''. In a simulation of a protostellar cluster with more than 400 stars, \citet{2003MNRAS.343..413B} demonstrated that dynamical three-body capture is common in protoclusters. %The typical initial state is a massive star with a wide lower mass companion. At a certain point, a third, more massive object joins the system as a wide companion star with nearly equal mass. The massive companion then absorbs most of the binding energy of the low mass component and becomes harder, with smaller separation. The low mass companion gets ejected from the system and a close equal mass binary remains. 

In order to gain a deeper understanding of massive star and cluster formation, the characteristics of binaries need to be determined. \citet{Lada2003,Briceno2007} found that massive star formation results in either dense OB clusters or unbound OB associations. 
The Orion Nebula Cluster, at a distance of 414~pc \citep{Menten2007, 2014ApJ...783..130R}, is one of the closest active star-forming regions. For a general overview see e.g. \citet{1989ARA&A..27...41G, Hillenbrand1997, Muench2008}.
The Orion Nebula Cluster comprises an expanding blister HII region with the Orion Trapezium Cluster ($\theta ^1$), an open cluster of young massive stars, at its center. These centrally concentrated young stars cause the ionization of the surrounding cloud.
The Orion Trapezium Cluster has six principal components, $\theta ^1$ Ori A to $\theta ^1$ Ori F. The stars $\theta ^1$ Ori (A, B, C, D) are all known to have companions. 

The Orion Nebula Cluster has already been thoroughly observed in recent decades. Deep spectroscopic surveys probed for close companions on scales $\lesssim$~1~AU \citep[e.g.][]{Morrell1991, Abt1991}. 
Adaptive optic assisted imaging and speckle interferometry resolved companions $\gtrsim 14$\textendash few 100~AU \citep[e.g.][]{Weigelt1999, Preibisch1999, Schertl2003}. 
While these ranges have been covered, there is still a gap in the separations where observations are scarce.
The region $\simsym 1$\textendash few 10~AU can only be resolved with long baseline interferometry.

In the following, we present the observational data obtained with GRAVITY \citep{GRAVITYCollaboration2017}, a K-band interferometric instrument at the Very Large Telescope Interferometer (VLTI). 
In \cref{chap:observation}, we describe our observations. \cref{chap:DataAna} introduces the data analysis. We present our results in \cref{chap:ONCstars}, discuss the results in \cref{chap:discussion}, and conclude in \cref{chap:conclusion}.

% * <pfuhl@mpe.mpg.de> 2018-05-02T11:58:14.952Z:
%
% ^.
% * <pfuhl@mpe.mpg.de> 2018-05-02T11:58:12.648Z:
%
% ^.
%\subfile{include/starformation.tex}
\section{Observations}\label{chap:observation}
Data were taken with GRAVITY, a novel instrument at the VLTI for $\simsym 10$ micro-arcsecond astrometric precision measurements with K-band interferometry \citep{GRAVITYCollaboration2017}. GRAVITY coherently combines the light of all four UTs (8.2~m diameter) or all four ATs (1.8~m diameter) with two interferometric beam combiners for fringe tracking and observing science objects, respectively. A star up to $10 $~mag in K-band can be used for fringe-tracking faint objects up to 17~mag in the science channel using the UTs. The spectrometers provide three spectral resolutions: low, medium, and high, with $R \simsym 22, 500, 4000$, respectively. %Additionally, GRAVITY has the capability to split the linear polarization of the incident star light. 

Table \ref{tab:OriObs} provides an overview of all observations. The 16 brightest objects in the Orion Nebula were selected for this study. Observations were primarily performed in medium resolution with the astrometric configuration of the ATs at the stations A0-G1-J2-K0. The detector integration time -- DIT -- depends on the source luminosity. A higher DIT is needed for fainter objects (e.g. a DIT of 30~s was used for $\theta ^1$ Ori F with a K-magnitude of 8.38) whereas shorter DITs are possible for bright objects (e.g. 3~s or 5~s for $\theta ^1$ Ori C with a K-magnitude of 4.57). 
The integration on the source is repeated several times, usually followed by a sky background observation with the same DIT and number of repetitions (NDIT). 
\longtab{
{\small
\begin{longtable}{cccccc}

\toprule
Object & Date [UT]  &Spectral resolution	&Baseline configuration & DIT [s] &NDIT  \\
  \endfirsthead
  
  \multicolumn{4}{c}%
  { \tablename\  \thetable{} \textemdash\ continued from previous page} \\
  \midrule
Object & Date [UT]  &Spectral resolution	&Baseline configuration & DIT [s] &NDIT  \\
\midrule
  \endhead
  
  \midrule
  \multicolumn{3}{r}{ \tablename\ \thetable{} \textemdash\ continued on next page}
  \endfoot

   \endlastfoot

\midrule

  $\theta ^1$ Ori A & 2016 Nov 25  & Medium &K0 G2 D0 J3 & 10 &60 \\
  $\theta ^1$ Ori A & 2017 Oct 12 &Medium &A0 G1 J2 K0 &5 &120 \\
  $\theta ^1$ Ori A & 2018 Jan 12 &Medium &A0 G1 J2 K0 &5 &120 \\
      \midrule
  $\theta ^1$ Ori B & 2017 Jan 11  & Medium &U1 U2 U3 U4 &5 &100 \\
  $\theta ^1$ Ori B & 2017 Feb 20  &Medium &A0 G1 J2 K0 &10 &25 \\
  $\theta ^1$ Ori B & 2017 Mar 18  &Medium &A0 G1 J2 J3 &10 &50 \\
  $\theta ^1$ Ori B & 2017 Mar 20  &Medium &A0 G1 J2 K0 &10 &100 \\
  $\theta ^1$ Ori B & 2017 Mar 21  &Medium &A0 G1 J2 K0 &10 &50 \\
  $\theta ^1$ Ori B & 2017 Oct 11 &Medium &A0 G1 J2 K0 &5 &240 \\
  $\theta ^1$ Ori B & 2017 Oct 12 &Medium &A0 G1 J2 K0 &5 &120 \\
  $\theta ^1$ Ori B & 2017 Oct 13 &Medium &A0 G1 J2 K0 &5 &240 \\
  $\theta ^1$ Ori B & 2018 Jan 11 &Medium &A0 G1 J2 K0 &5 &120 \\
    \midrule
  $\theta ^1$ Ori C & 2015 Nov 09  &Medium & A0 B2 D0 C1 &3 &200  \\
  $\theta ^1$ Ori C & 2016 Jan 09 &Medium &A0 G1 J2 K0 &3 &400  \\
  $\theta ^1$ Ori C & 2016 Oct 04 &Medium &A0 G1 J2 K0 &10 &45  \\
   $\theta ^1$ Ori C & 2016 Oct 04 &Low &A0 G1 J2 K0 &1 &  240\\
    $\theta ^1$ Ori C & 2016 Oct 04 &Low &A0 G1 J2 K0 &0.3 & 200\\
  $\theta ^1$ Ori C & 2016 Nov 25 &Medium &K0 G2 D0 J3 & 10 &30   \\
  $\theta ^1$ Ori C & 2017 Mar 18 &Medium &A0 G1 J2 J3 &10 &50   \\
  $\theta ^1$ Ori C & 2017 Oct 11 &Medium &A0 G1 J2 K0 &5 &120   \\
  $\theta ^1$ Ori C & 2017 Oct 12 &Medium &A0 G1 J2 K0 &5 &120  \\
   $\theta ^1$ Ori C & 2018 Jan 12 &Medium &A0 G1 J2 K0 &5 &120  \\
  \midrule
  $\theta ^1$ Ori D & 2016 Nov 26  &Medium &K0 G2 D0 J3 &10 &30 \\
  $\theta ^1$ Ori D & 2017 Mar 19  &Medium &A0 G1 J2 K0 &10 &60 \\
  $\theta ^1$ Ori D & 2017 Mar 21  &Medium &A0 G1 J2 K0 &10 &50 \\
  $\theta ^1$ Ori D & 2017 Oct 13 &Medium &A0 G1 J2 K0 &5 &120 \\
  $\theta ^1$ Ori D & 2018 Jan 10 &Medium &A0 G1 J2 K0 &5 &120 \\
  $\theta ^1$ Ori D & 2018 Jan 10 &Medium &A0 G1 J2 K0 &5 &120 \\
  \midrule
  $\theta ^1$ Ori E & 2017 Mar 20  &Medium &A0 G1 J2 K0 &10 &50 \\
  \midrule
  $\theta ^1$ Ori F & 2016 Jan 10  &Medium &A0 G1 J2 K0 & 30 &30 \\
  $\theta ^1$ Ori F & 2016 Jan 17  &Medium &A0 G1 J2 K0 &30 &10 \\
  $\theta ^1$ Ori F & 2016 Jan 21  &Medium &A0 G1 J2 K0 &30 &40 \\
  $\theta ^1$ Ori F & 2017 Jan 29  &Medium &A0 G1 J2 K0 &30 &10 \\
  $\theta ^1$ Ori F & 2017 Jan 30  &Medium &A0 G1 J2 K0 &30 &10 \\
  \midrule
  $\theta ^2$ Ori A & 2016 Nov 26  & Medium & K0 G2 D0 J3 & 10 &30  \\
  $\theta ^2$ Ori A & 2018 Jan 10  &Medium &A0 G1 J2 K0 &5 &120 \\
  $\theta ^2$ Ori A & 2018 Jan 12  &Medium &A0 G1 J2 K0 &5 &120 \\
  \midrule
  $\theta ^2$ Ori B & 2018 Jan 10  &Medium &A0 G1 J2 K0 &5 &120 \\
  \midrule
  $\theta ^2$ Ori C & 2018 Jan 12  &Medium &A0 G1 J2 K0 &5 &120 \\
   \midrule
  NU Orionis & 2017 Oct 13 &Medium &A0 G1 J2 K0 &5 &120  \\
  NU Orionis & 2018 Jan 10 &Medium &A0 G1 J2 K0 &5 &120  \\
  \midrule
  LP Ori & 2018 Jan 11 &Medium &A0 G1 J2 K0 &5 &120 \\
  \midrule
  Brun 862 & 2018 Jan 11 &Medium &A0 G1 J2 K0 & 5 &120 \\
  \midrule
  HD 37115 & 2018 Jan 11 &Medium &A0 G1 J2 K0 & 5 & 120 \\
  \midrule
  HD 37150 &2018 Jan 11 &Medium &A0 G1 J2 K0 & 5 &120 \\
  \midrule
  TCC 59 &2018 Jan 04 & Low &A0 G1 J2 K0 & 10 & 60 \\
  TCC 59 &2018 Jan 05 & Low &A0 G1 J2 K0  & 10 & 60 \\
  \midrule
  TCC 43 &2018 Jan 04 & Low &A0 G1 J2 K0  & 10 &30 \\
   
\bottomrule

\caption[GRAVITY observations of the Orion Nebula.]{{\small GRAVITY observations of the Orion Nebula. From left to right: name of the observed object, observation date, spectral resolution, baseline configuration, and the integration time (DIT) with the number of integrations (NDIT) }.}\label{tab:OriObs}

\end{longtable}
}
}
Data were reduced with the standard GRAVITY pipeline \citep{GRAVITYCollaboration2017}. The reduction algorithm follows the approach of \citet{2007A&A...464...29T} and creates a Pixel to Visibility Matrix (P2VM). The visibility is measured by combining the telescope beams with a relative phase shift of $0^{\circ}, 90 ^{\circ}, 180 ^{\circ}$, and $270 ^{\circ}$. Thus, we get four signals per baseline, resulting in $4 \cdot 6 = 24$ channels for the science object \citep{GRAVITYCollaboration2017}. The P2VM provides the phase relations, photometry, and coherence of the four incoming telescope beams and the 24 outgoing signals. A detailed description of the reduction is provided in \citet{2014SPIE.9146E..2DL} and an additional example of the use of GRAVITY to the study of massive multiple systems can be found in \citet{2017ApJ...845...57S}. 

For the instrument calibration, we also need to calibrate the wavelength of the fringe tracker and science channel, as well as to determine the dark field, bad pixels, and to compute the profile of the spectra with a flat field. Using the P2VM, we then compute real-time visibilities. Each file has several frames, one for each integration. The frames are averaged during reduction. 

For the calibration of the visibilities we observe point-like objects with a known diameter and which can be considered single stars. As we know the true shapes of the calibrator visibilities, we compute a visibility transfer function to adjust the measured visibilities to match the expected visibilities. This visibility transfer function is then applied to the visibilities of the science object.

\section{Data Analysis}\label{chap:DataAna}
In this section, we provide an overview of the modeling functions and tools used for data analysis. In the beginning, the observed data is fitted according to a binary model. For sufficiently well-sampled data, we are able to determine the orbital parameters of the binaries. In order to accurately determine the companion magnitude, we need to consider the effects of dust extinction. Finally, we provide a detection limit for our observations.

\subsection{Modeling a Binary Star}\label{sec:binary model}
We introduce the modeling functions that were used for analyzing the data. We assume a binary model and use it to fit the squared visibilities, closure phase and triple amplitude of our observational data. 
\paragraph{Visibility}\label{par:visibility}

Visibility of a binary model is described as:

\begin{equation}\label{eq:binary_model}
\nu _{\rm{bin}} = \dfrac{ \nu _{\rm{main}} + f \cdot \nu _{\rm{comp}}  \exp \left( -2i  \pi \dfrac{ u \cdot \Delta \alpha + v \cdot \Delta \delta }{ \lambda} \right) } {\left( 1 + f \right) },
\end{equation}
where $\nu_{\rm{main}}$ and $\nu_{\rm{comp}}$ are the complex visibilities for the primary and the companion star, respectively (see, for example, \citet{Lawson2000}). In the case of an unresolved star, $\nu_{\rm{main}} = \nu_{\rm{comp}} = 1$. The parameters $u$ and $v$ are the spatial frequencies of the telescope baselines; $\lambda$ is the observed wavelength; $\Delta \alpha $ and $\Delta \delta$ are the angular distances of the companion star from the primary star in R.A. and Dec., respectively; and $f = f_{\rm{comp}}/f_{\rm{main}}$ is the mean flux ratio over all wavelengths of the system, where $f_{\rm{comp}}$ and $f_{\rm{main}}$ are the flux of the companion star and primary star, respectively.

The parameters $u, v \text{, and}~ \lambda$ are provided by the observational data, while the distances $\Delta \alpha$ and $\Delta \delta$ together with the flux ratio $f$ are variable parameters to be fitted. To find starting values for these parameters, we use a grid-search algorithm. We scan $\Delta \alpha$ and $\Delta \delta$ between either $\pm$200~mas, $\pm$100~mas or $\pm$50~mas in steps of either 1 or 0.5~mas, and the flux ratio $f$ between 0 and 1 in steps of 0.1. The best result of the grid-search forms the starting values for the subsequent weighted least-squares optimization, using the Levenberg-Marquardt algorithm and the LMFIT package \citep{newville_2014_11813}. %The weights $\sigma$ are the measured uncertainties of the data.

% \begin{equation}\label{eq:chi2}
% \chi ^2 =\sum\limits_{i} \left( \dfrac{\nu_{Data,i} - \nu_{Model,i}}{\sigma _{Data,i}} \right) ^2
% \end{equation}

% An example for the observed visibility and the optimized model visibility is shown in Figure \ref{fig:vis_fit_example}. Data were taken on March 20th, 2017, the observed object was $\theta ^1$ Ori B.

% \begin{figure}
% \centering
% \includegraphics[width=0.8\textwidth]{"../../theta1_Ori_B1_2017-03-20T00:36:33_v2"}
% \caption[Visibility and fitted model function]{Observed visibilities (color) and the fitted model visibility (solid black line) for the observation of $\theta ^1$ Ori B at the 20th March 2017.}\label{fig:vis_fit_example}
% \end{figure}

\paragraph{Closure phase and triple amplitude}
The closure phase (CP) is computed by taking the argument of the bispectrum of the visibility function. The bispectrum is the triple product of complex visibilites for the telescopes $i,j,k$:
\begin{equation} 
CP(i,j,k) = \arg \left( \nu_{\rm{bin,ij}} \nu_{\rm{bin,jk}} \nu_{\rm{bin,ki}} \right)
\end{equation}
where the complex visibilites $\nu_{\rm{bin}}$ are computed using \cref{eq:binary_model}.

The triple amplitude is the modulus of the bispectrum. Most of the information is contained in the closure phase, thus, the triple amplitude is not always necessary for fitting the data.
The optimization is done in the same way as for the visibility model.% An example of the closure phase, triple amplitude, and the resulting fitted model is shown in \cref{fig:cp_t3_fit_example}. %The data were taken at the 20th March 2017 of $\theta ^1$ Ori B. 

% \begin{figure}
% \begin{minipage}{0.5\textwidth}
% \includegraphics[width=\textwidth]{"../../theta1_Ori_B1_2017-03-20T00:36:33_icp"}
% \end{minipage}
% \begin{minipage}{0.5\textwidth}
% \includegraphics[width=\textwidth]{"../../theta1_Ori_B1_2017-03-20T00:36:33_t3"}
% \end{minipage}
% \caption[Closure phase, triple amplitude and corresponding fit]{Left: observed closure phase and model fit. Right: observed triple amplitude and the optimized model function.}\label{fig:cp_t3_fit_example}
% \end{figure}

\subsection{Orbit Modeling}

For fitting the orbit, we use the KeplerEllipse class of PyAstronomy\footnote{https://github.com/sczesla/PyAstronomy} to determine the position at a given time for a set of orbital elements. This position is then compared with the positions obtained from the binary model fit (see \cref{sec:binary model}). The optimization is done with the Trust-region method, which supports boundaries of variables in contrast to a regular least-squares minimization.

The parameters of the orbit are defined as follows: $a$ is the semi-major axis of the Kepler ellipse, $P$ is the orbital period, $e$ the eccentricity, $\tau$ the time of periapsis passage, $\Omega$ the longitude of the ascending node, $\omega$ the argument of periastron, and $i$ the inclination of the orbit. The ascending node is defined as the point where the orbiting object passes the plane of reference in the direction of the observer. All parameters are in units of the respective initial guess. The semi-major axis $a$ is thus mostly reported in units of angular separation. For a more detailed discussion see e.g. \citet[pp. 22-24]{2005ormo.book.....R}.

As an observer, we see the 3D-orbit projected on a 2D-plane. The projection of different orbits can appear similar on the plane of reference, e.g. a highly inclined and eccentric orbit can appear as a not inclined circular orbit.

In order to not get stuck in a local minimum of $\chi ^2$, a good starting value is essential for optimization. If orbital elements are already determined in the literature, we take these elements as starting values for the optimization, e.g the values from \citet{Kraus2009} for the orbit of $\theta ^1$ Ori C$_2$ (see \cref{ssec:t1C}). If there are no previously determined orbital elements, we use the Basin-Hopping algorithm \citep{Wales1997}, to find a global minimum and use the result as starting value for the Trust-region method.

% \begin{figure}
% \centering
% \includegraphics[width=\textwidth]{./figures/thesis-figure8}
% \caption[Angular orbital parameter]{The angular orbital elements of a Kepler orbit. The plane of reference is in the observers' field of view.}\label{fig:orbital_elements}
% \end{figure}

\subsection{Dust Extinction}%
\label{sec:extinction}
For calculating the luminosity or absolute magnitude of a star, we need to consider its spectral type and extinction effects of the interstellar and circumstellar medium. $A(\lambda)$ is the extinction at wavelength $\lambda$ in magnitudes. The extinction of color is described by e.g. $E(B-V) = A(B) - A(V)$, with $B$ as the filter for $\simsym 440 \pm 90$~nm and $V$ as the filter for $\simsym 545 \pm 84$~nm. The interstellar reddening law is $R_{\rm{V}} = A(V) / E(B-V)$. For the diffuse interstellar medium, a typical value is $R_{\rm{V}} = 3.1$, whereas for dense clouds the value is $R_{\rm{V}} = 5$ \citep[p. 527]{allen2000allen}. In the K-band and for $R_{\rm{V}} = 3.1$, we get $A(K) = 0.108 A(V)$.%  \citep[p. 527, Table 21.6]{allen2000allen}. 

\subsection{Photometric Mass}
To get a mass estimate based on the magnitude, we use the isochrones in \citet[p. 151, p. 388, and references therein]{allen2000allen} and \citet[p. 130]{2005essp.book.....S}. With the first and second table, we get M$_{\rm{K}}$ for different spectral types and temperatures. With the effective temperatures in both tables, we link M$_{\rm{K}}$ to stellar masses using the third table. Now we can compare the values for M$_{\rm{K}}$ from the table to M$_{\rm{K}}(Star)$ and get an estimate of mass and spectral type. The tables are for class~V stars, meaning hydrogen-burning main sequence stars. For pre-main sequence (PMS) stars, the values are not accurate, merely representing a rough estimation.

\subsection{Companion Detection Limits}\label{sec:candid}
We use CANDID \citep{Gallenne2015} for determining detection limits of companion stars. CANDID is a Python tool which looks for high contrast companions. It provides two methods for computing limits.% \citep[described in][]{Gallenne2015}). 

The first method follows the approach of \citet{2011A&A...535A..68A}. \citet{2011A&A...535A..68A} inserted a binary model at different positions $(\alpha , \beta )$. They then compared the probability of the binary model with the probability of a uniform disk model, assuming the uniform disk is the true model. 

\begin{figure*}
\includegraphics[width=\textwidth]{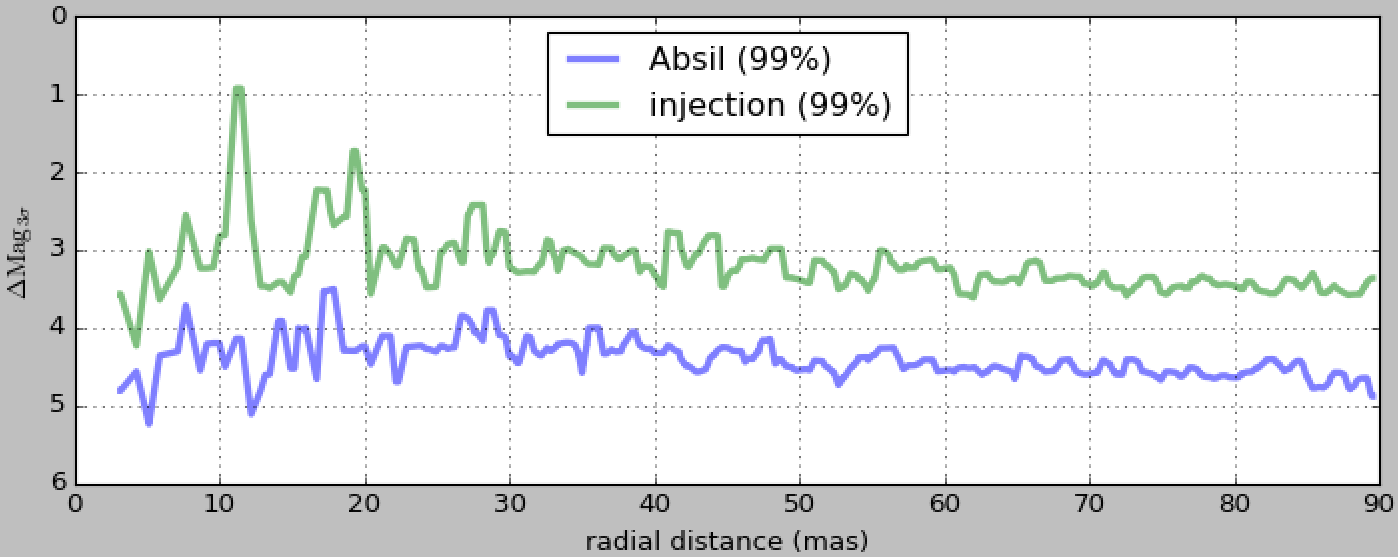}
\caption{Example for the companion detection limit for $\theta ^1$ Ori B, observed at the 12th October 2017, as determined with CANDID. The detection limit on the y-axis is denoted as a magnitude difference to the main star. The x-axis shows the separation in mas.}
\label{fig:ex_limit}
\end{figure*}

The second approach changes the null hypothesis. They injected companions at different positions with different flux ratios. Then, they determined the probability of the binary model being the true model, compared with the model of a uniform disk. In other words, the first approach tries to reconstruct a uniform disk model from binary data. The second approach tries to reconstruct a binary model from binary data. The second method yields more conservative results, which is why we choose it to determine our detection limits. Figure \ref{fig:ex_limit} shows an example of the companion detection limit for $\theta ^1$ Ori C. We take the worst limit as our companion detection limit.

\section{The Orion Nebula Cluster M42}\label{chap:ONCstars}
One of the closest active star forming regions is the Orion Nebula Cluster (ONC), at a distance of 414 $\pm$ 7 \citep{Menten2007, 2014ApJ...783..130R}. The ONC is located in a giant molecular cloud, at the sword of Orion. %It is an HII region, meaning that its hydrogen gas is ionized. 
A young star cluster (younger than 1~Myr) is located in the center of the Nebula -- see e.g. \citet{Muench2008}. This central region is called the ``Orion Trapezium Cluster'' (OTC) or $\theta ^1$ Orionis. The OTC is dominated by $\theta ^1$ Ori C, a young O-star with $\simsym 34$~\solmass . The radiation and outflow of $\theta ^1$ Ori C caused the ionization of its vicinity. The $\ion{H}{ii}$ region expanded into the surrounding molecular cloud and dissolved the molecular gas in which the young stars had been born. This process exposed large parts of the embedded star clusters and created the ONC. 

\subsection{Orion Trapezium Cluster Stars} 
In this study, we concentrate on the most luminous stars of the ONC. In the Trapezium Cluster, we observe stars with apparent K-magnitudes ranging from 4.57 to 8.38. In the following, we summarize previous results and present our findings for each of the observed targets.

\subsubsection{$\theta ^1$ Ori A (HD 37020, Brun 587, TCC 45, Parenago 1865)}\label{ssec:t1OriA}
$\theta ^1$ Ori A$_1$ is a B0.5-type star \citep{Levato1976, Simon-Diaz2006} with a K-band magnitude $m_{\mathrm{K}}$~=~5.67 \citep{2003yCat.2246....0C}. \citet{Hillenbrand1997} found a mass of 18.91~\solmass\  and an extinction of $A_{\rm{V}} = 1.89$~mag. \citet{Weigelt1999} calculated a mass of 20~\solmass\  and \citet{Schertl2003} assumed a mass of $16~M_{\odot}$, whereas \citet{Simon-Diaz2006} found a mass of $14 \pm 5~M_{\odot}$ and a radius of $6.3 \pm 0.9 R_{\odot}$.

\citet{1975IBVS..988....1L} found an eclipsing binary with a period of 65.43~days \citep{1976IAUC.3004....1M}, which we will refer to as A$_{3}$. \citet{Abt1991} derived a period of $65.09 \pm 0.07$~days from their measured radial velocities for $\theta^1$ Ori A$_3$. \citet{1989A&A...222..117B} concluded that the thermal spectrum and features correspond to a T~Tauri companion with a mass between 2.5 and 2.7~\solmass\ at a separation of 0.71~AU. However, \citet{Vitrichenko2001} determined a greater distance of $0.93 \pm 0.07$~AU.

\citet{Petr1998} discovered a third companion (A$_{2}$) at a separation of $\simsym 200$~mas, which corresponds to a projected distance of $\simsym 90$\textendash $100$~AU \citep[see e.g.][]{Weigelt1999,Preibisch1999,Close2012}. 
%If the orbit has a high inclination, the actual separation rises up to 500~AU \citep{Grellmann2013}.
\citet{Schertl2003} determined a mass of 4~\solmass\ for $\theta ^1$~Ori~A$_2$ and suggested a period of $P \sim 214$~yr. $\theta ^1$~Ori~A$_2$ is an F-type star extincted by $A_{\rm{V}} \sim 3.8$~mag \citep{Schertl2003}.

\begin{figure*}
\centering
\includegraphics[width=\textwidth]{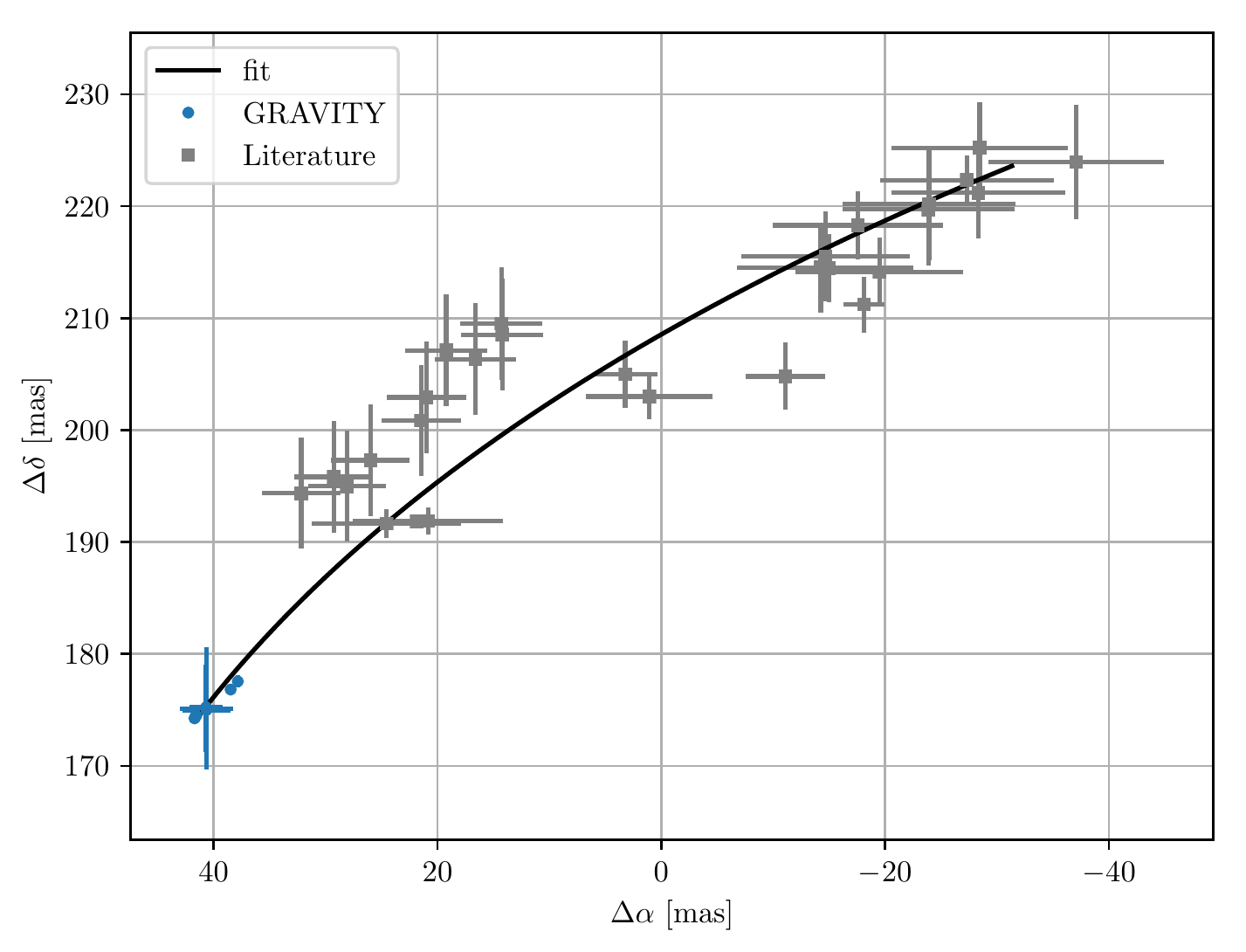}
\caption[Positions of $\theta ^1$ Ori A]{Measured positions of $\theta ^1$ Ori A$_2$, with east to the left. The primary star $\theta ^1$ Ori A$_1$ is located at position (0,0). Blue dotted positions were observed with GRAVITY. For some of the dots, the errorbar is smaller than the symbol displayed. Grey/square positions are taken from \citet{Close2012}, \citet{Schertl2003}, \citet{Petr1998}, \citet{Weigelt1999}, \citet{Balega2004}, \citet{Balega2007}, and \citet{Grellmann2013}.}\label{fig:t1Atrack}
\end{figure*}

Until now, it has not been entirely clear whether A$_2$ is gravitationally bound to $\theta ^1$ Ori A$_{1,3}$. We used GRAVITY data taken between November 2015 and January 2018 (Table \ref{tab:OriObs}) to get precise separation vectors. Our position measurements show an acceleration towards the primary star, proving that the system is gravitationally bound. In Figure \ref{fig:t1Atrack} one can see a motion of the companion star towards the main object. We use only the position measurements, because the spectral resolution is too low for precise radial velocity measurements. From the measured flux ratio $f \sim 0.23 \pm 0.05$ we infer a magnitude of $m_{\rm{K}} = 7.3 \pm 0.3$.

% where the indices correspond to the primary and secondary object, m is the magnitude of an object in K-band with m$_1 = m_{A_1} = 5.668$, $F$ is the total flux of an object, $L$ the respective luminosity and $r$ the distance to the object with $r_1 = r_2 = 414 \pm 7$~pc \citep{Menten2007, 2014ApJ...783..130R}.

\subsubsection{$\theta ^1$ Ori B (HD 37021, Brun 595, TCC 56, TCC 60, Parenago 1863)}\label{ssec:t1B}
$\theta ^1$ Ori B consists of at least six hierarchical components. 
$\theta ^1$ Ori B$_1$ is a B1V-type star \citep{1998AJ....115..821M} with a magnitude $m_{\rm{K}} = 6.00$ \citep{2003yCat.2246....0C}. \citet{Hillenbrand1997} determined a mass of $7.18$~ \solmass\  for an extinction of $A_{\rm{V}} = 0.49$. \citet{Weigelt1999} estimated a consistent mass of $m = 7$~\solmass . 

\begin{figure*}
\includegraphics[width=\textwidth]{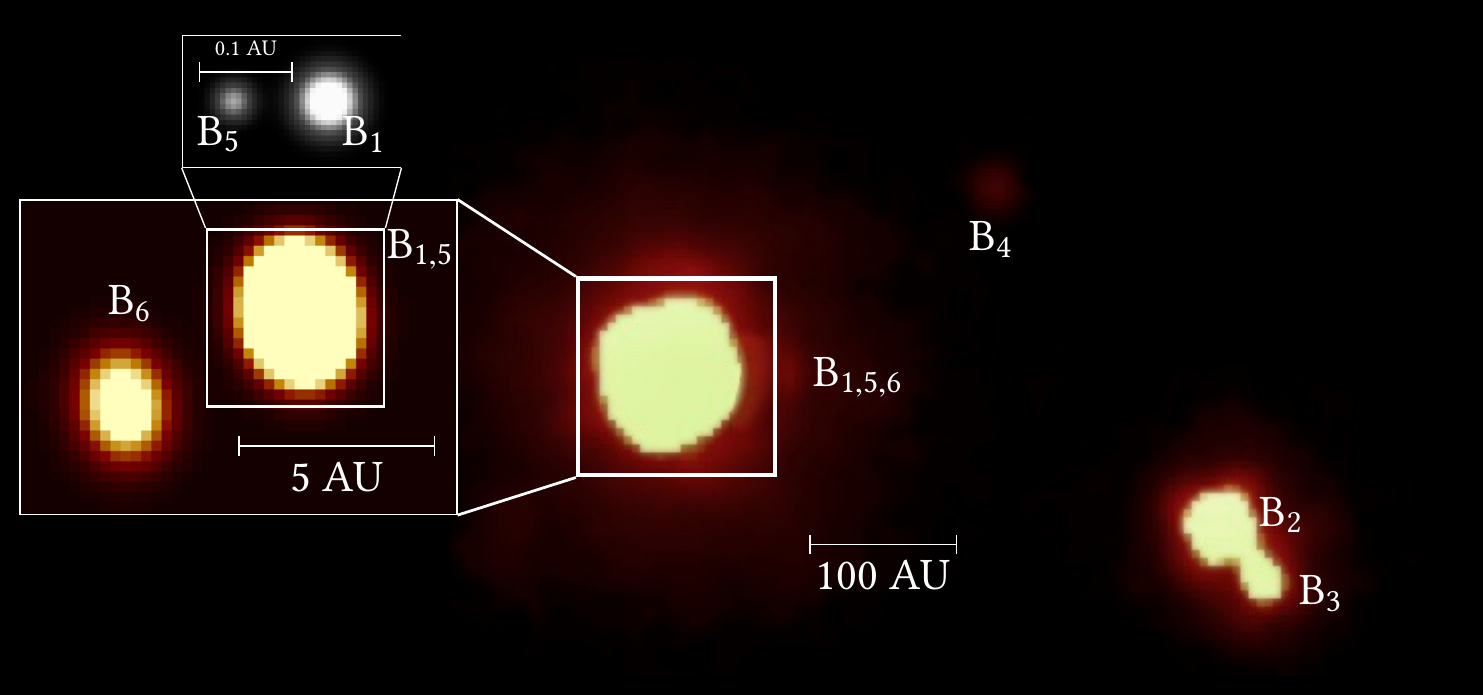}
\caption[The $\theta ^1$ Ori B system]{Main image: $\theta ^1$ Ori B group imaged in H-band. B$_1$ is an eclipsing binary B$_{1,5}$. It was created with NaCo data, based on data obtained from the ESO Science Archive Facility under request number 342335, ESO programme 60.A-9800(J). Additionally, with GRAVITY we detected another companion B$_6$ at a separation of $\simsym 13$~mas, shown in the zoomed image of B$_{1,5,6}$ (K-band). The image of B$_6$ orbiting B$_{1,5}$ is reconstructed from our observations. The zoom into the spectroscopic binary B$_{1,5}$ is only a representative image and was not created with observational data.}\label{fig:t1B_cluster}
\end{figure*}

\citet{Petr1998} found a visual companion at a separation of approximately \ang{;;1}, corresponding to 415~AU projected distance \citep{Close2012} at 450~pc. Taking the $414 \pm 7$~pc from \citet{Menten2007, 2014ApJ...783..130R}, the projected separation becomes $382 \pm 6$~AU. This companion itself is a resolved binary ($\theta ^1$ Ori B$_{2,3}$, see \cref{fig:t1B_cluster}) with $49 \pm 1$~AU projected separation and  a period of $\simsym 200$~yr \citep{Close2013}. Their K-magnitudes are estimated to be 7.6 and 8.6 \citep{Close2012}. According to \citet{Close2013}, B$_{2,3}$ has a system mass of $\simsym 5.5$~\solmass , which is in the same order of magnitude as the masses determined by \citet{Schertl2003} with $m_{B_2} = 4$~\solmass , and $m_{B_3} = 3$~\solmass . However, it differs from the values found by \citet{Preibisch1999}, which are $m_{B_2} = 1.6$~\solmass , and $m_{\rm{B_3}} = 0.7~M_{\odot}$. The inferred orbital period of B$_{2,3}$ around the main star depends on the inclination of the orbit. For a less inclined orbit, \citet{Close2012} determined a period of $P \sim 1920$~yr, whereas in \citet{Close2013} a highly inclined orbit was assumed and resulted in a period of $P \sim 11~000$~yr with an absolute separation of $\simsym 820 \pm 14$~AU.
%The primary is dominated by B2,3 (Lada et al 2004, Smith et al 2005b)

Located \ang{;;0.6} $\simsym 248 \pm 4$~AU north-west of B$_1$ is another faint companion (\cref{fig:t1B_cluster}), $\theta ^1$ Ori B$_4$ \citep{Simon1999} with $m_{\rm{K}} = 11.66$ \citep{Close2012}. \citet{Close2013} determined a period of $\simsym 2000 \pm 700$~yr. \citet{Preibisch1999} estimated the mass of B4 to be $m_{\rm{B_4}}$ = 0.2~\solmass\ and sets an upper mass limit of $< 2$~\solmass . In contrast, we estimate the mass of B$_4$ to be $m_{\rm{B_4}} \sim 1 ~M_{\odot}$, which is consistent with the limit of \citet{Preibisch1999}.

Considering an extinction of $A_{\rm{V}} = 0.49$ \citep{Hillenbrand1997}, we calculate an absolute magnitude M$_{K}(\rm{B_4}) = 3.57$ and compare the magnitude with the isochrones. %tables from \citet[p. 150, p. 388]{allen2000allen} and \citet[p. 130]{2005essp.book.....S}. 
This yields the mass estimate of $1$~\solmass . Because of its low mass, B$_4$ may be ejected from the system at a certain point, but appears temporarily stable \citep{Close2013}.

\citet{1921AN....212..383H} and \citet{1948AN....276..144S} found $\theta ^1$~Ori~B$_1$ to be an eclipsing binary ($\theta ^1$~Ori~B$_{1,5}$) with a 6.47~day period \citep{Abt1991}. \citet{Popper1976} determined $\theta ^1$~Ori~B$_5$ to be a late A-type star and a mass ratio of $q = m_{\rm{B_5}}/m_{\rm{B_1}} \sim 0.3$, which leads to $m_{\rm{B_5}} \sim 2$~\solmass. On the other hand, \citet{Close2003,Close2012,Close2013} assumed a mass of 7~\solmass\  for B$_5$, but did not justify their assumption. \citet{Close2012} determined a separation of 0.13~AU assuming a distance to the OTC of 450~pc. With the distance of $414 \pm 7$~pc determined by \citet{Menten2007, 2014ApJ...783..130R}, this converts to a separation of $0.120 \pm 0.002$~AU. 

\citet{2006Ap.....49...96V} claim the detection of a late type companion based on radial velocity anomalies. Further observations are needed to verify the detection.

With GRAVITY, we detect a previously unknown companion B$_6$ at separations between 8.5\textendash 17.2~mas, corresponding to a projected distance between $3.52 \pm 0.05$~AU and $7.12 \pm 0.12$~AU. The average flux ratio is $0.31 \pm 0.06$ and corresponds to an apparent K-magnitude of m$_{\rm{B_6}} = 7.3 \pm 0.5$.
Considering the absorption, the absolute K-magnitude is M$_{K}(\rm{B_6}) = -0.84 \pm 0.6$. The comparison with isochrones from \citet[p. 150, p. 388]{allen2000allen} and \citet[p. 130]{2005essp.book.....S} yields a mass $m_{\rm{B_6}} \sim 4$\textendash $6$~\solmass , which corresponds to a B-type star. 
\citet{Vasileiskii2000} suggested a close B-type companion to explain a secondary minimum observed in the eclipse of B$_1$. This supports our spectral classification of B$_6$.

\begin{figure*}
\centering
\includegraphics[width=\textwidth]{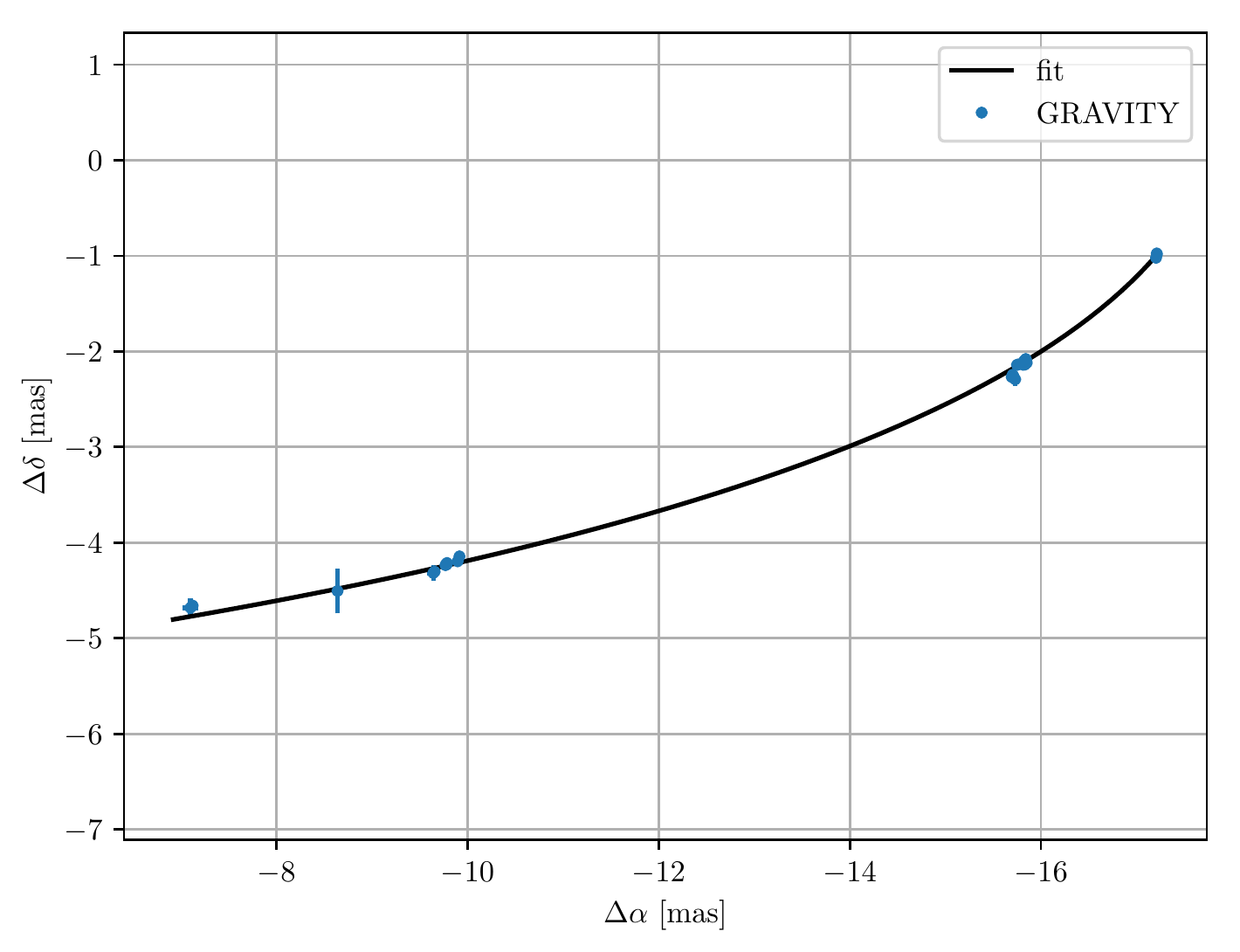}
\caption[Positons of $\theta ^1$ Ori B$_6$]{Positions of B$_6$ between January 2017 and January 2018. We see orbital motion around B$_{1,5}$ at (0,0), moving from east (left) to west (right).} \label{fig:orib_pos}
\end{figure*}

\begin{figure}
\centering
\begin{minipage}{0.49\textwidth}
\includegraphics[width=\textwidth]{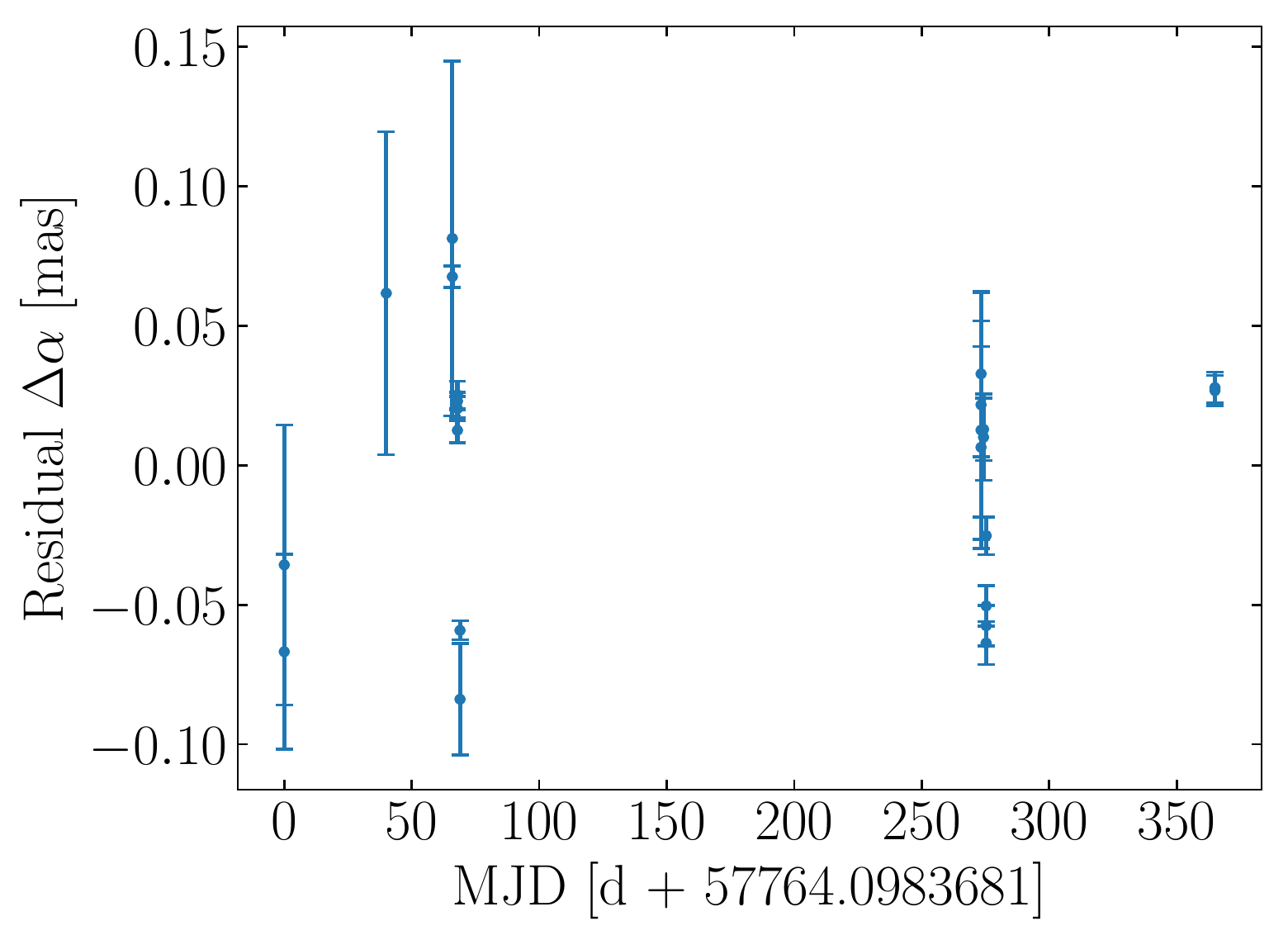}
\end{minipage}
\hfill
\begin{minipage}{0.49\textwidth}
\includegraphics[width=\textwidth]{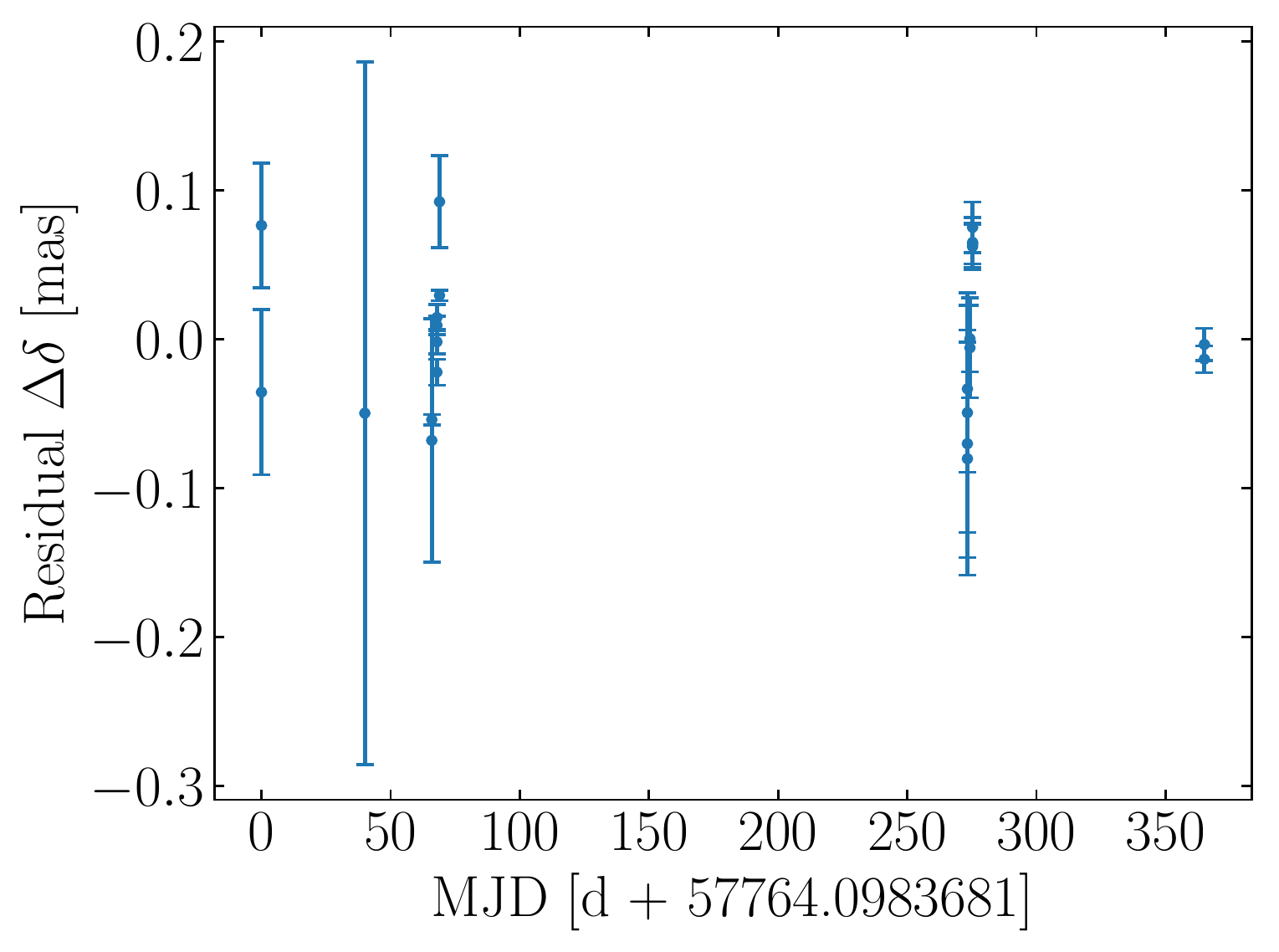}
\end{minipage}
\caption[Residuals of $\theta ^1$ Ori B best fit]{Residuals of the $\Delta \alpha$ (left) and $\Delta \delta$ (right) positions from the best fit. The RMS is $0.05$~mas in $\delta$ and $0.06$~mas in $\alpha$.}\label{fig:oriB_res}
\end{figure}

$\theta ^1$ Ori B$_6$ was observed between January 2017 and January 2018 (see Table \ref{tab:OriObs}). The position of the star relative to $\theta ^1$ Ori B$_{1,5}$ is presented in Figure \ref{fig:orib_pos} and shows orbital motion. For the determination of orbital parameters further observations are needed. We approximate the orbital path by fitting values for $\Delta \alpha$ and $\Delta \delta$ over time with a quadratic function. The resulting path is the black line in Figure \ref{fig:orib_pos}. The residuals to this fit scatter by a root mean square (RMS) of $0.05$~mas in $\delta$ and $0.06$~mas in $\alpha$ as can be seen in Figure \ref{fig:oriB_res}.

With CANDID \citep[see \cref{sec:candid}]{Gallenne2015}, we can exclude further companions at a $3\sigma$ level with $\Delta m < 3.5$, thus a mass $> 1.9$~\solmass\ for separations of $1.7$\textendash $8.3$~AU. For the range of $8.3$\textendash $16.6$~AU, the limit is 5~mag~($\approx 1.5$~\solmass ) and for $16.6$\textendash $46.8$~AU, we can exclude companions with $\Delta m = 5.2 ~(\approx 1.1$~\solmass ).

\subsubsection{$\theta ^1$ Ori C (HD 37022, Brun 598, TCC 68, Parenago 1891)}\label{ssec:t1C}
$\theta ^1$ Ori C$_1$ is a O7V-type star \citep{2011ApJS..193...24S} and the brightest and most massive member of the Trapezium Cluster with $m_{\rm{K}} = 4.57$ \citep{2002yCat.2237....0D}. Furthermore, it is one out of few O-stars with detected magnetic fields \citep{Stahl1996, 2002MNRAS.333...55D, doi:10.1093/mnras/stw2743}. \citet{1993A&A...274L..29S} discovered variations in the spectrum with a 15.43 day period. This variation also appears in X-ray, radial velocities, and magnetic fields, discussed by, for example, \citet{Stahl2008}, \citet{Wade2006}, \citet{Simon-Diaz2006}, and references therein. The magnetic field direction does not match the spin axis, which is an indication that $\theta ^1$ Ori C$_1$ was formed in a collision process \citep{Zinnecker2007}. 

\citet{Weigelt1999} discovered a close visual companion C$_2$ at a separation of 33~mas. \citet{Kraus2009} determined the orbital parameters in \cref{tab:orbitC} and a resulting mass ratio of $q(414 \text{pc}) = m_{\rm{C_2}} / m_{\rm{C_1}} = 0.23 \pm 0.05$. They estimate a total system mass of $44 \pm 7~M_{\odot}$ and a dynamical distance of $410 \pm 20$~pc.

\begin{table*}
\centering
 \begin{tabular}{cccc}
 \toprule
 & \citet{Kraus2009} & \citet{2015ASPC..494...57B} & This work  \\
 \midrule
 a [mas] &$43.61 \pm 3$ &$45 \pm 3$ & $45 \pm 2$\\
 $P$ [yr]& $11.26 \pm 0.5$ &$11.28 \pm 0.02$ & $11.4 \pm 0.2$\\
 e & $0.592 \pm 0.07$ &$0.59 \pm 0.01$ & $0.59 \pm 0.04$ \\
 $\tau$ & $2002.57 \pm 0.5$ &$2002.59 \pm 0.02$ & $2002.2 \pm 0.2$ \\
 $\Omega$ [$^{\circ}$]& $26.5 \pm 1.7$ & $28.3 \pm 0.3$ &$27.9 \pm 0.7$ \\
 $\omega$ [$^{\circ}$] & $285.8 \pm 8.5$ &$286.1 \pm 0.2$ & $283 \pm 2$ \\
 i [$^{\circ}$]&$99.0 \pm 2.6$ &$98.9 \pm 0.4$ & $98.6 \pm 0.6$ \\
    \bottomrule
 \end{tabular}
 \caption[Orbital parameters for $\theta ^1$ Ori C]{Orbital parameter for $\theta ^1$~Ori~C determined by \citet{Kraus2009}, \citet{2015ASPC..494...57B} and in this work including GRAVITY data. a is the semi-major axis in mas (44~mas = $18.2 \pm 0.3$~AU), $P$ the period in years, e the eccentricity, $\tau$ the time of the periastron passage, $\Omega$ the longitude of the ascending node, $\omega$ the argument of periapsis, and i the inclination of the orbit. The results of \citet{Kraus2009} and this work agree within the error bars. The results of \citet{2015ASPC..494...57B} differ in $\tau$ and $\omega$ with this work.
}\label{tab:orbitC}
\end{table*}
 
Using the calibration models from \citet{Martins2005}, \citet{Kraus2007} derived a  
mass of $m_{\rm{C_1}} = 34~M_{\odot}$, an effective temperature $T_{\rm{eff,C_1}} = 39~900$~K, and $\log L_{\rm{C_1}} / L_{\odot} = 5.41$. The resulting parameters for the companion star are $m_{\rm{C_2}} = 15.5~M_{\odot}$, $T_{\rm{eff,C_2}} = 31 900$~K and $\log L_{\rm{C_2}} / L_{\odot} = 4.68$, thus implying a O9.5-type star. The temperatures are in accordance with \citet{Simon-Diaz2006}, who found temperatures of $T_{\rm{eff,C_1}} = 39~000 \pm 1000$~K and derived a stellar radius of $R_{\rm{C_1}} = 10.6 \pm 1.5~R_{\odot}$. The spectroscopic mass of 45~\solmass\ and evolutionary mass of 33~\solmass\ for $\theta ^1$ Ori C$_1$ by \citet{Simon-Diaz2006} differ. \citet{1992A&A...261..209H} described a discrepancy between spectroscopic masses and masses determined using an evolutionary model. \citet{2017ApJ...847..120H} pointed out that spectroscopic masses are sensitive to $\log g$ and that an error of 0.1 \citep[as in][]{Simon-Diaz2006} in $\log g$ translates to a factor of $10^{0.1} \approx 1.26$, or an uncertainty of $126 \%$ for spectroscopic masses. \citet{2015ASPC..494...57B} added spectroscopic observations to the previous data and determined a total system mass of $45.5 \pm 10$~\solmass, a mass ratio of $q = 0.36 \pm 0.05$ and derived $m_{\rm{C_1}} = 33.5 \pm 5.2$~\solmass\ and $m_{\rm{C_2}} = 12 \pm 3$~\solmass . The separation of $44 \pm 3$~mas corresponds to $18.2 \pm 1.2$~AU.

\begin{figure*}
\centering
\includegraphics[width=\textwidth]{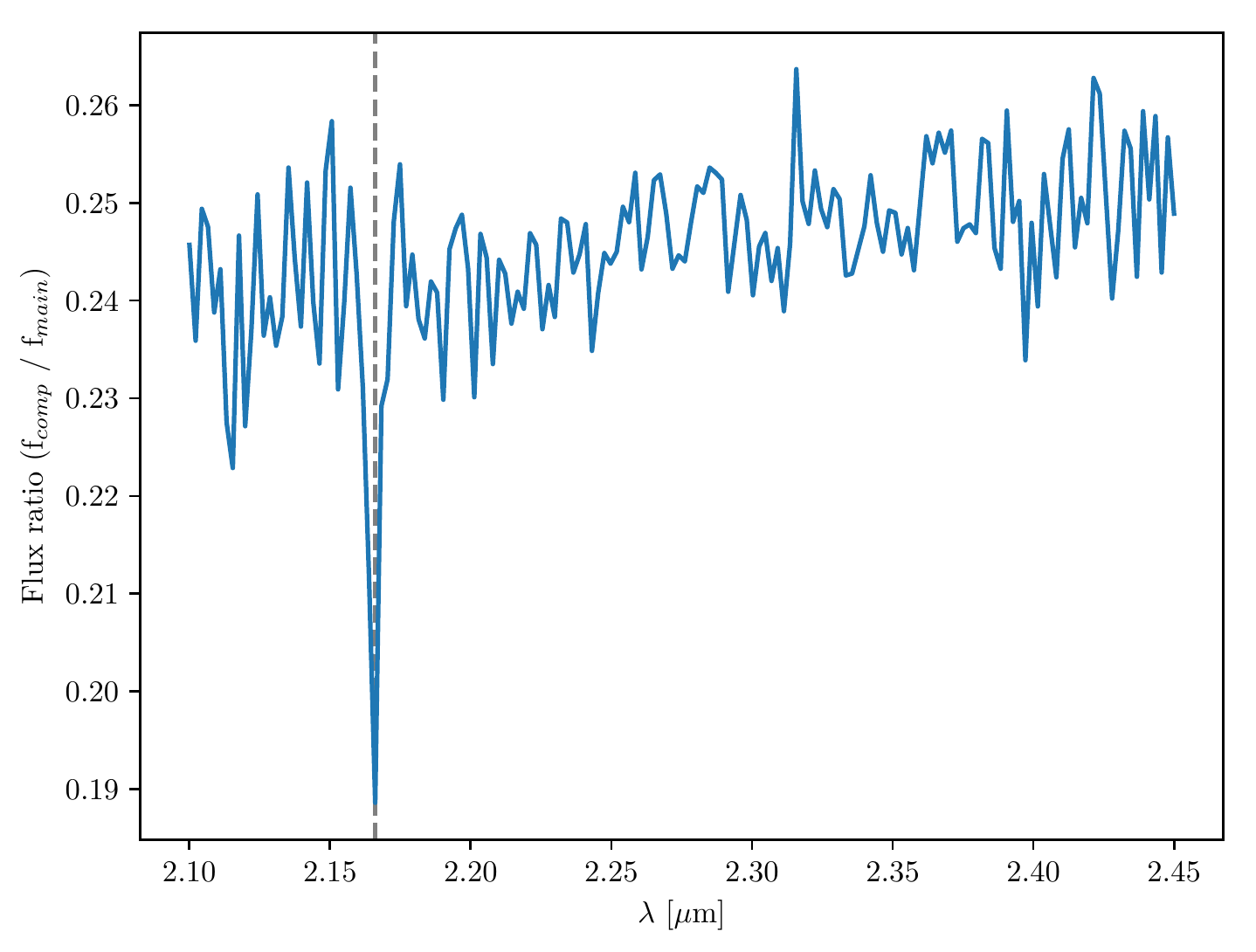}
\caption[Flux ratio versus wavelength of $\theta ^1$ Ori C]{Flux ratio $f = f_{\rm{C_2}} / f_{\rm{C_1}}$ as a function of observed wavelength. The vertical dashed grey line is at 2.166 microns, the Br-$\gamma$ line. The drop indicates that C$_1$ has a much higher flux at that wavelength than C$_2$. Data were observed at January 9th 2016.}\label{fig:OriC_spectrum}
\end{figure*}

\begin{figure*}
\centering
\includegraphics[width=\textwidth]{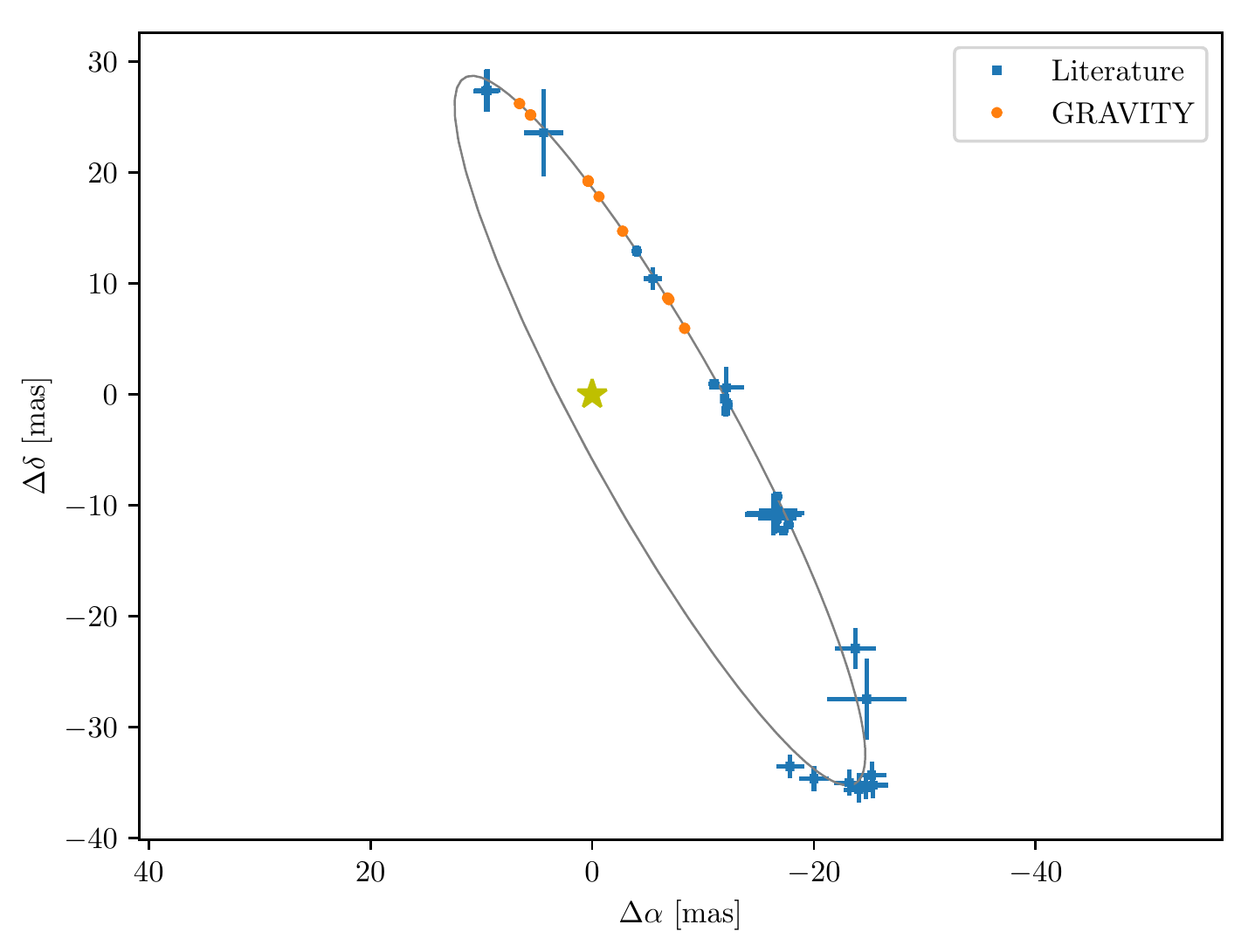}
\caption[Orbit of $\theta ^1$ Ori C$_2$]{Orbit of $\theta ^1$ Ori C$_2$. Orange dots are observed with GRAVITY, blue squares are positions taken from \citet{Weigelt1999}, \citet{Schertl2003}, \citet{Kraus2007}, \citet{2008ApJ...674L..97P}, \citet{Kraus2009}, and \citet{Grellmann2013}. The error bars of GRAVITY data are within the marker. The orbital parameters are listed in \cref{tab:orbitC}.}\label{fig:OriC_Orbit}
\end{figure*}

GRAVITY observations from November 2015 to January 2018 (for a list of observations see Table \ref{tab:OriObs}) show a variation of the flux ratio $f = f_{\rm{C_2}} / f_{\rm{C_1}}$ between 0.18 and 0.36, which points to a non-constant brightness of either the primary or the companion star or both stars. We compute an average K-magnitude of $6.0 \pm 0.4$. The extinction is $A_{\rm{V}} = 1.74$ \citep{Hillenbrand1997} and using the method described in the previous section, we infer a spectral type of B1 or younger and thus a stellar mass $> 10$~\solmass\ \citep[p. 389]{allen2000allen}. Looking at the flux ratio $f$ depending on wavelength (\cref{fig:OriC_spectrum}), we notice a drop at 2166~nm --- the Br-$\gamma$ line. This points to an absorption of C$_2$ in the Br-$\gamma$ line.
%Thus, the flux of C$_1$ is much stronger in the Br-$\gamma$ line than the flux of C$_2$. 
For the given example in Figure \ref{fig:OriC_spectrum}, C$_2$ is 1.3 times fainter at the Br-$\gamma$ line than at other wavelengths. With the new GRAVITY data, we fit the orbit of C$_2$ (see Figure \ref{fig:OriC_Orbit}) and find that our results agree with the parameters from \citet{Kraus2009}. \cref{tab:orbitC} presents both outcomes, and the orbital elements determined by \citet{2015ASPC..494...57B}.

\begin{figure}
\centering
\begin{minipage}{0.49\textwidth}
\includegraphics[width=\textwidth]{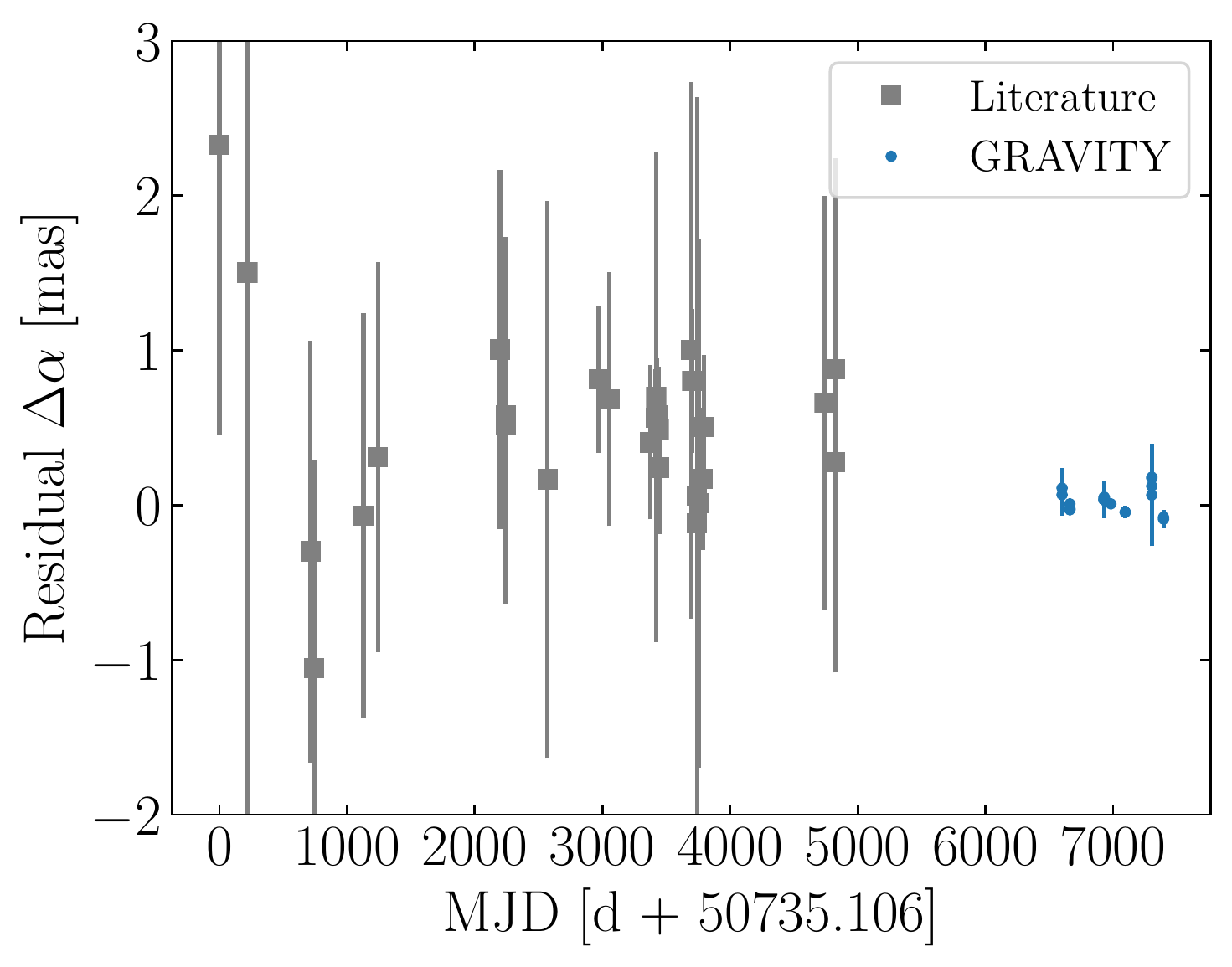}
\end{minipage}
\hfill
\begin{minipage}{0.49\textwidth}
\includegraphics[width=\textwidth]{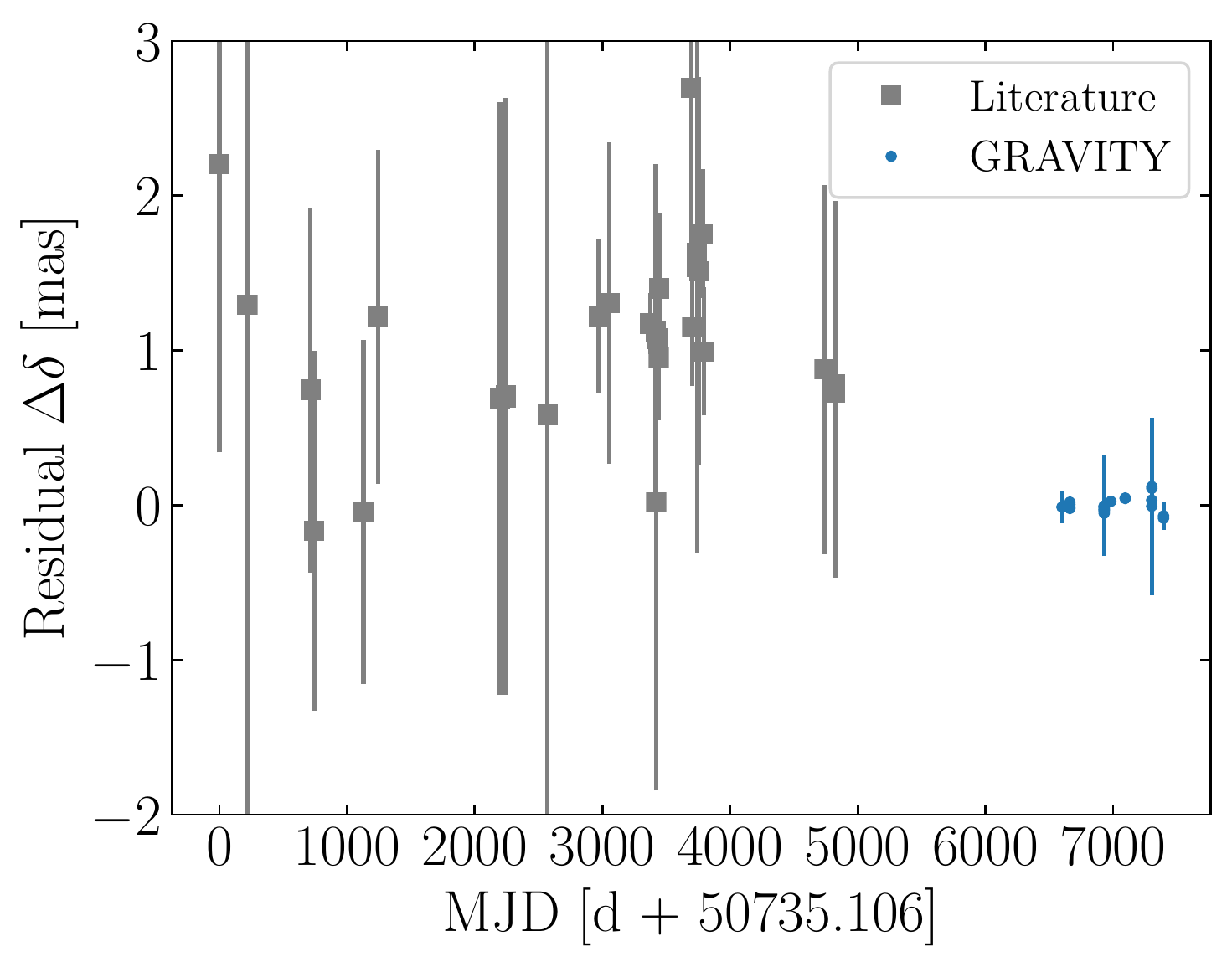}
\end{minipage}
\caption[Residuals of $\theta ^1$ Ori C$_2$ best fit]{Residuals of the $\Delta \alpha$ (left) and $\Delta \delta$ (right) positions from the orbit of $\theta ^1$ Ori C$_2$. The blue circles represent the residuals of GRAVITY data and their respective uncertainties. The grey squares are the residuals from non-GRAVITY observations. The RMS of the GRAVITY residuals is $0.07$~mas for $\Delta \alpha$, and $0.05$~mas for $\Delta \delta$.}\label{fig:oriC_res}
\end{figure}

\citet{Vitrichenko2002a} and \citet{Lehmann2010} found another spectroscopic companion C$_3$ with a period of 61.5 days, resulting in an estimated separation of $\simsym 1$~mas. They derived masses of 31~\solmass\ for C$_1$, 12~\solmass\ for C$_2$, and $1.0 \pm 0.2$~\solmass\  for C$_3$.
For a primary star with $33~M_{\odot}$ and a companion star with $1~M_{\odot}$, we expect a reflex motion of $\simsym 60~\mu$as for C$_1$. To this point, we have not detected such wobbling, probably because the position scattering was too large. The residuals from the orbit are shown in \cref{fig:oriC_res}. The RMS of the residuals is $0.07$~mas for $\Delta \alpha$ and $0.05$~mas for $\Delta \delta$.

With CANDID, we set a limit on further companions at a $3\sigma$ level. For separations of $1.70$\textendash $8.3$~AU, we compute $\Delta m < 3.2 ~(\approx 3$~\solmass ). For the range of $8.3$\textendash $46.8$~AU, we can exclude companions with $\Delta m < 4.2 ~(\approx 2.2$~\solmass ).

\subsubsection{$\theta ^1$ Ori D (HD 37023, Brun 612, Parenago 1889)}
$\theta ^1$ Ori D is a pre-main sequence B1.5V-type star \citep{2006MNRAS.371..252L} with a K-magnitude of 5.75 \citep{2003yCat.2246....0C}. \citet{Hillenbrand1997} found a mass of 16.6~\solmass, whereas \citet{Simon-Diaz2006} derived a mass of $18 \pm 6~M_{\odot}$. \citet{2006MNRAS.371..252L} found a mass of $11 \pm 1~M_{\odot}$ and \citet{2010A&A...520A..51V} determined a mass of 17.7~\solmass\ using rotating stellar models. 

\citet{Close2012} found a wide visual companion D$_2$ at a distance of $\simsym 1.4 '' = 580 \pm 10$~AU. Currently it is not clear if D$_2$ is physically bound to D$_1$. In order to estimate the mass of the companion, we used archival imaging data\footnote{Based on data obtained from the ESO Science Archive Facility under request number 338322, ESO programme 274.C-5036(A).} of the Trapezium. We retrieved a NaCo image from 2005 and the corresponding calibration files. After calibrating the image with the NaCo reduction pipeline, we extracted the total flux of D$_2$ and Ori F as a magnitude reference and compared the flux D$_2$ with the flux of $\theta ^1$ Ori F. With the flux ratio, we determine a magnitude of $m_{\rm{K}} = 11.69 \pm 0.06$. We assume that the extinction is comparable with the value for the primary \citep[$A_{\rm{V}} = 1.79$][]{Hillenbrand1997}) and get $M_{\rm{K}} = 3.4 \pm 0.1$. This corresponds to a mass of $\simsym 1 \pm 0.1$~\solmass .

A spectroscopic companion with a period of either 20.25 or 40.5~days was claimed by \citet{Vitrichenko2002b}.
Another indication for a companion at a separation of 18.4 mas ($\approx 7.6 \pm 0.2$~AU) and with a flux ratio of 0.14 was suggested by \citet{Kraus2007}. \citet{Grellmann2013} found indications for a structure at 2~mas or 4~mas, which is consistent with a close companion, but could not provide further constraints due to large uncertainties.

\begin{figure*}
\centering
\includegraphics[width=\textwidth]{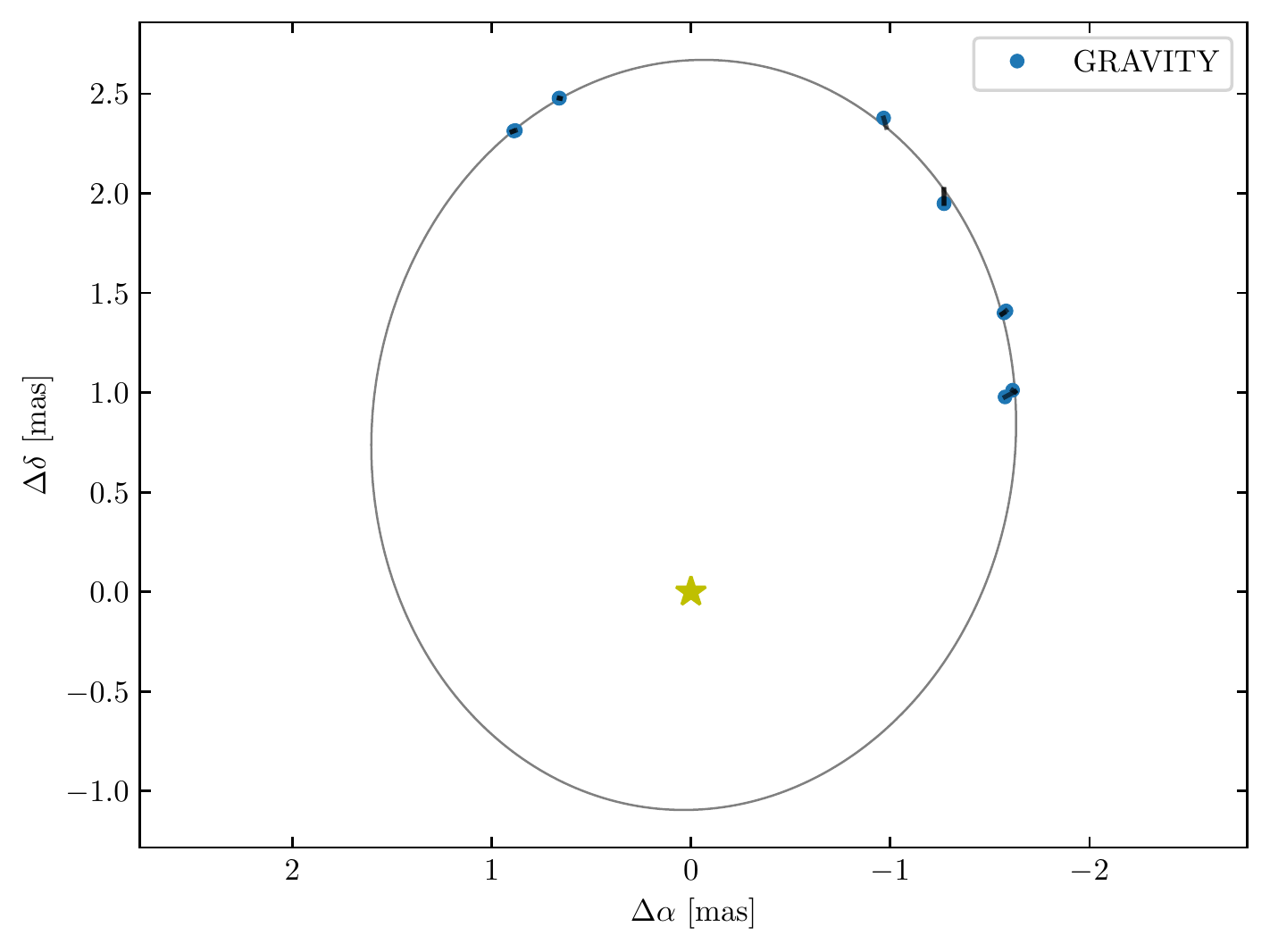}
\caption[Orbit of $\theta ^1$ Ori D$_3$]{Positions of the newly detected $\theta ^1$ Ori D$_3$ around the primary D$_1$ at (0,0). The orbital parameters are listed in \cref{tab:oriD_orbit}.}\label{fig:oriD_orbit}
\end{figure*}

\begin{figure}[tbh]
\centering
\begin{minipage}{0.49\textwidth}
\includegraphics[width=\textwidth]{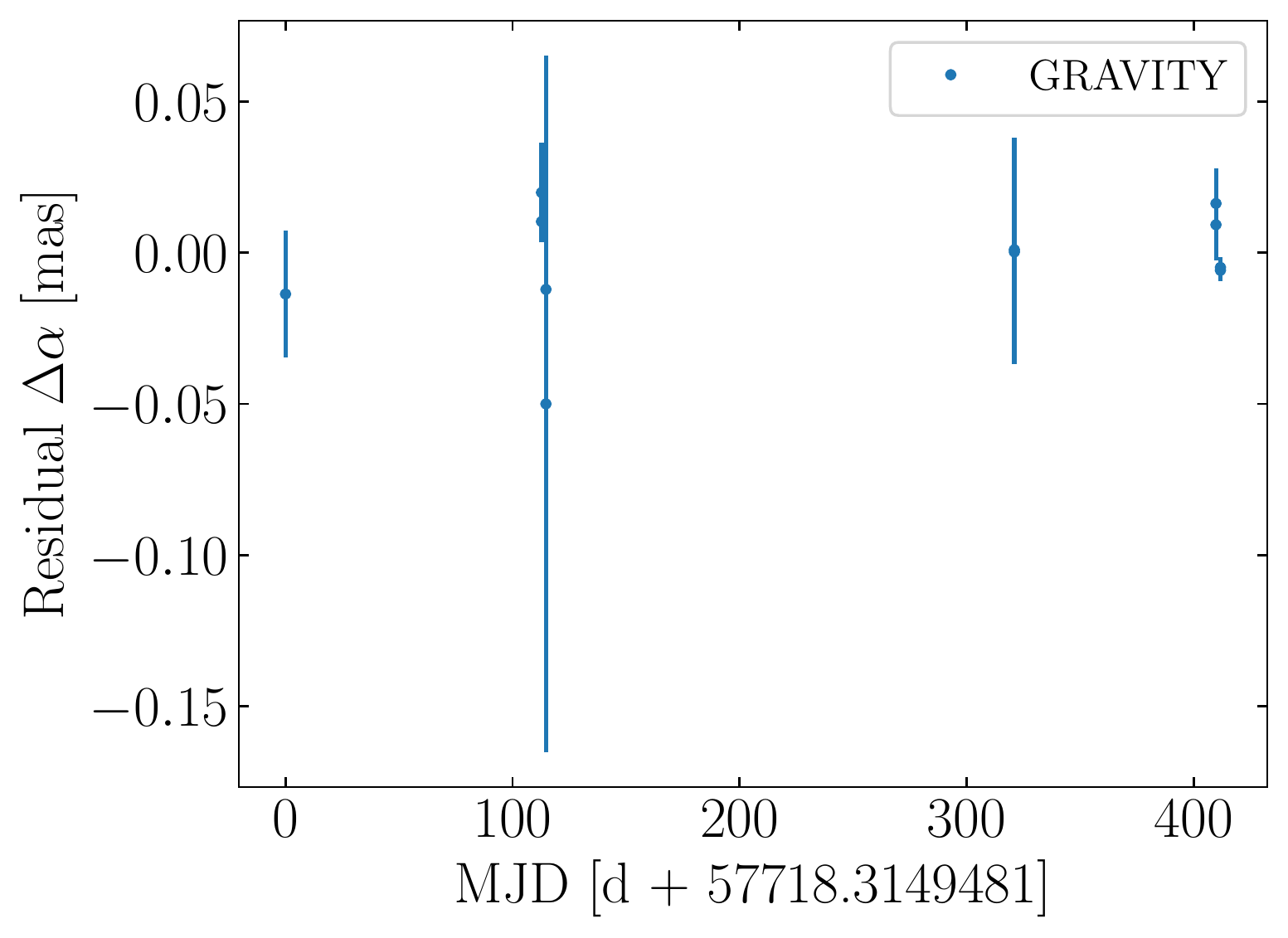}
\end{minipage}
\hfill
\begin{minipage}{0.49\textwidth}
\includegraphics[width=\textwidth]{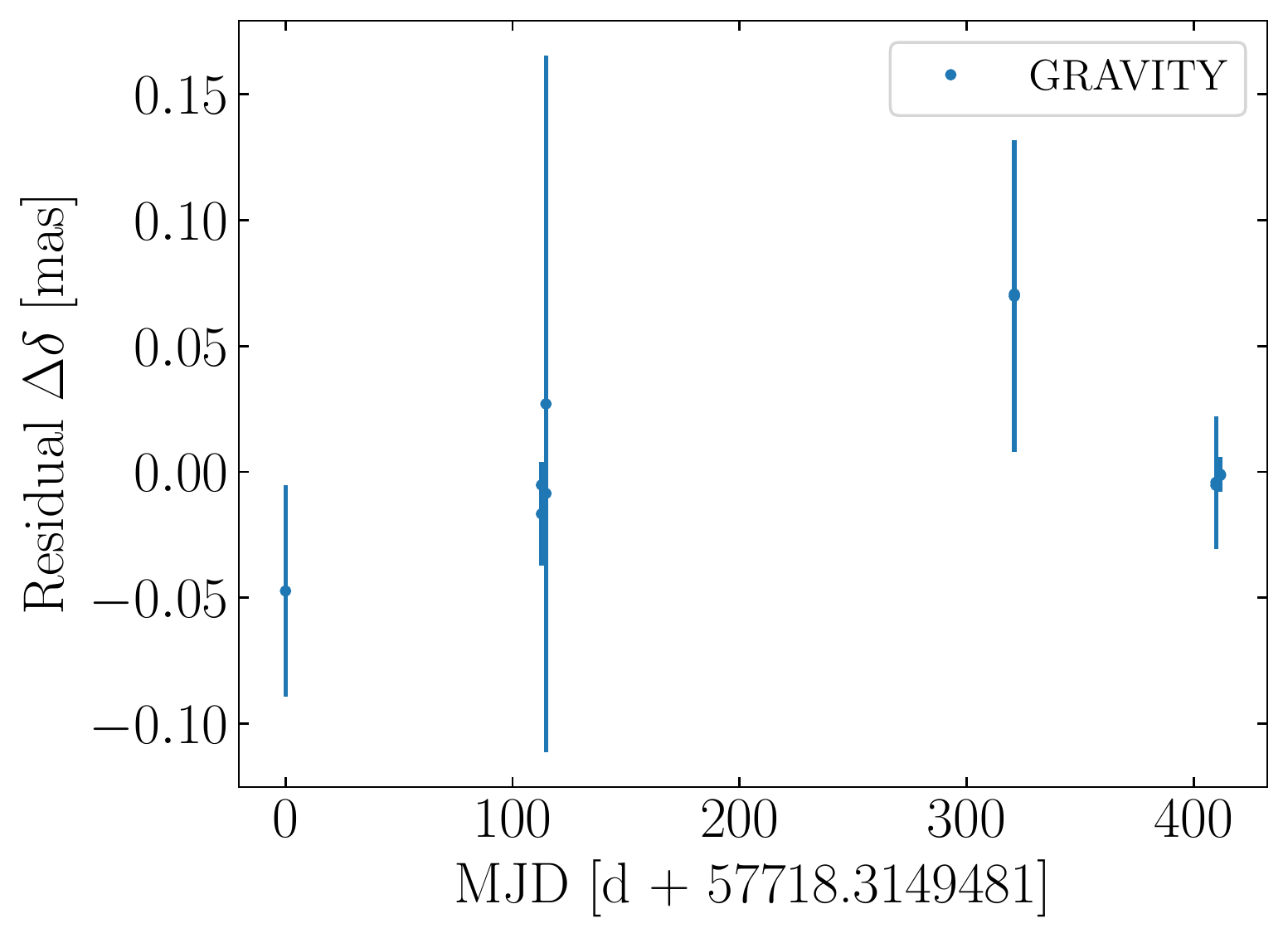}
\end{minipage}
\caption[Residuals of $\theta ^1$ Ori D best fit]{Residuals of the $\Delta \alpha$ (left) and $\Delta \delta$ (right) positions from the fitted orbit of $\theta ^1$ Ori D$_3$. The RMS of the residuals is $0.02$~mas for $\Delta \alpha$ and $0.03$~mas for $\Delta \delta$.}\label{fig:oriD_res}
\end{figure}

With GRAVITY, we detected a star $\theta ^1$ Ori D$_3$ with a flux ratio $f = 0.34 \pm 0.04$ at separations between $1.9$~mas~$\approx 0.79 \pm 0.2$~AU and $2.6$~mas~$\approx 1.08 \pm 0.2$~AU. The observed separations could correspond to the spectroscopic companion reported by \citet{Vitrichenko2002b}, since the inferred separations match quite well. However, we find no evidence for a companion at 18~mas. The trajectory of the detected companion does not favor a very eccentric orbit, i.e. it cannot be related to the detection claim of \citet{Kraus2007}. The positions of D$_3$ are plotted in \cref{fig:oriD_orbit}. We calculate the apparent magnitude of D$_3$ to be $6.9 \pm 0.3$~mag and use $A_{\rm{V}} = 1.79$ \citep{Hillenbrand1997} to estimate a mass of $6\pm 1$~\solmass\ and a B spectral type. This agrees with \citet{doi:10.1093/mnras/stx076}, who determined that the temperature of the spectroscopic companion of D$_1$ has to be $\simsym 20 000$~K, which corresponds to $\simsym 7$~\solmass . The determined orbital parameters are shown in \cref{tab:oriD_orbit}. The angles $\omega$ and $\Omega$ are not well constrained. With the orbit and a distance of $414 \pm 7$~pc \citep{Menten2007, 2014ApJ...783..130R}, we get a system mass of $21.68 \pm 0.05$~\solmass . This corresponds to a companion mass of $\simsym 6 \pm 1$~\solmass\ and a primary mass of $\simsym 16 \pm 1$~\solmass . Our data scatter with an RMS of $0.02$~mas for $\Delta \alpha$ and $0.03$~mas for $\Delta \delta$ (see \cref{fig:oriD_res}).

\begin{table}[tbh]
\centering
\begin{tabular}{cc}
\toprule
Orbital parameter &This work \\
\midrule
a [mas] &$1.86 \pm 0.06$ \\
$P$ [yr] &$0.1452 \pm 0.0002$\\ 
e &$0.43 \pm 0.03$ \\
$\tau$ & $2017.101 \pm 0.001$ \\
$\Omega ~[^{\circ}]$ &$346 \pm 24$ \\
$\omega ~[^{\circ}]$  &$166 \pm 27$ \\ 
i [$^{\circ}$] &$160 \pm 12$\\
\bottomrule
\end{tabular}
\caption[Orbital parameters for $\theta ^1$ Ori D$_3$]{Orbital parameters of the best fit for the positions of $\theta ^1$ Ori D$_3$. $1.86 \pm 0.06$~mas correspond to $0.77 \pm 0.03$~AU at $414 \pm 7$~pc distance.}\label{tab:oriD_orbit}
\end{table}

We set a $3\sigma$ detection limit with CANDID, excluding companions with $\Delta m < 2.5 ~(\approx 2.8$~\solmass ) in a range of $1.70$\textendash $8.3$~AU. For the $8.3$\textendash $16.6$~AU range the detection limit is $\Delta m = 3.9 ~(\approx 1.9$~\solmass ) and for $16.6$\textendash $46.8$~AU we set a limit of $\Delta m = 4.4 ~(\approx 1.9$~\solmass ).

\subsubsection{$\theta ^1$ Ori E (Brun 584, TCC 40, Parenago 1864)}
$\theta ^1$ Ori E is the second-strongest X-ray source in the Trapezium, exceeded only by $\theta ^1$ Ori C \citep{1982Sci...215...61K}. Its K-magnitude is 6.9 \citep{Muench2002} and \citet{Morales-Calderon2012} determined a spectral type of G2IV. The extinction is $A_{\rm{V}} = 3.8$ \citep{2002ApJ...574..258F}.

\citet{2006IAUC.8669....2C} and \citet{Herbig2006} discovered $\theta ^1$ Ori E to be a double lined spectroscopic binary, which consists of two approximately identical stars. \citet{Herbig2006} concluded that the components of $\theta ^1$ Ori E are located in the G-K region, but otherwise do not resemble typical T~Tauri stars. They assumed masses of $ 3$\textendash $4~M_{\odot}$. \citet{Costero2008} determined a period of $P = 9.89520 \pm 0.00069$~d and a mass ratio $q = 1.004 \pm 0.018$. The period corresponds to a semi-major axis of 0.22~mas or $0.091 \pm 0.001$~AU. They also concluded that the binary system is escaping the Trapezium Cluster. \citet{Morales-Calderon2012} determined masses of $2.81 \pm 0.05$~\solmass\ and $2.80 \pm 0.05$~\solmass.
The eclipsing companions cannot be resolved with GRAVITY. We did not find evidence for further companions.

With CANDID, we set a $3\sigma$ detection limit of $\Delta m = 2.8 ~(\approx 1.8$~\solmass ) for $1.70$\textendash $8.3$ ~AU. For $8.3$\textendash $16.6$~AU the limit is $\Delta m = 3.9 ~(\approx 1.4$~\solmass ). In the range of $16.6$\textendash $46.8$~AU, we exclude companions with $\Delta m = 4 ~(\approx 1.4$~\solmass ).

\subsubsection{$\theta ^1$ Ori F (Brun 603, TCC 72, Parenago 1892)}
$\theta ^1$ Ori F is a B8-type star \citep{Herbig1950} with a magnitude in K-band of 8.38 \citep{Muench2002}. Studies by \citet{Petr1998} and \citet{Simon1999} did not detect companions with a separation $\ge 55$~AU. We did not find any values for the extinction of $\theta ^1$ Ori F in the literature. With the method described in \cref{ssec:t1B}, we estimate a lower mass limit of 2.2~\solmass . The typical mass for an B8-star is $\simsym 2.8$~\solmass\ ((\citeauthor{allen2000allen}, \citeyear{allen2000allen}, p. 150, p. 388; \citeauthor{2005essp.book.....S}, \citeyear{2005essp.book.....S}, p.~130), thus we estimate a mass range of 2.2\textendash 2.8~\solmass\ for $\theta ^1$ Ori F.

With the recent GRAVITY data, we can place a $3\sigma$ detection limit of $\Delta m = 1.75 ~(\approx 1.5$~\solmass ) in the range of $1.70$\textendash $8.3$~AU using CANDID. For the range $8.3$\textendash $16.6$~AU, we set a limit of $\Delta m = 2.6 ~(\approx 1.4$~\solmass ), and for $16.6$\textendash $46.8$~AU we get $\Delta m = 2.89 ~(\approx 1.2$~\solmass ).

\subsection{Orion Nebula Cluster stars}
The following stars are not strictly members of the Trapezium Cluster. However, they reside within 2.6~pc and belong to the youngest and most massive stars of the ONC. The apparent K-magnitudes are in the range of 4.49 to 11.05.

\subsubsection{$\theta ^2$ Ori A (HD 37041, Brun 682, Parenago 1993)} 
$\theta ^2$ Ori A$_1$ is of spectral type O9.5IV \citep{2011ApJS..193...24S} with a K-magnitude of 4.94 \citep{2002yCat.2237....0D}. \citet{Preibisch1999} estimated a mass of $\simsym 25$~\solmass . \citet{Simon-Diaz2006} determined a mass of $39 \pm 14~M_{\odot}$, an effective temperature of 35~000 K and a stellar radius of $8.2 \pm 1.1~R_{\odot}$. 

$\theta ^2$ Ori A is a hierarchical system comprising a spectroscopic companion A$_2$ with a period $P = 20.9741 \pm 0.0028$~days \citep{1974JRASC..68..205A, Abt1991}. Assuming a circular orbit, this corresponds to a separation of $\simsym 0.46 \pm 0.04$~AU ($\simsym 1$~mas). With the mass ratio $q \approx 0.35$ \citep{Abt1991} the companion should be in the range of $\simsym 9$\textendash 19~\solmass .

\citet{Preibisch1999} discovered a visual companion A$_3$ at a separation of \ang{;;0.38}, corresponding to $157 \pm 3$~AU, and a mass ratio $q \approx 0.25$ ($\approx 6$\textendash 13~\solmass). \citet{doi:10.1093/mnras/stw2743} claimed the detection of another spectroscopic companion, but provided no orbital period or any further constraints on the companion.

\begin{figure*}
\centering
\includegraphics[width=\textwidth]{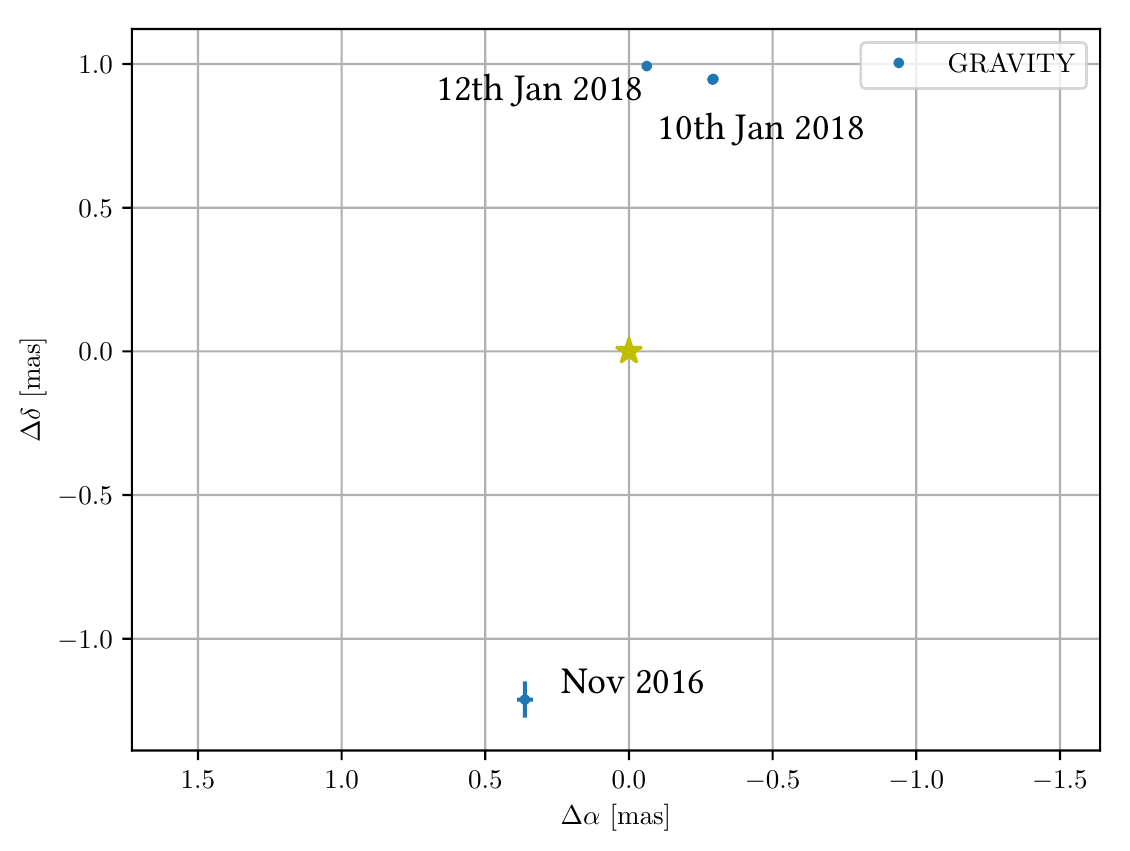}
\caption[Positions of $\theta ^2$ Ori A$_2$]{Positions observed with GRAVITY of $\theta ^2$ Ori A$_2$, relative to A$_{1,3}$ at (0,0).}\label{fig:2OriA_pos}
\end{figure*}

With GRAVITY, we detect a companion at a separation of 1.3~mas ($\approx 0.538 \pm 0.011$~AU), likely the spectroscopic companion A$_2$. The observed positions are displayed in Figure \ref{fig:2OriA_pos}. We observe a flux ratio of $f = 0.52 \pm 0.04$, which corresponds to $m_{\rm{K}} = 5.7 \pm 0.2$. Using $A_{\rm{V}} = 1.12$ \citep{Hillenbrand1997} we find an absolute magnitude in K-band of $M_{\rm{K}} = -2.5 \pm 0.2$. Comparing this magnitude with isochrones %from \citet[p. 150, p. 388]{allen2000allen} and \citet[p. 130]{2005essp.book.....S}
, we suggest a stellar mass of $\simsym 10 \pm 2$~\solmass\ and an early B spectral type. This result is consistent with previous estimates.

With CANDID, we set a $3\sigma$ detection limit for the range of $1.70$\textendash $8.3$~AU of $\Delta m = 5.25~(\approx 1.6$~\solmass ) and a limit of $\Delta m = 6.2~(\approx 1.5$~\solmass ) for the range $8.3$\textendash $16.6$~AU. For $16.6$\textendash $46.8$~AU, we place a limit of $\Delta m = 6.47~(\approx 1.1$~\solmass ).

\subsubsection{$\theta ^2$  Ori B (HD 37042, Brun 714, Parenago 2031)}
$\theta ^2$ Ori B is a B2-B5 PMS star \citep{Hillenbrand1997} with a K-magnitude of 6.41 \citep{2002yCat.2237....0D}. \citet{Simon-Diaz2006} determined a mass of $9 \pm 3$~\solmass\ and a temperature of $29~000 \pm 1000$~K together with a radius of $4.5 \pm 0.6~R_{\odot}$. The values agree with the results of \citet{Nieva2014}, who obtained $M = 14.8 \pm 3.4~M_{\odot}$, $T_{\rm{eff}} = 29~300 \pm 300$~K and $R = 4.3 \pm 0.4$.

Previous observations, e.g. \citet{Abt1991} or \citet{Preibisch1999}, did not find indications for a companion star. GRAVITY observations made in January 2018 (see Table \ref{tab:OriObs}) allowed the detection of a companion at a separation of 95.8~mas~$\approx~40 \pm 1$~AU with a small flux ratio of $f = 0.02 \pm 0.01$. This yields an apparent magnitude of $10.6 \pm 1.3$. Using $A_{\rm{V}} = 0.73$ \citep{Hillenbrand1997}, we obtain $M_{\rm{K}} = 2.4 \pm 1.3$. A comparison with isochrones from \citet[p. 150, p. 388]{allen2000allen} and \citet[p. 130]{2005essp.book.....S} yields a mass estimate of $1.6 \pm 0.7$~\solmass\ and thus a late-A/early-F-type star. %TODO add new values to table until this magnitude.

Using CANDID, we set a $3\sigma$ detection limit of $\Delta m = 2.7~(\approx 1.9$~\solmass ) for the $1.70$\textendash $8.3$~AU and $\Delta m = 3.9~(\approx 1.6$~\solmass ) for $8.3$\textendash $46.8$~AU.

\subsubsection{$\theta ^2$ Ori C (HD 37062, Brun 760, Parenago 2085)}
$\theta ^2$ Ori C$_1$ is a B5V-type star \citep{2017ARep...61...80S} with $m_{\rm{K}} = 7.54$ \citep{2003yCat.2246....0C}. \citet{2005ApJS..160..557S} determined an effective temperature $T_{\rm{eff}} = 13 800$. Comparing $T_{\rm{eff}}$ with typical values for main sequence stars \citep[p. 130]{2005essp.book.....S}, we get a mass estimate of $4 \pm 1$~\solmass , which agrees with the mass for B5-type stars.% in \citet[p. 150, p. 388]{allen2000allen} and \citet[p. 130]{2005essp.book.....S}.

\citet{Corporon1999} detected a spectroscopic binary C$_2$ with a period $P \approx 13$~days. This corresponds to a separation of 0.4~mas or $0.165 \pm 0.003$~AU, assuming a circular orbit. 

With GRAVITY, we resolved for the first time a third companion C$_3$ at 38~mas, a projected separation of $15.7 \pm 0.2$~AU. The detected flux ratio is $f = 0.115 \pm 0.003$. With $A_{\rm{V}} = 0.92$ \citep{Hillenbrand1997}, this results in an apparent magnitude of $9.89 \pm 0.07$. The absolute magnitude is $M_{\rm{K}} = 1.7 \pm 0.1$.
Thus, a comparison with isochrones %\citet[p. 150, p. 388]{allen2000allen} and \citet[p. 130]{2005essp.book.....S} 
yields an estimate of $1.7 \pm 0.2$~\solmass\ and A spectral type. 

\subsubsection{NU Ori (HD 37061, Brun 747, Parenago 2074)}
NU Ori$_1$ is a O9V-type star \citep{2012AJ....144..130B} with a K-magnitude of 5.49 \citep{2003yCat.2246....0C}. \citet{Hillenbrand1997} determined a stellar mass of $16.3~M_{\odot}$, whereas \citet{2017A&A...601A.129L} estimated $\simsym 13$~\solmass\ using effective temperatures but flagged it as particularly uncertain. \citet{0004-637X-601-2-979} estimated a mass of $\simsym 14$~\solmass , using the luminosity and the effective temperature but stated that there are systematic uncertainties from the evolutionary tracks of PMS stars. Thus, we assume that the mass of NU Ori$_1$ is in the range of $16 \pm 3$~\solmass .

NU Ori has a spectroscopic companion NU Ori$_2$, discovered by \citet{Morrell1991}. Its orbital elements were determined by \citet{Abt1991}, who found a period of $P = 19.1387 \pm 0.0028$~d. and the lower limit for the mass ratio $q = 0.19$. With a primary mass between 13 and 19~$M_{\odot}$, the lower limit for NU Ori$_2$ is 2.5\textendash 3~$M_{\odot}$. Assuming a circular orbit, we get a separation of $0.35 \pm 0.03$~AU.

\citet{Preibisch1999} discovered a companion star NU Ori$_3$ at \ang{;;0.47}. At a distance of $414 \pm 7$~pc \citep{Menten2007, 2014ApJ...783..130R}, this corresponds to $195 \pm 4$~AU. The mass estimate is 1~\solmass , with an upper limit $<4$~\solmass . \citet{Kohler2006} also detected a companion at $\ang{;;0.47 \pm 0.01}$ with $\Delta m_{\rm{K}} = 3.23 \pm 0.1$~mag. With this magnitude we are now able to estimate the stellar mass using the method described in \cref{ssec:t1B}. For an apparent K-magnitude of $8.7 \pm 0.1$, we get an absolute magnitude of $M_{\rm{K}} = 0.4 \pm 0.1$ using the extinction $A_{\rm{V}} = 2.09$ \citep{Hillenbrand1997}. This yields a mass estimate of $2.4 \pm 0.6$~\solmass\ and thus an early A or late B-type star. 

\begin{figure*}
\centering
\includegraphics[width=\textwidth]{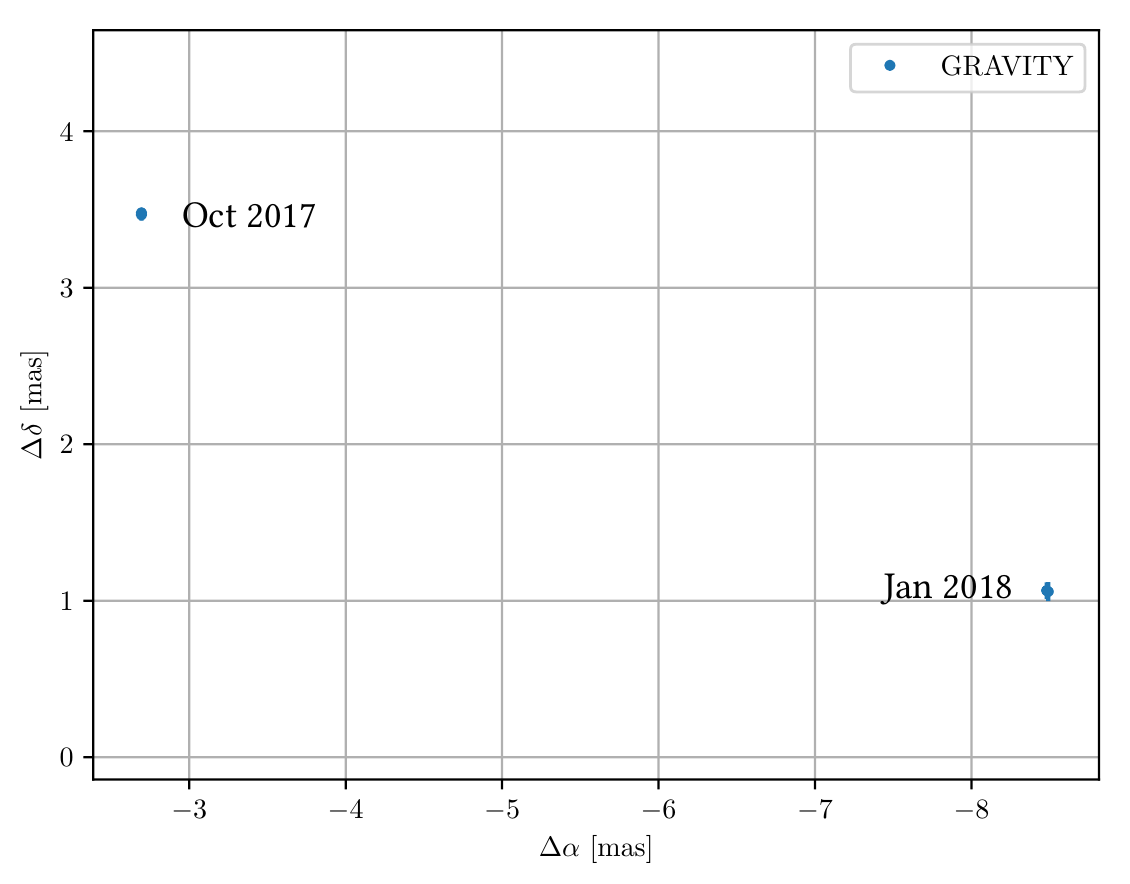}
\caption[Positions of NU Ori$_4$]{Positions of NU Ori$_4$ with respect to NU Ori$_{1,2}$ at (0,0).}\label{fig:NUO_pos}
\end{figure*}

\citet{Grellmann2013} presumed another companion at either 20~mas or 10~mas separation. With our interferometric data, we found a companion NU Ori$_4$ at a distance of $d = 8.6$~mas~$ \approx 3.6 \pm 0.1$~AU with a flux ratio $ f = 0.184 \pm 0.009$ (see Figure \ref{fig:NUO_pos}). This new detection is most likely a different star than the spectroscopic companion, because a period of 19~days translates to a distance of $\approx 0.9$~mas, a factor of 10 smaller than the newly discovered separation of 8.6~mas. % This could be the companion predicted by \citet{Grellmann2013}. 
With the flux ratio of $0.184 \pm 0.009$, we get an apparent magnitude of $7.3 \pm 0.1$ and an absolute K-magnitude of $-1 \pm 0.1$. This results in a mass estimate of $4 \pm 1$~\solmass\ and B spectral type.

We set a $3\sigma$ detection limit of $\Delta m = 3.8~( \approx 2$~\solmass ) for separations of $1.70$\textendash $8.3$~AU. For the range $8.3$\textendash $46.8$~AU, we determine a limit of $\Delta m = 4.6~( \approx 1.8$~\solmass ).

\subsubsection{Brun 862 (Parenago 2208)}
Brun 862 is a K3\textendash\ M0I-type star \citep{Hillenbrand1997} with $m_{\rm{K}} = 4.49$ \citep{2003yCat.2246....0C}. To get a mass estimate, we take the calibration of MK spectral types for supergiants (luminosity class I) from \citet[p. 390, Table 15.8.]{allen2000allen}. For spectral type K3-M0, the corresponding mass is 13~\solmass .

With GRAVITY observations from January 2018 (Table \ref{tab:OriObs}), we can either fit a companion Brun 862$_2$ at a separation of $0.29 \pm 0.01$~AU or fit a single star with a diameter of $0.33 \pm 0.01$~AU~($\simsym 71 ~R_{\odot}$), represented by a uniform disk. Both models fit the data equally well. 
For the first model the resulting flux ratio is $f = 0.26 \pm 0.04$. With $A_{\rm{V}} = 6.78$ \citep{Hillenbrand1997}, this would result in an absolute magnitude of  $M_{\rm{K}} = -2.9 \pm 0.4 $. Assuming a main sequence star, we could estimate a mass of $\simsym 10$~\solmass\ and suggest a late O or early B spectral type. On the other hand, the latter model of a single extended star is more plausible, considering that Brun 862 is classified as a supergiant. We compare the radius of $71~R_{\odot}$ with values from \citet{2005ApJ...628..973L}, who list K2 and K2.5 stars with $\simsym 100~R_{\odot}$. Thus, our determined radius agrees well with \citet{2005ApJ...628..973L}.
We determine companion detection limits on a $3\sigma$ level for $1.70$\textendash $46.8$~AU of $\Delta m = 4.75 ~(\approx 2.2$~\solmass ).

\subsubsection{TCC 59}
TCC 59 is a Young Stellar Object (YSO) with a protoplanetary disk \citep{1996AJ....111..846O}. It has a K-magnitude of 11.05 \citep{Muench2002}. For a lower mass limit, we compare the absolute K-magnitude with isochrones % \citet[p. 150, p. 388]{allen2000allen} and \citet[p. 130]{2005essp.book.....S}, 
as described in \cref{ssec:t1B}, and get 1.5~\solmass . We found no extinction measurements for this star, but as a YSO, its reddening in K-band is supposedly non-negligible. The color of (J-K)~=~1.35 \citep{Muench2002} is very red compared to the other stars in our sample. This indicates significant dust extinction or intrinsics infrared excess due to, e.g., a circumstellar disk. Thus, we expect TCC 59 to be intrinsically brighter and more massive and only provide a lower limit. \citet{Close2012} claimed the detection of a companion star with $136 \pm 3$~mas~($\approx 56 \pm 2$~AU) separation.

In the data taken with GRAVITY in January 2018, we find signatures in visibilities and closure phases, but cannot find a good fit. Thus, it is not clear whether there is a companion star or whether the signatures result from a potential disk.

\subsubsection{TCC 43}
TCC 43 has a K-magnitude of 10.44 \citep{Muench2002}. \citet{Petr1998} and \citet{Simon1999} observed TCC 43 and did not find a companion star. With GRAVITY we see minor signatures in visibilities and closure phase, but cannot find a good fit. We set a lower mass limit of $\simsym~1.5$ \textendash 1.7~\solmass , i.e. an A or F-type using the method described in \cref{ssec:t1B}. We have no value for the extinction, therefore the star might be brighter and more massive. 

With GRAVITY we can exclude companions in the $1.70$\textendash $16.6$~AU range with $\Delta m = 1.4~ (\approx 0.9$~\solmass ) on a $3\sigma$ level. For separations of $16.6$\textendash $46.8$~AU, we place a limit of $\Delta m = 0.61~(\approx 1.1$~\solmass ).

\subsubsection{LP Ori (HD 36982, Brun 530, Parenago 1772)}
LP Ori is a B1.5V-type star \citep{2017ARep...61...80S} with $m_{\rm{K}} = 7.47$ \citep{2003yCat.2246....0C}. \citet{Hillenbrand1997} found an extinction $A_{\rm{V}} = 1.47$ and a mass of 7.15~\solmass . \citet{0004-637X-852-1-5} determined a mass of $6.70 ^{+0.64} _{-0.37}$~\solmass .
\citet{Preibisch1999} and \citet{Abt1991} observed LP Ori but found no companion. 

With GRAVITY we set a $3\sigma$ companion limit of $\Delta m = 2.12 ~(\approx 1.9$~\solmass ) for separations of $1.70$\textendash $8.3$~AU and $\Delta m = 2.87 ~(\approx 1.5$~\solmass ) for $8.3$\textendash $46.8$~AU. 

\subsubsection{HD 37115 (Brun 907, Parenago 2271)}
HD 37115$_1$ is a B5-type star \citep{1994A&AS..105..301R} with a K-magnitude of 7.13 \citep{2003yCat.2246....0C}. \citet{Preibisch1999} estimated a mass of 5~\solmass , \citet{Hillenbrand1997} of 5.7~\solmass\ and \citet{0004-637X-601-2-979} estimated a mass of 5.5~\solmass . We take the mean mass $5.4 \pm 0.4$~\solmass . \citet{0004-637X-818-1-59} determined $A_{\rm{V}} = 5.9 \pm 0.3$.

\citet{Preibisch1999} found a companion at $\simsym 890$~mas separation, which corresponds to $368 \pm 6$~AU. The mass ratio is $\simsym 0.29$ and the estimated mass is $\simsym 1.5$~\solmass\ with an upper limit of $< 5$~\solmass .

We do not detect a companion with GRAVITY but set a $3\sigma$ detection limit of $\Delta m = 2~( \approx 2.2$ \solmass) for the range of $1.70$\textendash $8.3$~AU. For $8.3$\textendash $16.6$~AU we get a limit of $\Delta m = 2.42~( \approx 1.9$~\solmass ) and limit of $\Delta m =3.24~( \approx 1.5$~\solmass ) for separation range $16.6$\textendash $46.8$~AU. 

\subsubsection{HD 37150 (Brun 980, Parenago 2366)}
HD 37150 is a B3III/IV-type star \citep{1999MSS...C05....0H} with $m_{\rm{K}} = 7.11$ \citep{2003yCat.2246....0C}. We estimate a lower mass limit of 7~\solmass , using the calibration table for MK spectral types from \citet[p. 390, Table 15.8]{allen2000allen}.

We do not detect a companion with GRAVITY. We set a $3\sigma$ detection limit of $\Delta m = 2.29~( \approx 1.9$~\solmass ) in the range of $1.70$\textendash $8.3$~AU and $\Delta m = 3.08 ~(\approx 1.5$~\solmass ) for separations of $8.3$\textendash $46.8$~AU.

\subsection{Summary}

% \begin{figure}
% \centering
% \includegraphics[width=\textwidth]{"../../plots/comp_per_star"}
% \caption[Number of companion stars versus mass of primary star]{The number of companions versus the mass of the primary star. The values are reported in Table \ref{tab:overview}. Blue stars are late B, A, F or G-type stars, orange triangles and green squares are B-type stars and red circles represent O-type stars.}\label{fig:comp_per_mass}
% \end{figure}

\begin{figure}[p]
\centering
\includegraphics[width=\textwidth]{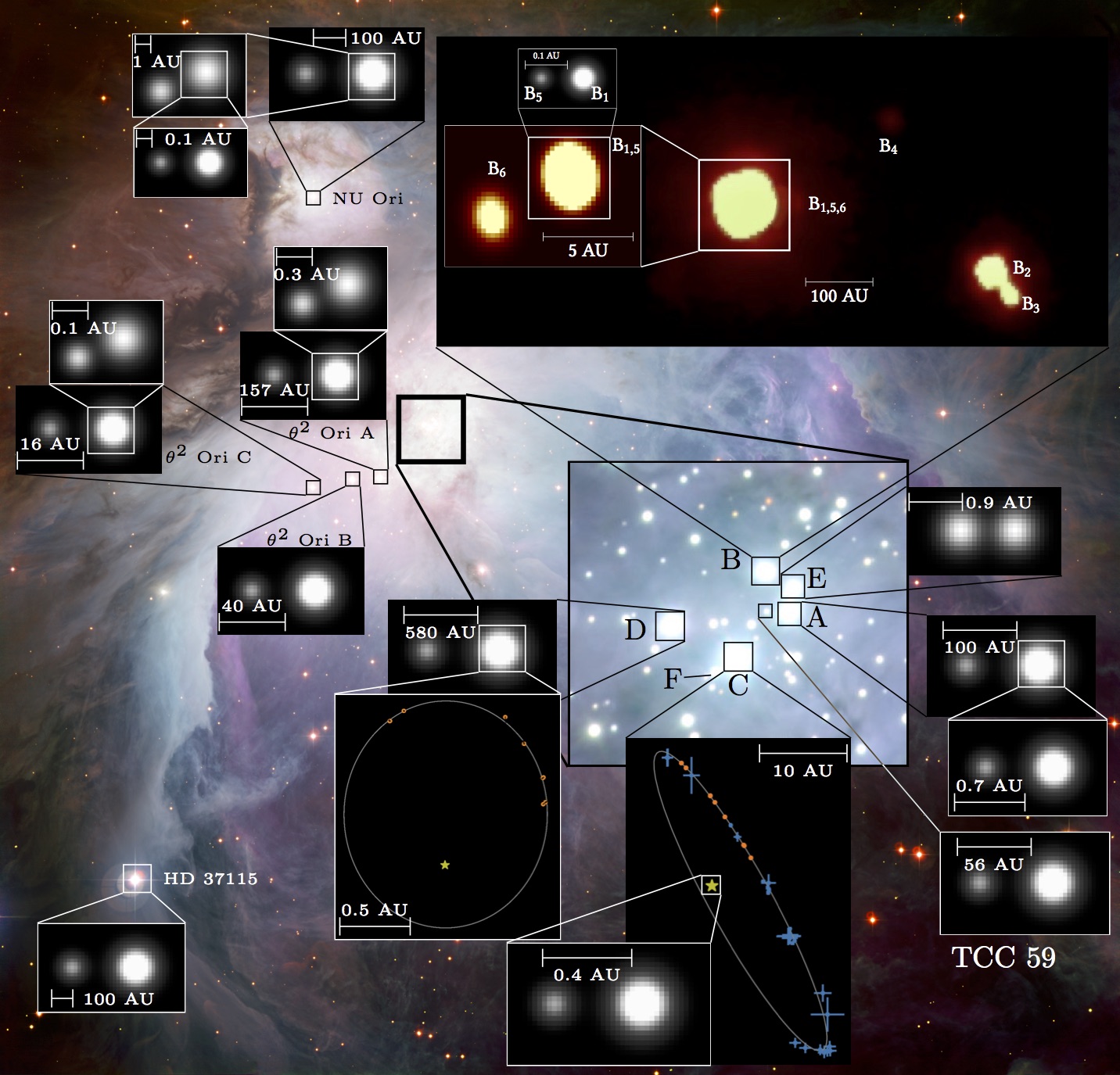}
\caption[Summary of the observed multiple systems]{Summary of the observed multiple systems in the Orion Nebula. We observed 16 multiple systems with a total number of 22 companion stars. The respective scales are indicated. The images of $\theta ^1$ Ori B are from actual obervational data, except for the spectroscopic B$_1$, B$_5$ system, which is only a representation. The orbital positions for $\theta ^1$ Ori D and $\theta ^1$ Ori C are the positions obtained in this work and from the literature. All remaining close up depiction of stars (gray) are only for illustrative purposes and were not created with observational data. The background image of the Orion Nebula was created by ESO/Igor Chekalin. The zoom of the Trapezium Cluster ($\theta ^1$) is a cut from ESO/M. McCaughrean et al. (AIP).}\label{fig:Orion_summary}
\end{figure}

We illustrate the observed companion systems in \cref{fig:Orion_summary} and provide a summary of all stellar systems and their properties in Table \ref{tab:overview}. Bold objects were observed with GRAVITY. For an overview of all observations, refer to Table \ref{tab:OriObs}. In total, 16 objects were observed, out of which eleven are confirmed multiple systems. This leads to a multiplicity fraction of $11 / 16 = 0.688$. All multiple systems combined have a total number of 22 confirmed companion stars. Thus, we get a companion fraction of $22 / 16 = 1.375$.

Brun 862 is a supergiant with no clear detection of a companion star. The evolutionary stage of Brun 862 differs greatly from the remaining stars in our sample. Additionally, the Gaia parallax of Brun 862 \citep[$1.690 \pm 0.094$, ][]{2016A&A...595A...1G, 2018arXiv180409365G, 2018arXiv180409376L} diverges significantly from the Gaia parallax of, e.g., $\theta ^1$ Ori C ($2.472 \pm 0.082$). We will not include Brun 862 in the following discussion.% because of its evolved evolutionary stage.
%The resulting companions per star are summarized in Figure \ref{fig:comp_per_mass}.

\newgeometry{bottom=2cm, top=2cm, right=2cm, left=2cm}
\restoregeometry
\longtab{
\newgeometry{a4paper, bottom=2cm, top=2cm, right=2cm, left=2cm}
\begin{landscape}

\setlength{\tabcolsep}{4pt}
{\small
\begin{longtable}{ccccccccc}

\toprule 
Object &Component &Sep [AU] &Period &Spec. type &$m_{\rm{K}}$ &Mass [$M_{\odot}$] &Limit [$M_{\odot}$]  &Reference 
  \endfirsthead

  \multicolumn{4}{c}%
  { \tablename\ \thetable{} \textemdash\ continued from previous page} \\
  \midrule
Object &Component  &Sep [AU] &Period &Spec. Type &$m_{\rm{K}}$ &Mass [$M_{\odot}$] &Limit [$M_{\odot}$]  &Reference  \\
\midrule
  \endhead
  
  \midrule
  \multicolumn{3}{r}{ \tablename\ \thetable{} \textemdash\ Continued on next page}
  \endfoot

   \endlastfoot

\midrule

\textbf{$\theta ^1$ Ori A} &\textbf{A$_1$} & & &\textbf{B0.5V} & \textbf{5.67} & \textbf{14 $\pm$ 5} &  & [1, 2, 3]\\
&\textbf{A$_2$}  & \textbf{90 \textendash 100} &\textbf{214~yr} & \textbf{F PMS} &\textbf{7.3 $\pm$ 0.3} & \textbf{4} & & [This work, 4, 5, 6, 10, 26]\\
&A$_3$ & 0.71 & $65.09 \pm 0.07$~d &T Tauri & &$2.6 \pm 0.1$ &  & [7, 8, 9, 11, 12, 13]\\
\midrule
\textbf{$\theta ^1$ Ori B} & \textbf{B$_1$}  & & &\textbf{B1V} &\textbf{6.00} & \textbf{7.2 $\pm$ 0.2} &\textbf{$<$ 1.9}  &[This work, 2, 16, 22, 33]\\  %TODO  Parenago \&Kukarkin 1947
&B$_2$ & 382 $\pm$ 6& 1920 yr & &7.6 &4 &  &[4, 5, 10, 26]\\
&B$_3$ & 49 $\pm$ 1& 200 yr & &8.6 & 3 &  &[4, 5, 10, 26]\\
%&B$_{2,3}$ & 48 & 160 yr & & & &  &[5, 6, 10, 26] \\
&B$_4$ & 248 $\pm$ 4 &$2000 \pm 700$ yr & &11.66  &1 &  &[15, 26, 32, 42]\\
&B$_5$ & 0.120 $\pm$ 0.002 & 6.47~d & & & 2 &  & [13, 14, 26]\\
%&B$_6$  &&&K7III&&1.8&&[47]\\
&\textbf{B$_6$} & \textbf{3.5-7.2} & & \textbf{B} &\textbf{7.3 $\pm$ 0.5} & \textbf{4 \textendash 6} & & [This work, 19]\\
\midrule
\textbf{$\theta ^1$ Ori C} &\textbf{C$_1$} & & &\textbf{O7V} &\textbf{4.57} &\textbf{33.5 $\pm$ 5.2}& \textbf{$<$ 3}  & [This work, 17, 18, 20, 21, 45]\\
&\textbf{C$_2$}  &\textbf{18.1 $\pm$ 1.7} & \textbf{11.4~yr} &\textbf{O9.5} & \textbf{6.0 $\pm$ 0.4} & \textbf{12 $\pm$ 3} & & [This work, 5, 7, 20, 21, 22, 23]\\
&C$_3$. & $0.41 \pm 0.01$ & 61.49~d & & & $1.0 \pm 0.2$ &  & [23, 24]\\ % Donati et al 2002 nur vielleicht
\midrule
\textbf{$\theta ^1$ Ori D} &\textbf{D$_1$}  & & &\textbf{B1.5V} & \textbf{5.75} & \textbf{16 $\pm$ 1} &\textbf{$<$ 2.8}  & [2, 3, 27]\\
&D$_2$ & 580 $\pm$ 10& & &11.69 $\pm$ 0.06 &1 $\pm$ 0.1 & & [This work, 26] \\
%&D$_3$  & 0.58 & $\sim20.268 \pm 0.001$~d  & & &  &  & [25]\\   %vielfache der periode: / $\sim40.528 \pm 0.002$~d                     
&\textbf{D$_3$} & \textbf{0.77 $\pm$ 0.03} & \textbf{53.03 $\pm$ 0.07 d} &\textbf{B} & \textbf{6.9 $\pm$ 0.3} & \textbf{6 $\pm$ 1} &  & [This work, 6, 25, 43]\\
\midrule
\textbf{$\theta ^1$ Ori E} &\textbf{E$_1$}  & &  &\textbf{G2IV} &\textbf{6.9} & \textbf{2.81 $\pm$ 0.05} &\textbf{$<$ 1.8}  & [28, 29, 30, 31]\\
&E$_2$ &0.091 $\pm$ 0.001 & $9.8952 \pm 0.0007$~d & G0IV \textendash G5III & & 2.80 $\pm$ 0.05 &  & [28, 30, 31]\\
\midrule
\textbf{$\theta ^2$ Ori A} &\textbf{A$_1$}  & & &\textbf{O9.5IV} & \textbf{4.94} & \textbf{39 $\pm$ 14} & \textbf{$<$ 1.6} & [3, 17, 18]\\
  &\textbf{A$_2$}  & \textbf{0.42 $\pm$ 0.01} &\textbf{20.974 $\pm$ 0.003 d}  & &\textbf{5.7 $\pm$ 0.2} & \textbf{10 $\pm$ 2} & & [This work, 13]\\ 
&A$_3$ & 157 &  & & &6\textendash 13 &  & [16, 32]\\
\midrule  
\textbf{$\theta ^2$ Ori B} &\textbf{B$_1$} & & & \textbf{B2 \textendash B5} & \textbf{6.41} & \textbf{14.8 $\pm$ 3.4} & \textbf{$<$ 1.9} & [3, 18, 33, 34]\\
 &\textbf{B$_2$} &\textbf{40 $\pm$ 1} & &\textbf{A \textendash F} &\textbf{10.6 $\pm$ 1.3} & \textbf{1.6 $\pm$ 0.7} & & [This work]\\
\midrule 
\textbf{$\theta ^2$ Ori C} & \textbf{C$_1$} & & &\textbf{B5V} &\textbf{7.54} &\textbf{4 $\pm$ 1} & & [2, 39]\\
&C$_2$ & 0.165 $\pm$ 0.003 & 13~d & & & &  & [35] \\
 & \textbf{C$_3$} & \textbf{15.7 $\pm$ 0.2} & &\textbf{A}& \textbf{9.89 $\pm$ 0.07} & \textbf{1.7 $\pm$ 0.2} &\textbf{} & [This work]\\
\midrule           
\textbf{NU Ori} & \textbf{NU Ori$_1$} & & &\textbf{O9V} &\textbf{5.49} &\textbf{16 $\pm$ 3}&\textbf{$<$ 2}  &[2, 6, 33, 37]\\
& NU Ori$_2$ &0.36 $\pm$ 0.01 & $19.1387 \pm 0.0028$~d & &&$> 2.5$ &  & [This work, 13]\\
& NU Ori$_3$ & 195 $\pm$ 4& & A or B &$8.7 \pm 0.1$ &2.4 $\pm$ 0.6 &  & [This work, 32, 36]\\
& \textbf{NU Ori$_4$} & \textbf{3.6 $\pm$ 0.1} & &\textbf{B} & \textbf{7.3 $\pm$ 0.1} &\textbf{4 $\pm$ 1} & &[This work]\\
\midrule

\textbf{HD 37115} &\textbf{HD 37115}$_1$ & & &\textbf{B5} &\textbf{7.13} &\textbf{5.4 $\pm$ 0.4} &\textbf{$<$ 2.2}  &[This work, 2, 32, 40]\\
 &HD 37115$_2$ &368 $\pm$ 6 & & & & 1.6 $\pm$ 0.1&  &[32]\\
\midrule 
      \textbf{TCC 59} &\textbf{TCC 59$_1$} & & &\textbf{YSO} &\textbf{11.05} & \textbf{$\geq$ 1.5} &  &[This work, 29] \\
      &TCC 59$_2$ &56 $\pm$ 2 & & & & &  &[26] \\
\midrule
\textbf{Brun 862} & &  & &\textbf{K3-M0I} &\textbf{4.49} &\textbf{13} &\textbf{$<$ 2.2}  & [This work, 2, 33] \\
%&\textbf{Brun 862$_2$} & \textbf{0.29} & & & \textbf{6.0 $\pm$ 0.4} &\textbf{10} & & [This work] \\
\midrule
     \textbf{$\theta^1$ Ori F} & & & &\textbf{B8} &\textbf{8.38} &\textbf{2.2 \textendash 2.8} &\textbf{$<$ 1.5}  & [This work, 29, 38]\\
     \midrule
     \textbf{TCC 43} & & & &\textbf{A \textendash F} &\textbf{10.44} &\textbf{$>$ 1.5 \textendash 1.7} &\textbf{$<$ 1.1}  &[This work, 29]\\
     \midrule
\textbf{LP Ori} & & & &\textbf{B1.5V} &\textbf{7.47} &\textbf{6.70 $^{+0.64} _{-0.37}$} &\textbf{$<$ 1.9}  &[This work, 2, 33, 39, 44]\\
\midrule
\textbf{HD 37150} & & & & \textbf{B3III/IV} & \textbf{7.11} &\textbf{$\geq$ 7} &\textbf{$<$ 1.9}  &[This work, 2, 41]\\

\bottomrule

\caption[Properties of all observed stars]{\small{Overview for all the observed stars. From left to right starting with the object, its component, the separation or, if known, the semi-major axis, the period, spectral type, magnitude in K-band, mass of component, detection limit and the reference. Objects in bold have been observed with GRAVITY. The uncertainties for the mass are taken from the literature. Values without uncertainties are not necessarily exact but were published without an error estimate. \\
\textbf{References:} [1] \citet{Levato1976}, [2] \citet{2003yCat.2246....0C}, [3] \citet{Simon-Diaz2006}, [4] \citet{Petr1998}, [5] \citet{Schertl2003}, [6] \citet{Grellmann2013}, [7] \citet{1976IAUC.3004....1M}, [8] \citet{1975IBVS..988....1L}, [9] \citet{1989A&A...222..117B}, [10] \citet{Close2003}, [11] \citet{Vitrichenko1998}, [12] \citet{Vitrichenko2001}, [13] \citet{Abt1991}, [14] \citet{Popper1976}, [15] \citet{Simon1999}, [16] \citet{1998AJ....115..821M}, [17] \citet{2011ApJS..193...24S}, [18] \citet{2002yCat.2237....0D}, [19] \citet{Vasileiskii2000}, [20] \citet{Kraus2007}, [21] \citet{Kraus2009}, [22] \citet{Weigelt1999}, [23] \citet{Lehmann2010}, [24] \citet{Vitrichenko2002a}, [25] \citet{Vitrichenko2002b}, [26] \citet{Close2012}, [27] \citet{2006MNRAS.371..252L}, [28] \citet{Herbig2006}, [29] \citet{Muench2002}, [30] \citet{Morales-Calderon2012}, [31] \citet{Costero2008}, [32] \citet{Preibisch1999}, [33] \citet{Hillenbrand1997}, [34] \citet{Nieva2014}, [35] \citet{Corporon1999}, [36] \citet{Kohler2006}, [37] \citet{2012AJ....144..130B}, [38] \citet{Herbig1950}, [39] \citet{2017ARep...61...80S}, [40] \citet{1994A&AS..105..301R}, [41] \citet{1999MSS...C05....0H}, [42] \citet{Close2013}, [43] \citet{doi:10.1093/mnras/stx076}, [44] \citet{0004-637X-852-1-5}, [45]  \citet{2015ASPC..494...57B}}
}\label{tab:overview}

\end{longtable}
}

\end{landscape} 
\restoregeometry
}
\section{Discussion}\label{chap:discussion}

Our sample comprises 15 objects, excluding Brun 862. Eleven of them are multiple systems with up to six members. With long baseline interferometry observations performed using GRAVITY, we close the gap between close spectroscopic companions and more distant visual companions. This yields a complete sample of companions for our observed systems down to our detection limit of 1.5\textendash 3~\solmass . 

\subsection{Multiplicity}

\citet{Duchene2013} provided an overview of the multiplicity of stars. The multiplicity fraction $MF$ is defined as
\begin{equation}\label{eq:mf}
MF = \dfrac{N_{\rm{mult}}}{N_{\rm{mult}} + N_{\rm{single}}},
\end{equation}
where $N_{mult}$, the number of multiple systems, is divided by the total number of systems, namely the sum of multiple star and single star systems.
The companion frequency or companion fraction $CF$ is the average number of companions per target
\begin{equation}
CF = \dfrac{N_{\rm{comp}}}{N_{\rm{prim}} + N_{\rm{single}}},
\end{equation}
with $N_{\rm{comp}}$ as the number of companion stars, $N_{\rm{prim}} = N_{\rm{mult}}$ as the number of primary stars and $N_{\rm{single}}$ as the number of single stars. 

\begin{figure*}
\centering
\includegraphics[width=\textwidth]{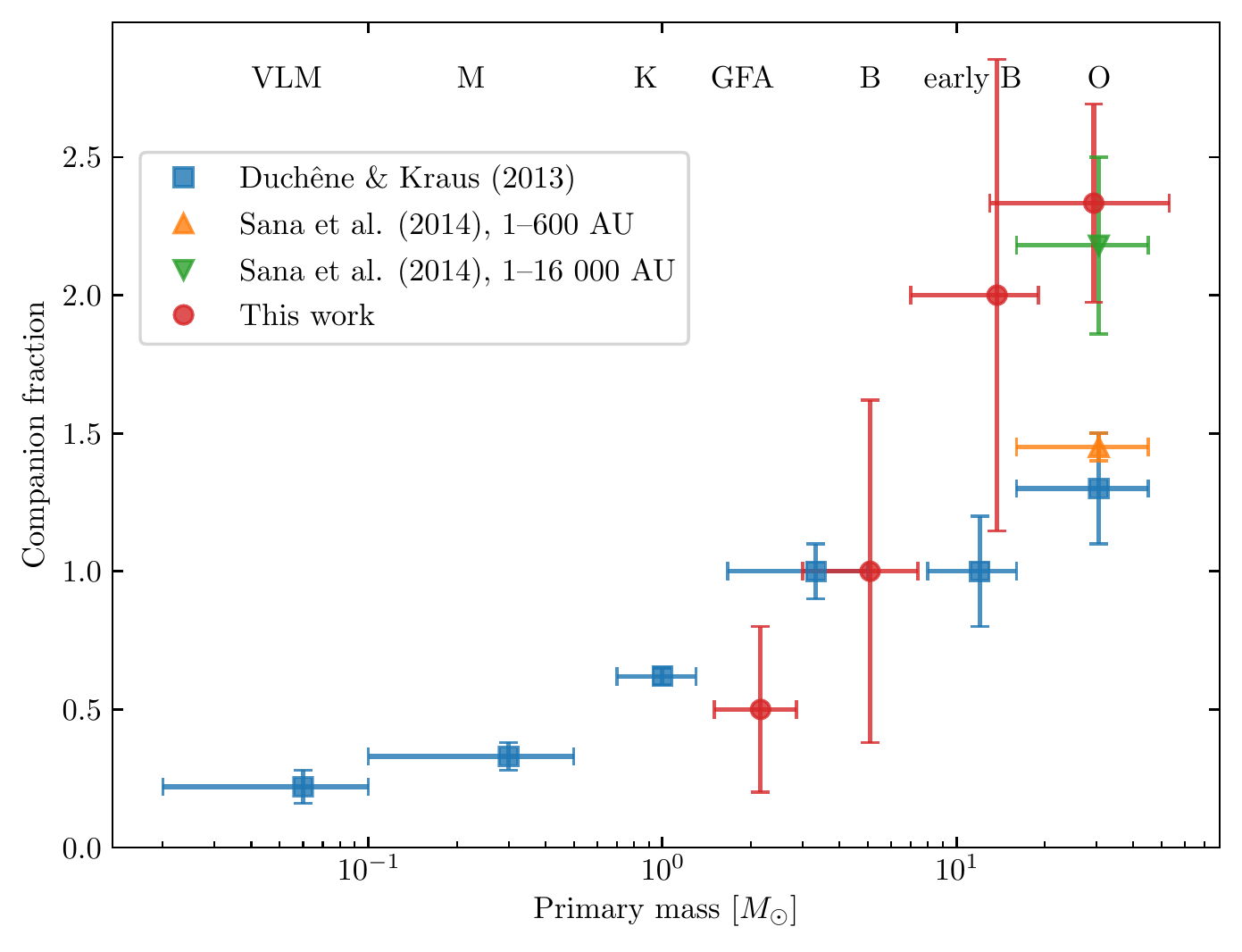}
\caption[Companion fraction per mass]{The companion fraction $CF$ as measured with GRAVITY for mass ranges $< 3$~\solmass , 3 \textendash 7~\solmass , 7 \textendash 16~\solmass , $> 16$~\solmass\ (red circles). The values from \citet{Duchene2013} for very low mass stars (VLM), spectral types M, K, G, F, A, B, early B and O (from left to right), and the values from \citet{2014ApJS..215...15S} for O stars are plotted for reference. The companion fraction of \citet{2014ApJS..215...15S} depends on the considered separation range for companions. A range $\lesssim 600$~AU is similar to the separations in our sample.}\label{fig:companionfraction_spt}
\end{figure*}

We compare our resulting companion fraction with the values from \citet{Duchene2013}, \citet{2014ApJS..215...15S}, and \citet{2017ApJS..230...15M}, as shown in Figure \ref{fig:companionfraction_spt}. %TODO plot
Generally, we notice a rising companion fraction for higher masses. For stars with 3--7~\solmass , we obtain a companion fraction of $1 \pm 0.6$, which agrees well with \citet{Duchene2013}, and \citet{2017ApJS..230...15M}. For 7--16~\solmass , our companion fraction is $2.0 \pm 0.9$ and agrees with $1.0 \pm 0.2$ by \citet{Duchene2013}, and $1.6 \pm 0.2$ by \citet{2017ApJS..230...15M} within the error bars. We get a companion fraction of $2.3 \pm 0.4$ for stars $> 16$~\solmass , which is nearly by a factor of two larger than the result $1.3 \pm 0.2$ obtained by \citet{Duchene2013}. \citet{2014ApJS..215...15S} presented the result of a survey of O stars and derived a companion fraction of $1.45 \pm 0.5$ for main sequence stars with companions at separations of 1\textendash 600~AU, which is comparable to our observed companion separations. Their survey covered companions up to separation ranges of 1\textendash 16~000~AU and yields a companion fraction of $2.18 ^{+0.3} _{-0.32}$ for O main sequence stars. We cannot assign companion stars at such large separations in the ONC, because the system would be unstable due to interactions with other cluster members. Still, our observed companion fraction of $2.3 \pm 0.4$, for separation ranges up to $\simsym 600$~AU, agrees with the result of $2.18 ^{+0.3} _{-0.32}$ by \citet{2014ApJS..215...15S}, for their full separation range of 1\textendash 16~000~AU. Our result is also in good agreement with the companion fraction of $2.1 \pm 0.3$, determined by \citet{2017ApJS..230...15M} for O-type stars with masses > 16~\solmass . 

\begin{figure*}
\centering
\includegraphics[width=\textwidth]{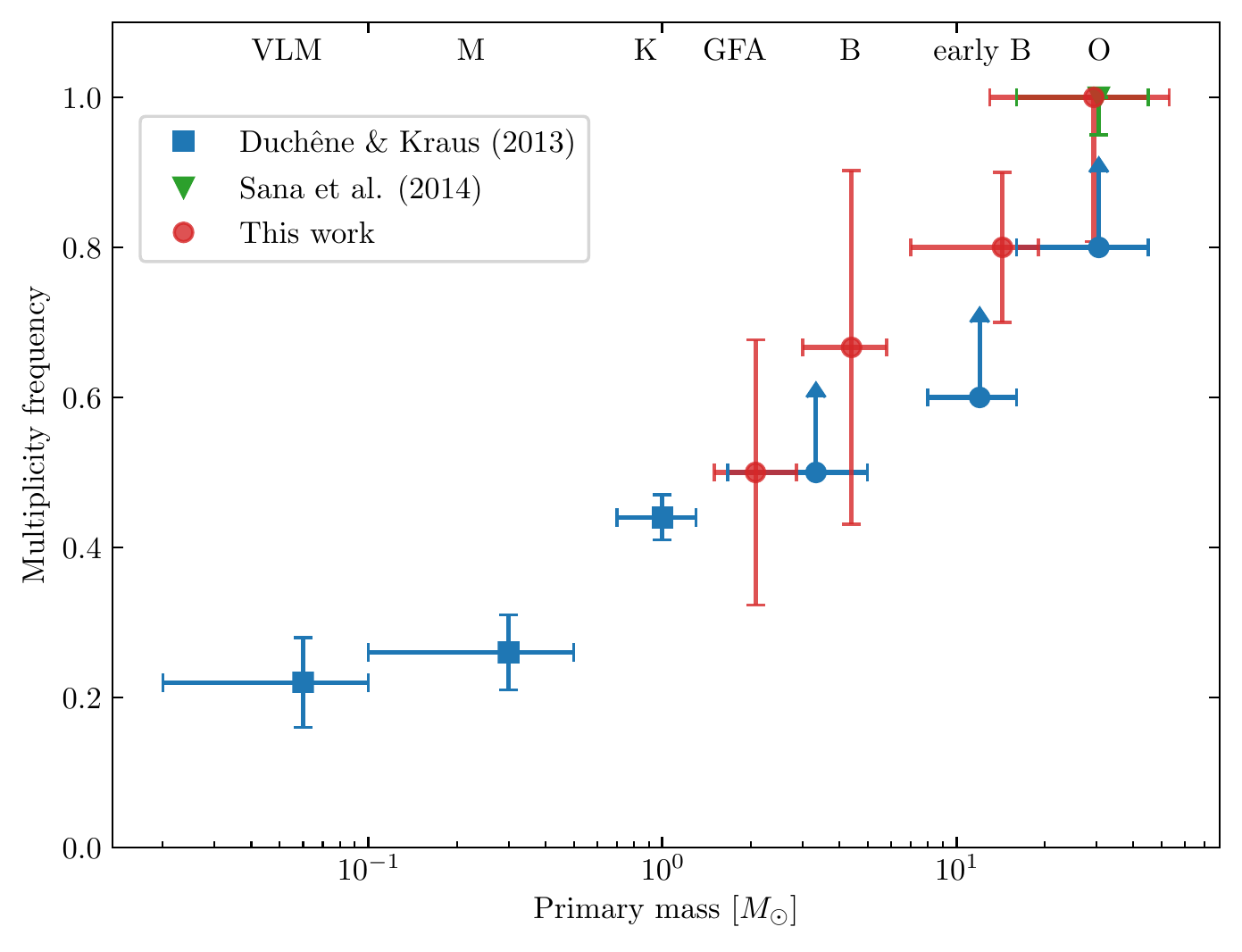}
\caption[Multiplicity fraction per mass]{The multiplicity fraction $MF$ as measured with GRAVITY for mass ranges $< 3$~\solmass , 3 \textendash 7~\solmass , 7 \textendash 16~\solmass , $> 16$~\solmass\ (red circles). The values from \citet{Duchene2013} for the whole mass range of $\lesssim 1$~\solmass\ to O stars and the values from \citet{2014ApJS..215...15S} for O stars are plotted for reference. For each mass range we use $1 / \sqrt{N}$ as uncertainty estimate.}\label{fig:mf}
\end{figure*}

The multiplicity fraction or multiplicity frequency (MF) is the number of multiple systems divided by the number of targets, as in \cref{eq:mf}. We compare our MF with previous results in \cref{fig:mf}. For stars $> 3$~\solmass , the MF of our sample is higher than the lower limits from \citet{Duchene2013}. For the mass range $\geq 16$~\solmass\ we obtain a MF of 100\%. This agrees with the MF $1 ^{+0.00} _{-0.05}$ from \citet{2014ApJS..215...15S} and $1 ^{+0.00}_{-0.2}$ from \citet{2017ApJS..230...15M} for main sequence stars of type O. We consider our sample to be complete for stars $> 3$~\solmass . For each mass range we use $1 / \sqrt{N}$ as uncertainty estimate. Thus, the actual MF of stars $< 3$~\solmass\ is probably higher than our biased value.

\subsection{Initial Mass Function}\label{sec:IMF}

In our discussion of the initial mass function, we will only consider stars $\geq 3$~\solmass . This corresponds to our most conservative detection limit (see \cref{tab:overview}). Down to this limit, we consider our sample to be complete for separations $\lesssim 600$~AU.

The initial mass function (IMF) describes the frequency of stars with masses in a given mass bin $(m + dm)$ at birth. For a detailed review of the IMF see \citet{2010ARA&A..48..339B, 1986FCPh...11....1S} or \citet{Kroupa2013} and references therein. \citet{1955ApJ...121..161S} suggested a power law of the form:
\begin{equation}
\Phi (\log m) = \dfrac{dN}{d \log m} \propto m^{-\Gamma},
\end{equation}
with $m$ as the mass, $N$ as the number of stars in the mass range $\log m + d \log m$ and $\Gamma = 1.35$. The IMF can also be written in the form of 
\begin{equation}
\xi (m) = \dfrac{dN}{dm} \propto m^{-\alpha},
\end{equation}
with the relation
\begin{equation}
\xi (m) = \dfrac{dN}{d m} = \dfrac{1}{m \log 10} \dfrac{dN}{d \log m} = \dfrac{\Phi (\log m)}{m},
\end{equation}
yielding $\alpha = \Gamma + 1$. For $> 1$~\solmass , \citet{Muench2002} suggested $\Gamma = 1.21$ for the Trapezium. More generally, \citet{2001MNRAS.322..231K} estimated $\Gamma = 1.3 \pm 0.7$, which agrees with \citet{2003PASP..115..763C} and their result of $\Gamma = 1.3 \pm 0.3$ for young clusters. \citet{Kroupa2013} noted that a good estimate for the intermediate mass regime $1 < m < 8$~\solmass\ is difficult, but suggested $\Gamma = 1.3$ as the best estimate. E.g. \citet{Muench2002, 2010ApJ...722.1092D} concluded that the IMF of the ONC is described by a Salpeter IMF for stars $\geq 0.6$~\solmass . Further measurements of the exponent focused mainly on the low mass regime ($\leq 1$~\solmass ), see e.g. \citet{doi:10.1093/mnras/stx1906}, \citet{2016MNRAS.461.1734D}, \citet{2012ApJ...748...14D} or \citet{Muench2002}.

\begin{figure*}
\centering
\includegraphics[width=\textwidth]{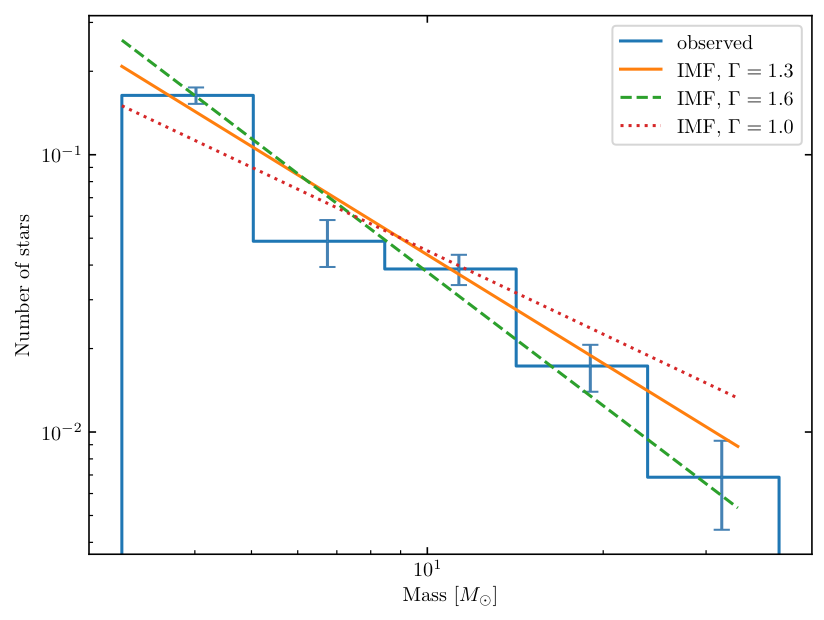}
\caption[Number of stars per mass]{Normalized histogram of stars per mass. The observed distribution is compared with distribution functions of \citet{2003PASP..115..763C}. We estimate the uncertainties to be $\pm 1/ \sqrt{N}$ per bin.}\label{fig:number_mass}
\end{figure*}

\begin{figure*}
\centering
\includegraphics[width=\textwidth]{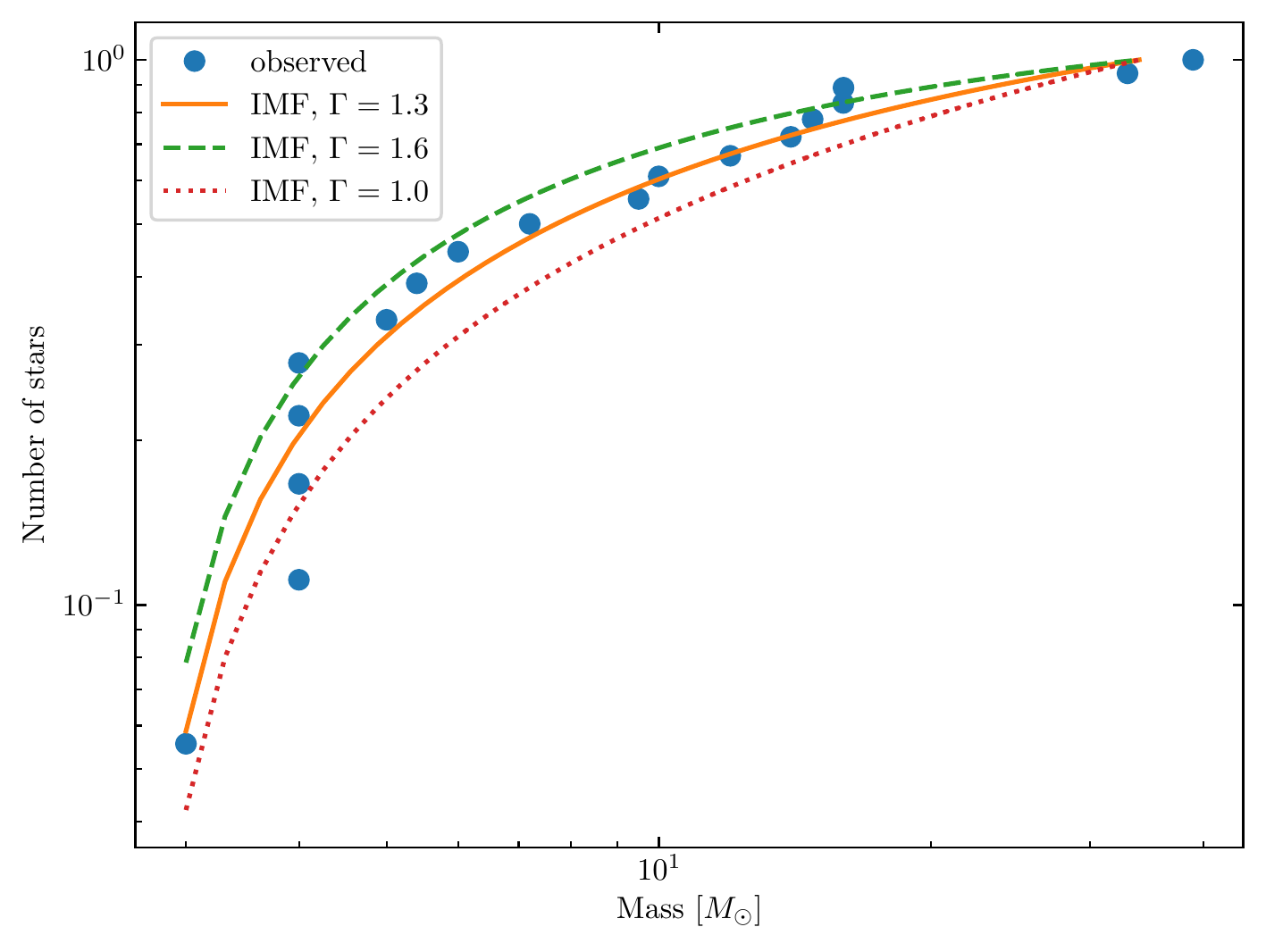}
\caption[Cumulative distribution of stars per mass]{Cumulative distribution of stars per mass with the respective distribution functions.}\label{fig:number_mass_cum}
\end{figure*}

We bin our observed distribution of masses for all stars and compare the histogram with a model IMF suggested by \citet{2001MNRAS.322..231K} and \citet{2003PASP..115..763C}, as shown in \cref{fig:number_mass}. The distribution is calculated by integrating the probability density $f(m) = dN / dm \propto m^{-2.3 \pm 0.3}$. The cumulative distribution is depicted in \cref{fig:number_mass_cum}.
The observed mass distribution of the most massive ONC stars --- including their companions --- agrees remarkably well with the IMF for field stars proposed by \citet{2001MNRAS.322..231K} and \citet{2003PASP..115..763C}. This illustrates the importance of resolving companions to get a complete sample. Counting only the primary component would result in a smaller power-law index and yield the impression that massive stars were distributed differently than in the field. 
We estimate the uncertainties in Figure \ref{fig:number_mass} to be $\pm 1/\sqrt{N}$ for each bin. 

\subsection{Distribution of Masses and Mass Ratios}
We compare stellar masses, mass ratios $q$, and separations of companions by plotting different quantities against each other. It is important to stress that our sample is only complete for masses $\geq 3$~\solmass . This leads to a bias, especially for the mass ratio, since we will be missing stars at the low mass end. We present our results here based on the observed stars, keeping in mind that we are missing a part of the population. Furthermore, photometric mass estimates rely strongly on evolutionary models and can comprise many uncertainties. It is only possible to provide a precise system mass if the Orbit is known.

\begin{figure*}
\centering
\includegraphics[width=\textwidth]{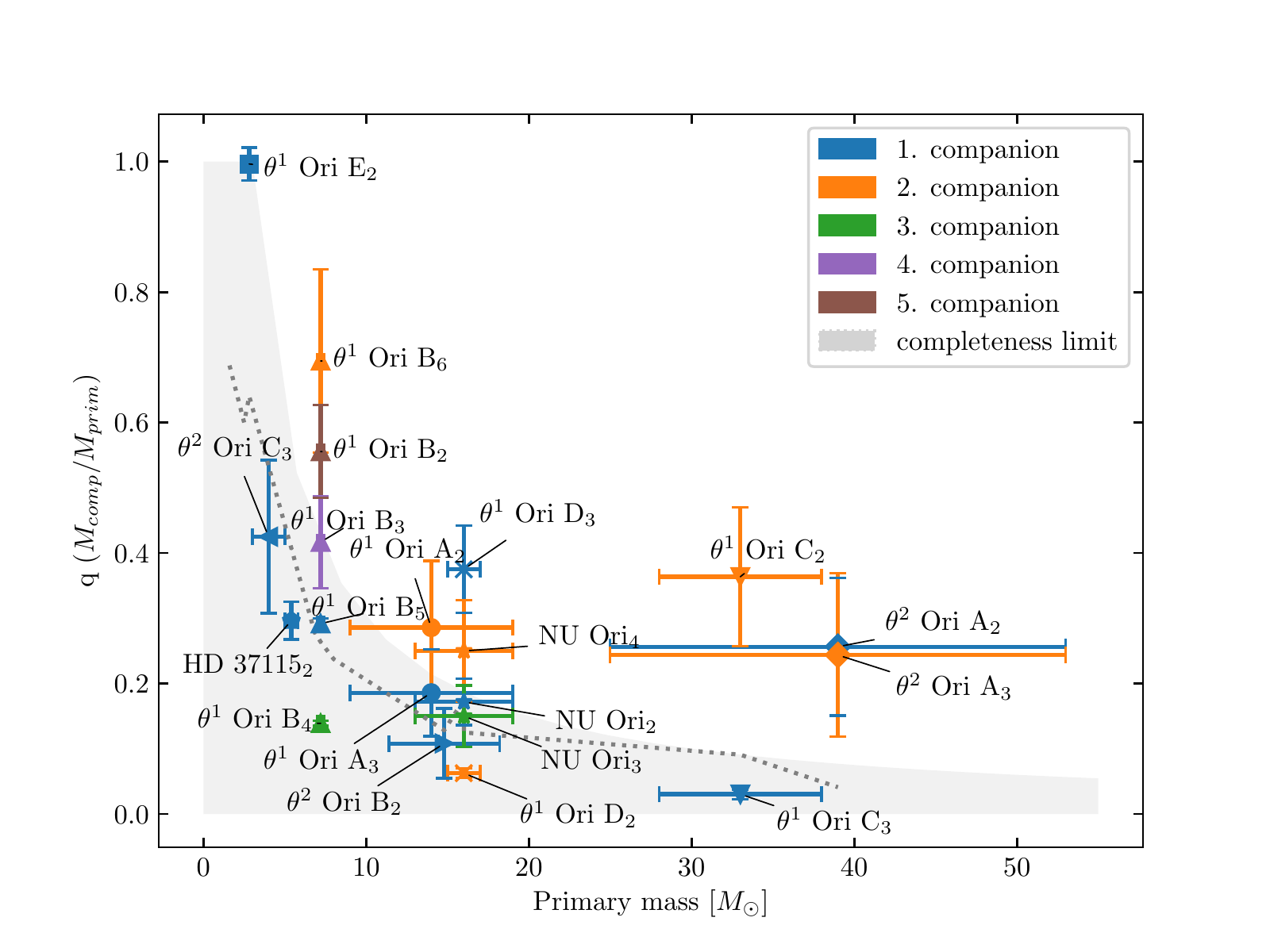}
\caption[Primary mass vs q sorted by separation]{Mass ratio per primary mass; each system is represented by a different marker. The colors indicate the companion order, starting from the innermost. Thus, the first companion (blue) means the companion closest to the primary, the second companion (orange) means the second-closest companion, etc. The gray area indicates the mass range $\leq 3$~\solmass , for which we are no longer complete. The gray dotted line marks the individual mass limits for each system as given in \cref{tab:overview}.}\label{fig:prim_mass_vs_q_sep}
\end{figure*}
 
First, we look at the primary mass and the corresponding mass ratios for all companions. In Figure \ref{fig:prim_mass_vs_q_sep}, the mass ratio per primary mass is displayed. Different markers represent different systems. Additionally, the color categorizes the companions according to their order, going from the innermost to the outermost. We find high mass ratios $\gtrsim 0.6$ only for primary masses $\lesssim 8$~\solmass . For stars with masses $\gtrsim 8$~\solmass , the mass ratio is $\lesssim 0.5$. In every system, the most massive companion is either the closest or second-closest companion to the primary star. 

The most massive companions $\gtrsim 8$~\solmass\ belong to the most massive primary stars with $\geq 30$~\solmass . They have a mass ratio in the range 0.3--0.4. We see no tendency for the companion to have a high mass similar to the primary --- e.g. $q \approx 1$ --- if the primary is a high mass star. This disagrees with the result of \citet{Chini2012}, who found that most massive stars have companions of similar mass. In our sample, only intermediate mass stars with masses between 3\textendash 8~\solmass\ have a $q \gtrsim 0.5$.

\begin{figure*}
\centering
\includegraphics[width=\textwidth]{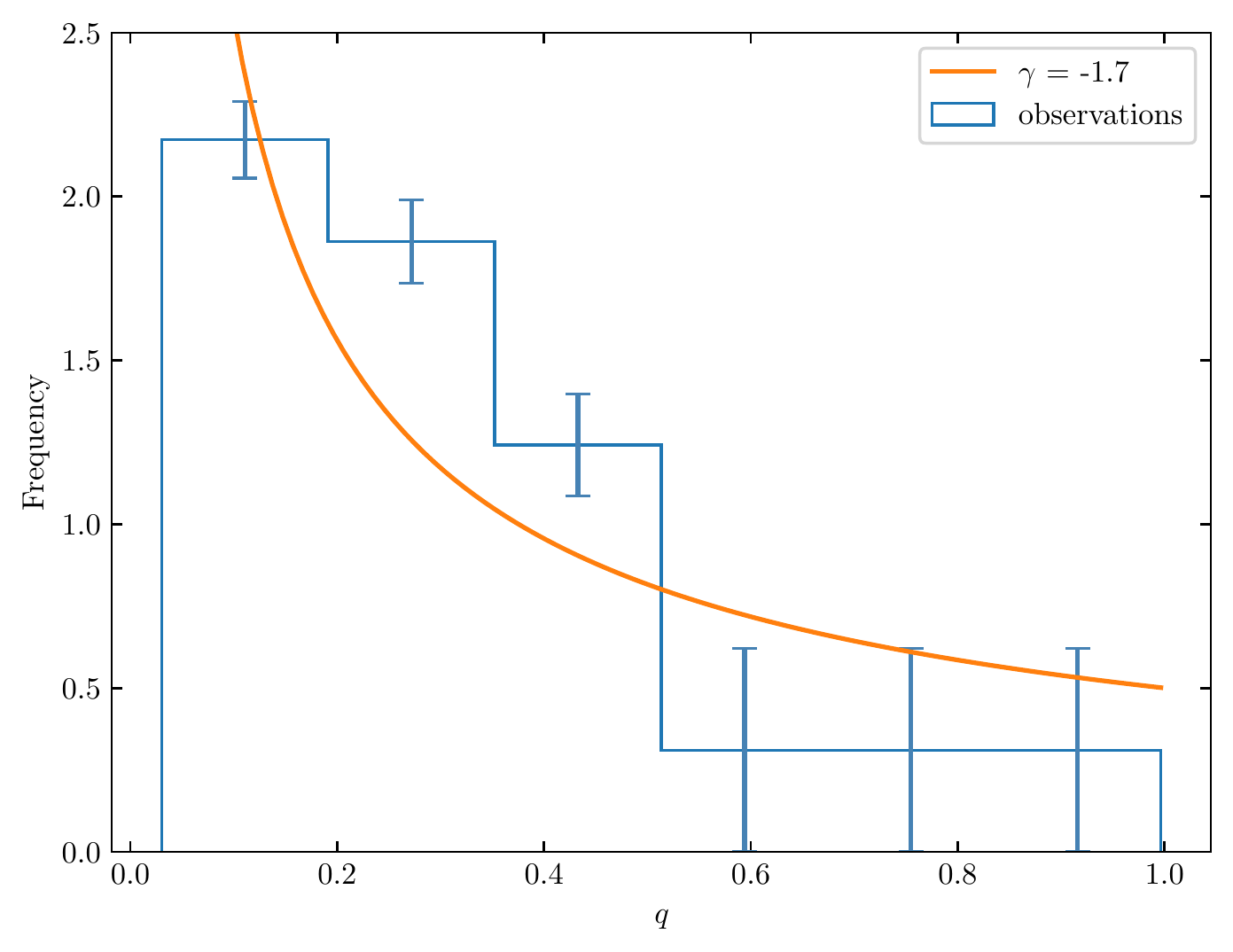}
\caption[Mass ratio distribution and best power-law fit]{Normalized histogram of all mass ratios $q$ and the best fit curve of a power law $\propto q^{\gamma} = q^{-1.7 \pm 0.3}$. }\label{fig:mass_q_dist}
\end{figure*}

\begin{figure*}
\centering
\includegraphics[width=\textwidth]{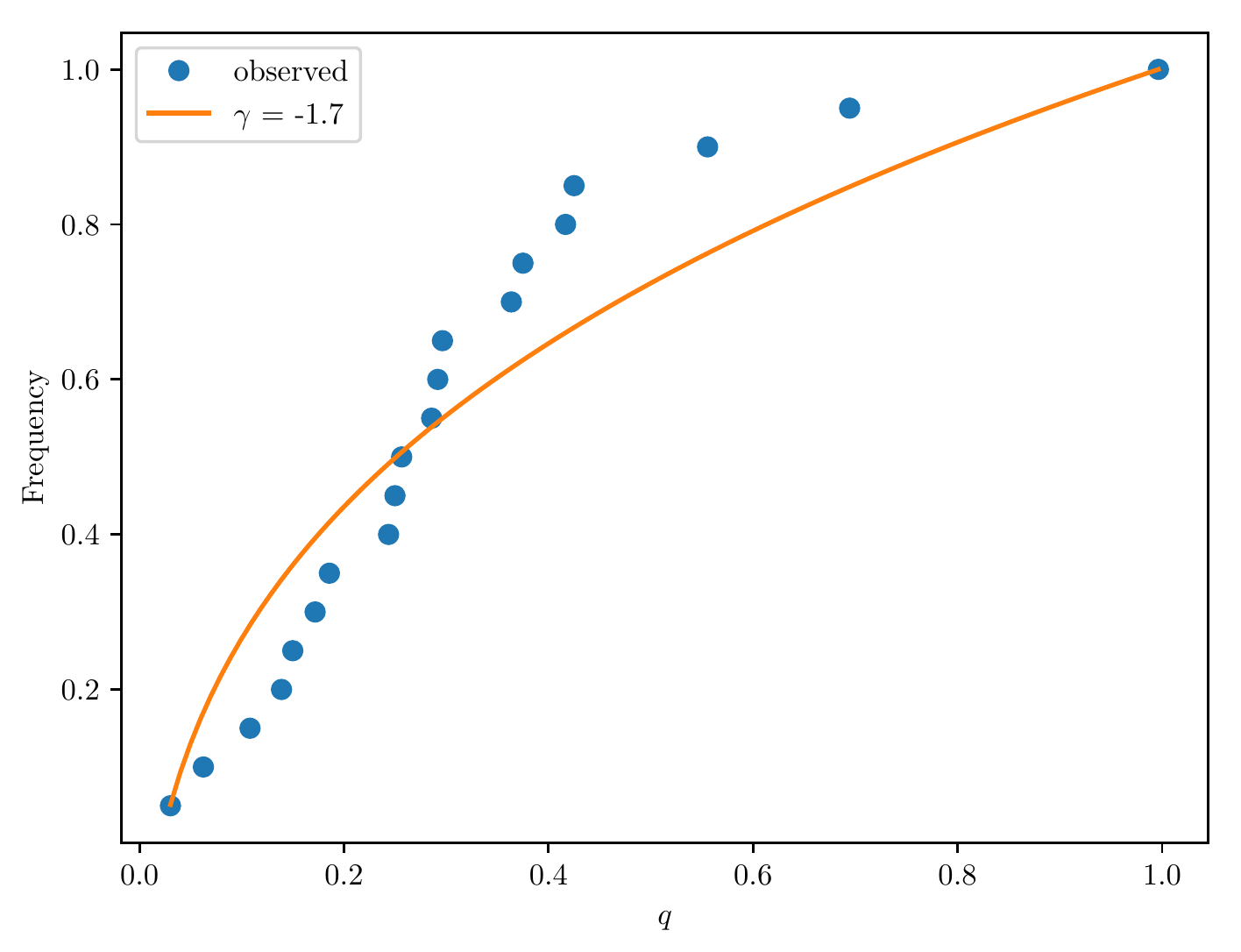}
\caption[Cumulative mass ratio distribution and best power-law fit]{Cumulative distribution of all mass ratios $q$ and the best fit curve of a power law $\propto q^{\gamma} = q^{-1.7 \pm 0.3}$.}\label{fig:mass_q_dist_cum}
\end{figure*}

A histogram of the mass ratio $q$ is shown in \cref{fig:mass_q_dist} and the cumulative distribution in \cref{fig:mass_q_dist_cum}. We find a clear preference for small mass ratios, hence large mass differences. The most common values are $\leq 0.2$. Our distribution is biased, because our sample is incomplete for stars with masses $< 3$~\solmass . Furthermore, small mass ratios ($q < 0.01$) are hard to detect with interferometry and imaging due to the extreme contrast ratios. In other words, any survey will be incomplete at low $q$. The underlying distribution is likely to show an even stronger preference for low $q$. We want to compare our distribution with other samples. \citet{Duchene2013} stated that a power law is not an ideal representation for most distributions of $q$, but it is still the best way to compare multiple systems with different mass ranges. Thus, we follow their approach and fit our distribution with a power law $d N / dq \propto q^{\gamma }$, similar to \cref{sec:IMF}. Our distribution of $q$ is best fitted by  $\gamma = -1.7 \pm 0.3$ (see \cref{fig:mass_q_dist}). \citet{2017ApJS..230...15M} found values for $\gamma$ of $-0.5 \pm 0.3$ for stars with a period of 10 days, $\gamma = -1.7 \pm 0.3$ for stars with a period of 1000 days and masses > 5~\solmass , and $\gamma = -2.0 \pm 0.3$ for stars with a period $\gtrsim 270 $~years and masses > 5 \solmass . Our sample consists of stars with periods ranging from a few days to > 1000~yr (see \cref{tab:overview}). These values agree well with our resulting $\gamma$ of $-1.7 \pm 0.3$. Thus, considering the uncertainties, our companion mass distribution can be described by an IMF with $\alpha = 2.3 \pm 0.3$. In \cref{fig:mass_q_dist}, we calculate $\gamma$ for \citet{2003PASP..115..763C} by taking a primary mass and considering companions down to $q = 0.1$. For stars $\leq 1$~\solmass the IMF follows a lognormal distribution. As soon as the primary mass is $< 10$~\solmass , the lognormal distribution changes the companion mass distributions of companions $\leq 1$~\solmass . Hence, $\gamma$ is no longer constant.

\begin{figure*}
\centering
\includegraphics[width=\textwidth]{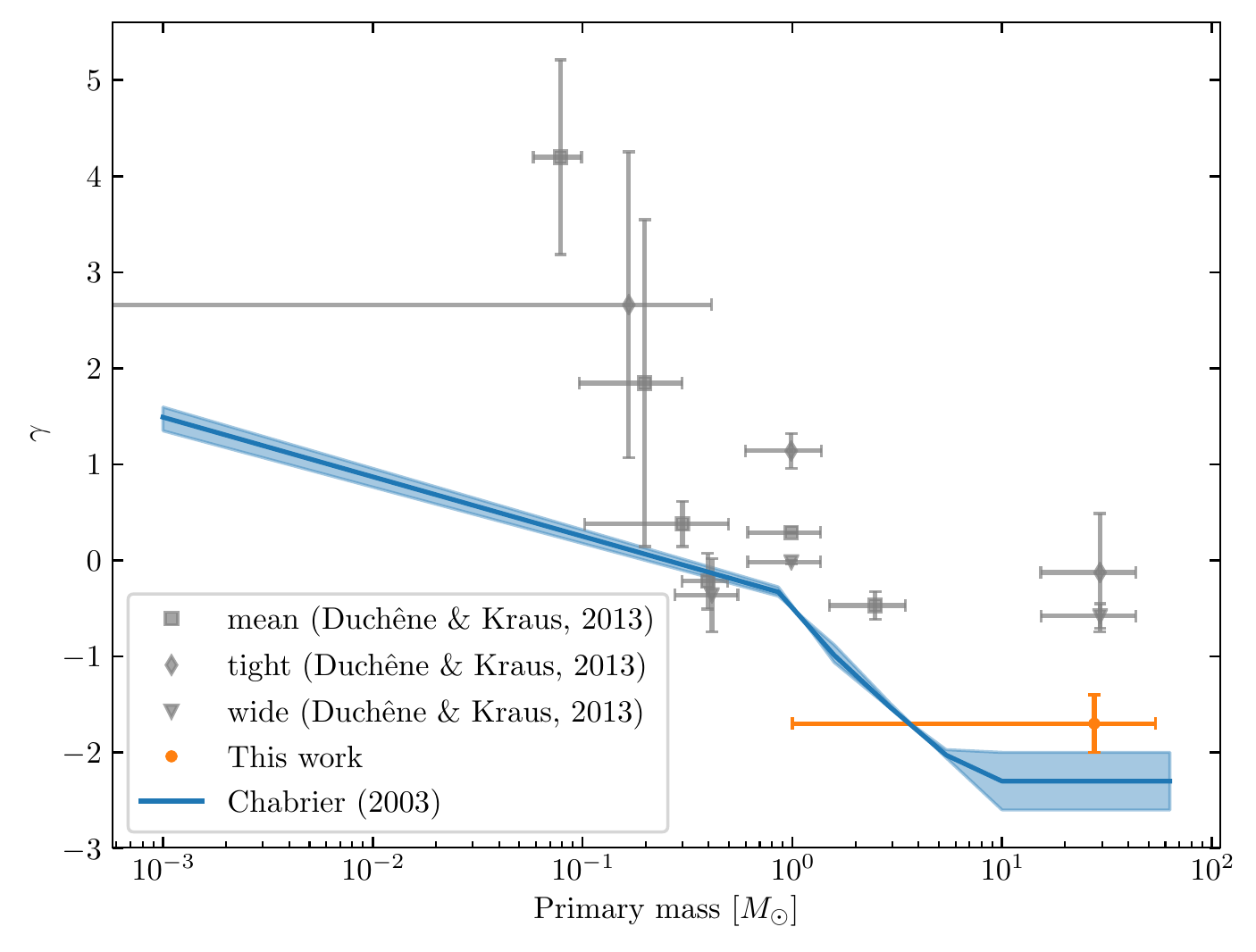}
\caption[Exponent of $q$ distribution per primary mass range]{The power-law index $\gamma$ of the mass ratio distribution with $d N / d q \propto q^{\gamma}$ for different masses. The squares represent the fits of the overall population of multiple systems by \citet{Duchene2013}. The diamonds are the results for tight binaries (smaller separations than average) and the down-facing triangles the result for wide binaries (larger separations than average) by \citet{Duchene2013}. Our data is represented by a distribution with an index of $-1.7\pm 0.3$ (red circle). If the companion stars follow the IMF of \citet{2003PASP..115..763C}, the index of $q$ is equal to the index of the IMF with $\gamma = -2.3 \pm 0.3$, for companions with masses $1 \leq m_{\rm{comp}} \leq m_{\rm{primary}}$. The shaded area indicates the uncertainties of the IMF.}\label{fig:q_gamma_mass}
\end{figure*}

We compare our distribution of $q$ with the distributions in \citet{Duchene2013}. \cref{fig:q_gamma_mass} shows $\gamma$ for systems with different primary mass ranges. Our sample is in the range $\gtrsim 1$~\solmass . The distributions by \citet{Duchene2013} follow a nearly flat distribution $\gamma \lesssim 0.5$ for masses $\gtrsim 0.3$~\solmass . We notice a steeper distribution in our sample with $\gamma = -1.7$ than \citet{Duchene2013} for masses $\geq 3$~\solmass . If we assume that the companion mass follows an IMF, the index $\gamma$ will correspond to the power-law index $\alpha = - \gamma$. Our result agrees within the uncertainties of an IMF according to \citet{2001MNRAS.322..231K} with $\alpha = 2.3 \pm 0.6$ and is consistent the exponent $1.90 ^{+0.37} _{-0.36}$ obtained by \citet{Schneider69}. %However, we are not in the range of \citet{2003PASP..115..763C} with $\alpha = 2.3 \pm 0.3$.

\begin{figure*}
\centering
\includegraphics[width=\textwidth]{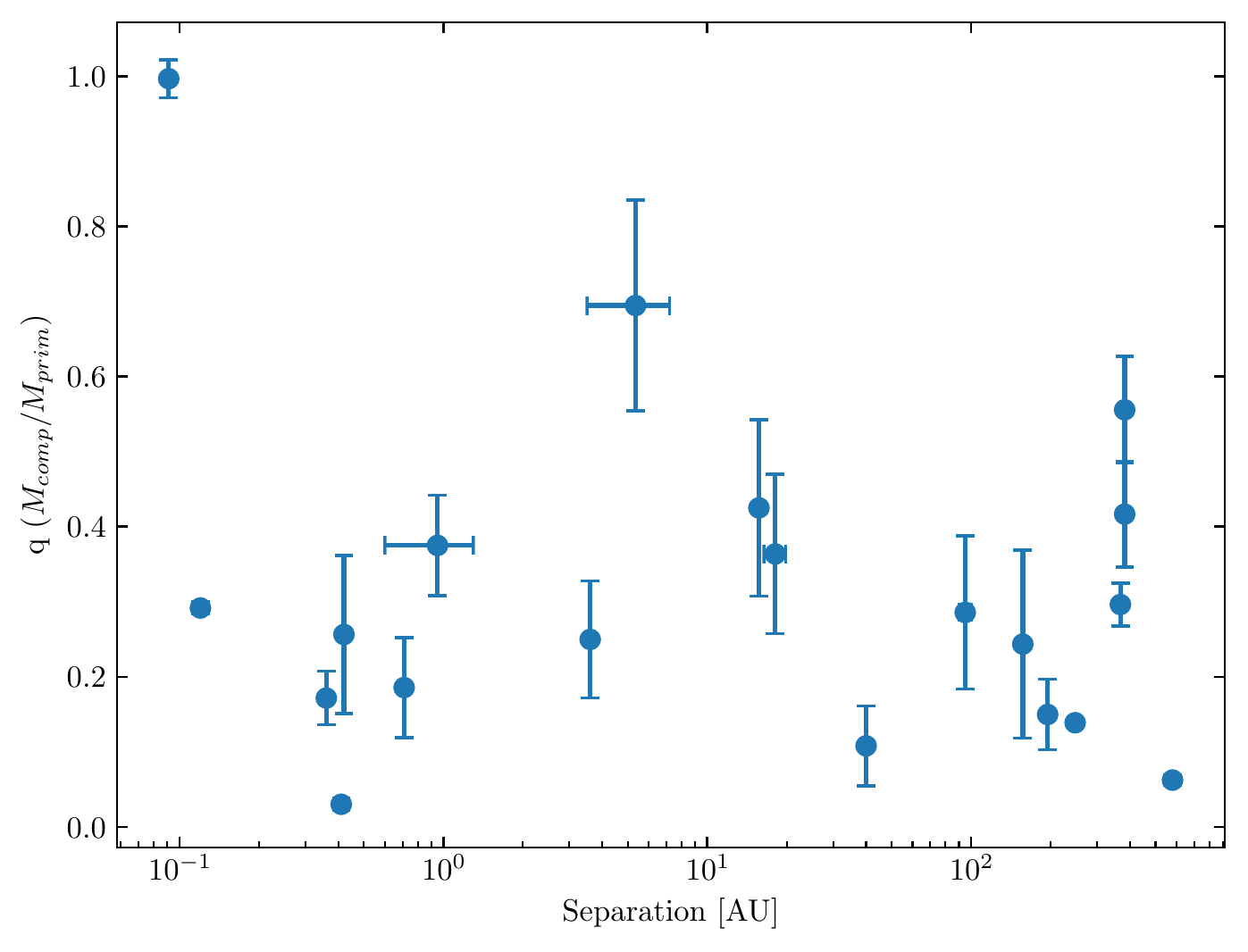}
\caption[Mass ratio per separation]{Separation in AU in a logarithmic scale per mass ratio $q$. }\label{fig:q_vs_comp_mass_sep}
\end{figure*} 

In \cref{fig:q_vs_comp_mass_sep}, we compare the resulting mass ratios $q$ with the corresponding companion separation. We notice no significant correlation between q and separation. The only system with $q \approx 1$ has a small separation of $\approx 0.09$~AU. But we also find systems with a separation of $\simsym 400$~AU and $q\sim 0.6$, which is one of the highest mass ratios in our sample.

\begin{figure*}
\centering
\includegraphics[width=\textwidth]{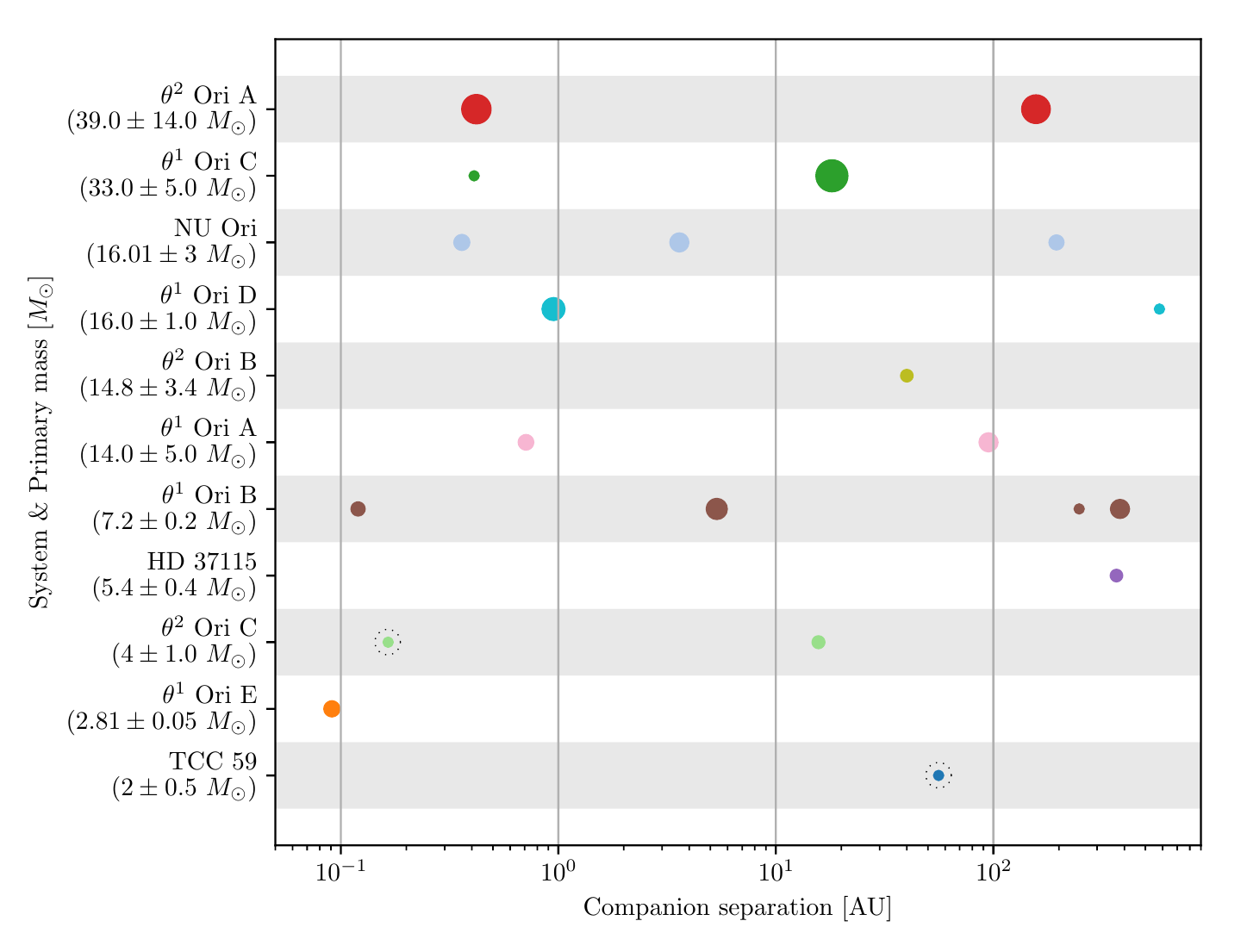}
\caption[Primary mass vs separation]{The companion separation per system with given primary mass in \solmass . In the range between 1\textendash 100~AU, we are sensitive down to 3~\solmass . There are equally many companion stars in the range 0.1\textendash 1~AU as in the range 1\textendash 100~AU. The colors indicate different systems, the marker size scales with the square root of the companion mass. The dotted circles indicate missing information about the mass of the first companion of $\theta ^2$ Ori C and TCC 59. }\label{fig:prim_mass_sep}
\end{figure*}

Figure \ref{fig:prim_mass_sep} shows the companion separation per system. There appears to be a preferred separation range for companion stars. We find eight companions within 1~AU, thus we notice a tendency for binaries with separation $< 1$~AU, which are typically spectroscopic binaries. Then there are only two companions within 1\textendash 10~AU. The next five companions cover the range 10\textendash 100~AU. This means, there are as many companions within 1\textendash 100~AU as there are within 0.08\textendash 1~AU.

\begin{figure*}
\centering
\includegraphics[width=\textwidth]{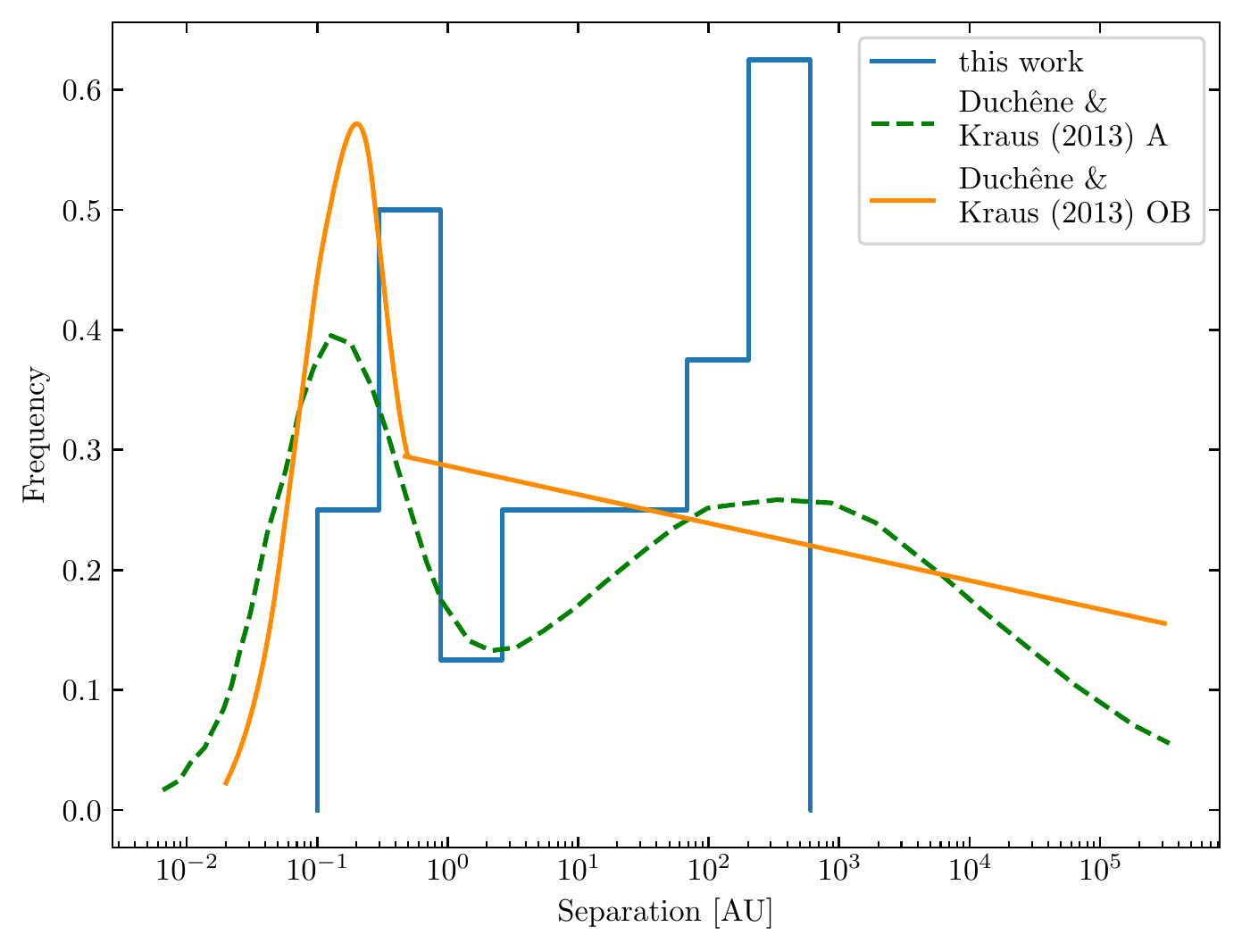}
\caption[Histogram of separation for companions with distributions from \citet{Duchene2013}]{Normalized histogram of companion separation.  For comparison, we plot the qualitative distribution of \citet{Duchene2013} for A (dotted) and OB stars (solid). The distribution of our separations for mainly B and O stars is bimodal and resembles more that of the \citet{Duchene2013} distribution of A-type stars. We notice a decrease of companions in the range of 1\textendash 100~AU.}\label{fig:hist_sep}
\end{figure*}

We present the distribution of companion separation in \cref{fig:hist_sep}. \citet{Duchene2013} described the orbital period distribution of OB stars with a peak at $\simsym 0.2$~AU and with a decreasing power-law tail for $> 1$~AU. They also determined the distribution for A stars, which is bimodal. A qualitative sketch is provided in \cref{fig:hist_sep} in comparison to the distribution of our observed separations. We also find a peak for companions in short distance but in a range up to $\simsym 4$~AU. Furthermore, we notice a decrease between 1\textendash 100~AU and a second peak at 400\textendash 600~AU, i.e. a bimodal distribution. Our distribution more closely resembles that of the \citet{Duchene2013} distribution for A stars, even though our sample consists mainly of O and B stars (see \cref{tab:overview}). 

\begin{figure}
\begin{minipage}{0.49\textwidth}
\centering
\includegraphics[width=\textwidth]{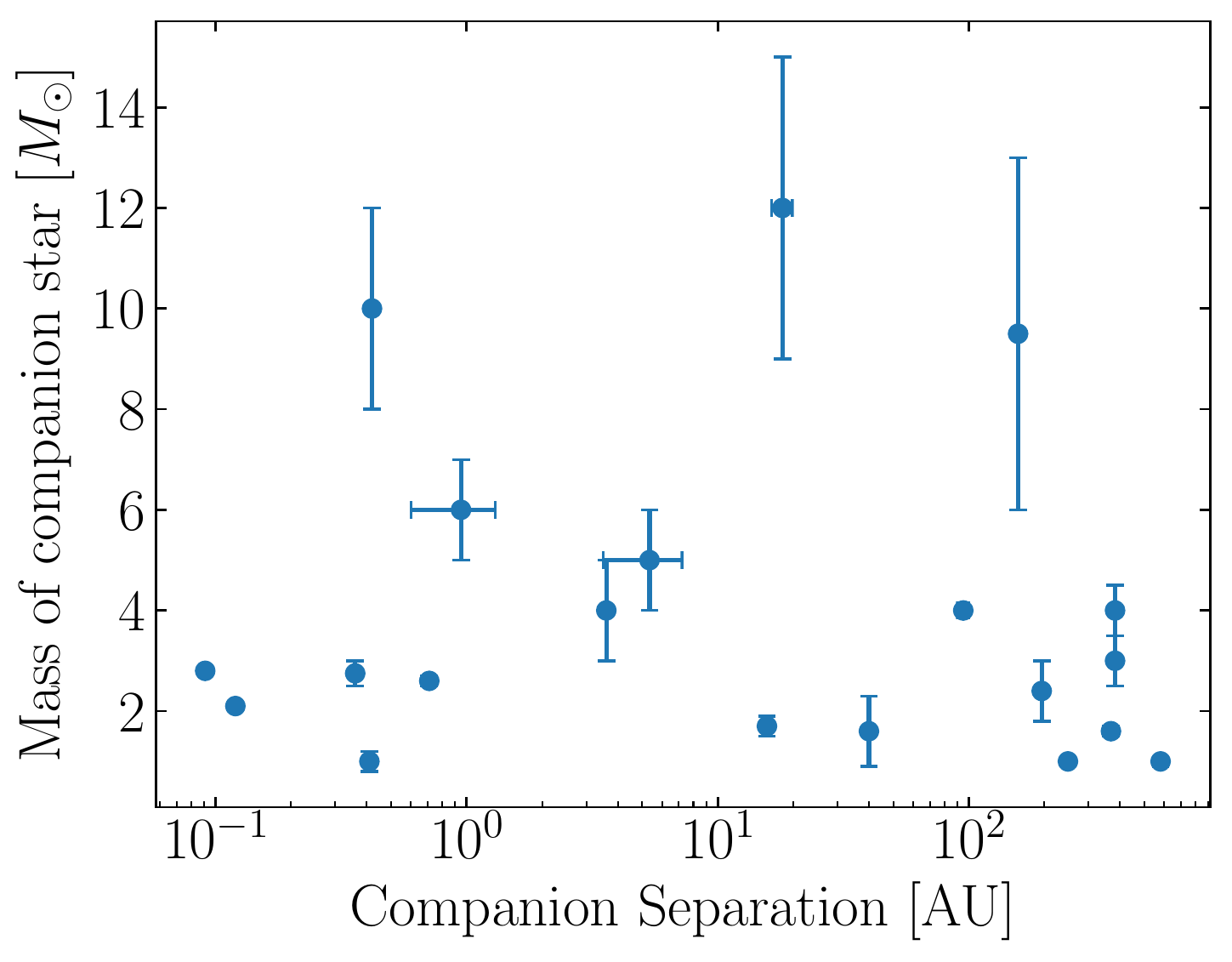}
\end{minipage}
\begin{minipage}{0.5\textwidth}
\includegraphics[width=\textwidth]{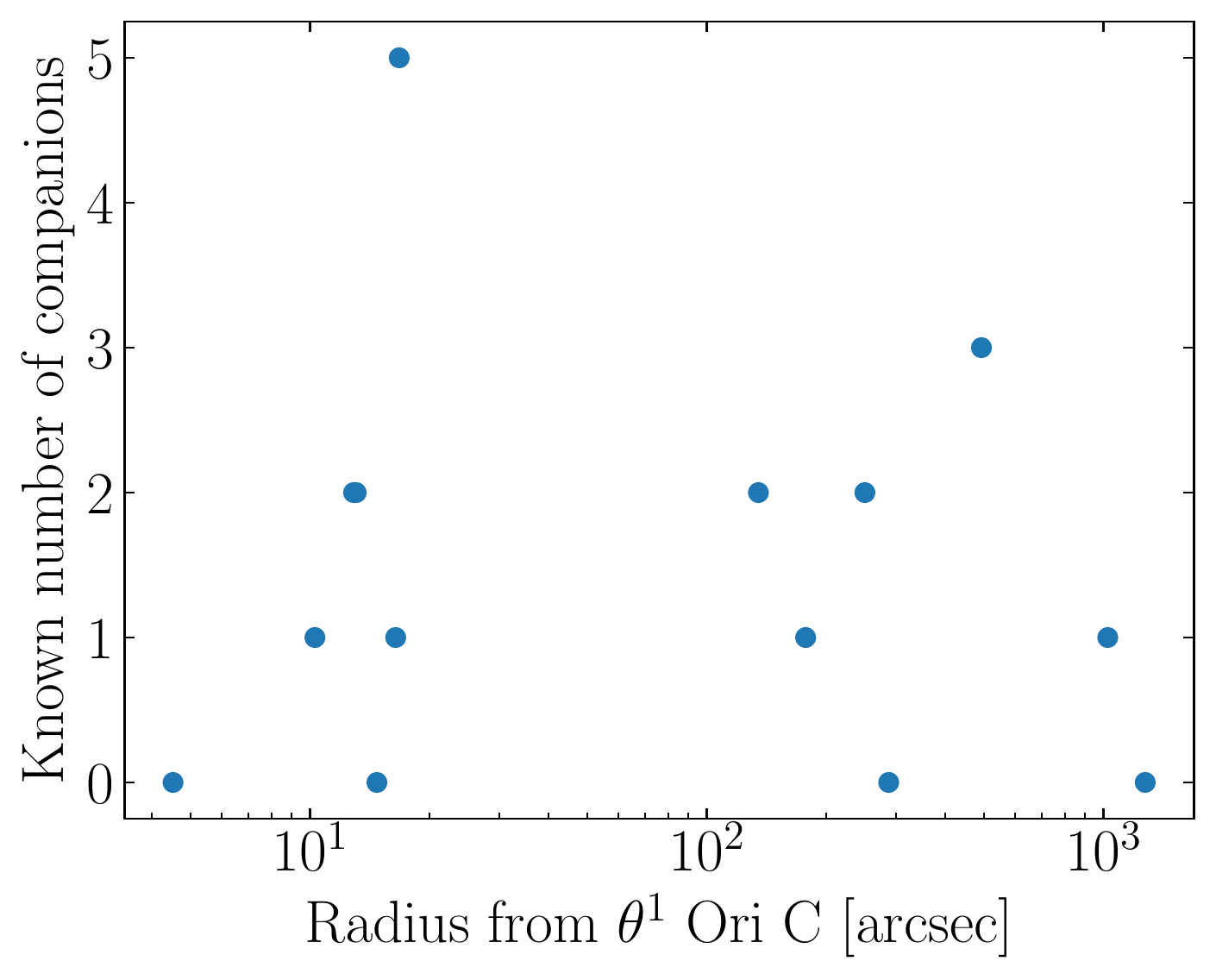}
\end{minipage}
\caption[Companion mass vs separation and Multiplicity vs radial distance from $\theta ^1$ Ori C]{Left: The separation versus the companion mass. Right: Multiplicity of objects plotted as a function of distance from $\theta ^1$ Ori C in arcseconds.}\label{fig:comp_mass_sep}
\end{figure}

On the left of \cref{fig:comp_mass_sep}, we compare the companion mass with separation. There appears to be no trend for the companion mass over the separations.
Finally, we compare the multiplicity with the system distance from $\theta ^1$ Ori C as shown in Figure \ref{fig:comp_mass_sep} on the right. In a comparable plot by \citet{Preibisch1999}, there was a trend towards fewer companions for more distant objects from $\theta ^1$ Ori C. We do not observe such a trend in our sample. It is possible that stars move away from their original birthplace, so that their location and multiplicity differs from their primordial distribution. It is also possible that the multiplicity was not triggered by winds and outflows of $\theta ^1$ Ori C, but is universal. It could also depend on the stellar density of the cluster in general.

\subsection{Comparison with Star Formation Models}
We compare our results with star formation models. The core accretion model predicts that the multiplicity and companion fraction rises with stellar mass \citep{2001IAUS..200..346C}. Disk fragmentation predicts low mass companions at 100-1000 AU \citep{2006MNRAS.373.1563K, Krumholz2016}. \citet{2006ApJ...641L..45K} suggested that high temperatures stabilize the core, leading to less fragmentation, even for high mass cores. This results in a small amount of massive protostars and a preference for binaries with high mass components. These massive components of high mass stars make the mass distribution of companion stars top-heavy. The massive core fragment also needs a larger volume to form, thus, we expect a correlation between system mass and separation. %After dynamical interactions, the typical outcome is 

Competitive accretion predicts a dependence of the separation $r$ on the system mass $M$. In a turbulent medium, the relation is $r \propto M^{-2}$ \citep{2005MNRAS.362..915B}. The result of competitive accretion is a cluster with a large range of masses, where high mass stars are formed at the center of the cluster core \citep{2005IAUS..227..266B} with close high mass companions. The companion frequency rises with stellar mass, see e.g. \citet{2012A&A...538A..74P}.

Binaries with low masses and wide separations continue to accrete mass and evolve to a close high mass binary. The result is a close system with two massive components \citep{2005MNRAS.362..915B}. The fragmentation of clouds leads to clusters similar to the Trapezium Cluster \citep{2005MNRAS.362..915B, 2005ASSL..327..425B, 2005IAUS..227..266B}. The model also predicts three-body captures, where a high mass primary star with a low mass companion captures a massive wide companion star \citep{2005IAUS..227..266B}. The massive companion star absorbs most of the binding energy of the low mass star and the separation between the two high mass components shrinks. The low mass companion is either ejected or evolves to a wide binary. The resulting system is a close high mass binary with a high mass companion and a third low mass component at larger separations \citep{2002MNRAS.336..705B}. This preference for high mass companions also yields a top-heavy companion mass distribution. Another important factor is the low mass accretion rate of competitive accretion. \citet{2010ApJ...709...27W} showed that star formation with competitive accretion requires $10 ^6$~yr when considering protostellar outflows and magnetic fields. Dynamical processes also need time to take place.

For massive star formation through stellar collisions/mergers, the stellar mass distribution does not result in a Salpeter IMF. The collisions lead to runaway growth of a few objects, which does not produce a smooth mass distribution \citep{2011MNRAS.410.2799M, Krumholz2015}. The collision of stars requires high densities $> 10^6$ stars/pc$^3$. Massive stars are a merger product and become less likely to be close binary systems \citep{2005MNRAS.362..915B}. 

Our observations of the stars in the ONC yield a Salpeter IMF for all stars, including the companions. We do not observe massive binaries with equally massive companion stars. Also, we find no preference for close massive systems. The mass of the companion star is not correlated with the separation. We only find high mass ratios ($\gtrsim 0.5$) for primary stars with $\lesssim 7$~\solmass . In our sample, we observe fewer companions in the range of 1\textendash 100~AU. This indicates different formation mechanism for different separation ranges. This could be the transition of a mechanism responsible for tight binaries, e.g. failed mergers or accretion onto binaries, and e.g. disk fragmentation. \citet{doi:10.1146/annurev-astro-081915-023307} concluded that the disk of a star with 8~\solmass\ is sufficiently cool for fragmentation at $\geq 50$~AU. This agrees with our gap between 1\textendash 10~AU and a slowly rising number of companions within 10\textendash 100~AU.
\cref{tab:comparison_models} provides an overview of the various models and the predicted correlations. We compare the predictions with our findings. 

{\renewcommand{\arraystretch}{1.9}
\begin{table*}
\centering
{\small
\begin{tabular}{ccccc}
\toprule
Parameter &\thead{Core \\ accretion}&\thead{Competitive \\ accretion} &Collisions &This work \\
\midrule
CF &CF $\propto$ mass 	&CF $\propto$ mass &--  & CF $\propto$ mass \\

 IMF &top heavy	& top heavy & \makecell{strong deviation \\ from Salpeter IMF} &Salpeter IMF \\

$m_2$ and $m_1$ &correlated &-- &-- &\makecell{uncorrelated/ \\ slightly correlated} \\

 $r$ and $M$ &correlated & $r \propto M^{-2}$ &-- &uncorrelated \\

$q$ and $r$ &-- & anti-correlated &-- &\makecell{ most $q \leq 0.5$ \\ uncorrelated}  \\
\bottomrule
\end{tabular} 
}
\caption[Comparison of star formation model properties with observed properties]{Comparison of observable quantities from the star formation models with our observations.}\label{tab:comparison_models}
\end{table*}
}

We can exclude collisions as the main star formation process, because it does not represent our IMF and the density of the Trapezium \citep[$\simsym 10^4$ stars/pc$^3$][]{Hillenbrand1998a} is lower than the required $\simsym 10^6$ stars/pc$^3$. $\theta ^1$ Ori C$_1$ might still be the result of stellar merging, but the collision of stars is not a dominant process in the ONC. We find several aspects in favor of competitive accretion in our sample, for example, the formation of massive stars at the cluster center and the variety of the mass range. However, other features do not support competitive accretion, e.g. we see no tight massive binaries or a preference for equal masses. 

An important factor is that competitive accretion needs $\simsym 1$~Myr to form massive stars \citep{2010ApJ...709...27W, 2010ARA&A..48..431P}, whereas the ONC has an average age of $< 1$~Myr with a spread of less than 2~Myr. Core accretion provides a mechanism of gaining mass without tightening the binary separation. Both models have difficulties with reproducing our observations. However, we clearly do not find a strong dependence of binary separation with system mass $r \propto M^{-2}$ and an anti-correlation of mass ratios and separation. This argues against a dominant mode of competitive accretion. We also need to consider that some dynamic processes may not be completed yet, which could change the masses or companion fraction of the ONC. 

\citet{2012ApJ...754...71K} showed in a simulation that a combination of core accretion and competitive accretion is also possible. Massive stars started formation in distinct massive cores, according to the core accretion model. But the formed stars engaged in dynamical interactions while accreting mass, similar to the competitive accretion model. This resulted in hierarchical systems like the Trapezium Cluster. This combination could thus also be a possible scenario for the Trapezium, but needs further examination.

One has to note that both accretion scenarios initially form companions at large radii, which become eventually tight binaries, e.g. \citet{2007ARA&A..45..565M}. This tightening process occurs through dynamical interactions with the disk or cluster members, or through ambient gas accretion and energy loss. It is therefore plausible that the observed fraction of close equal mass binaries depends on the cluster age. The ONC --- with an age of ~1 Myr \citep{Hillenbrand1997} --- is one of the youngest massive star clusters in the Milky Way. Our finding of a large number of wide binaries with high mass ratios  might reflect the fact that the binary population in the ONC did not have enough time to be altered by dynamical interactions e.g. (\citet{2018ApJ...854...44M} show different simulated scenarios of how close binaries can be formed, 60\% form by unstable triples). \citet{2017A&A...599L...9S} also notice a lack of close companions and conclude their findings may support a theory in which binaries form initially at large separations and then harden to closer systems. %In other words, close systems did not have enough time to form. 
This could explain why the companion separations and masses in the ONC are different than the distributions in more evolved clusters.

\section{Conclusion}%
\label{chap:conclusion}
%Sterne
In order to gain a deeper understanding of massive star and cluster formation, this work presents an interferometric study of massive stars in the Orion Trapezium Cluster and its vicinity. The outstanding resolution of the VLTI ($\simsym 2$~mas) and the sensitivity of GRAVITY allowed us to probe stars for companions in the widely unexplored range of 1--100~AU. We observed the 16 most massive stars with masses of 2\textendash 44 \solmass . We detected three new companions for the systems $\theta ^1$ Ori B, $\theta ^2$ Ori B, and $\theta ^2$ Ori C. We confirmed the suspected companion for NU Ori and determined a separation of $3.6 \pm 0.1$ AU. Combined with the companions reported in the literature --- based on speckle/AO imaging and spectroscopic surveys --, we find a total of 22 companion stars. $\theta ^1$ Ori B$_6$ is at a separation of 3.5\textendash 7.2 ~AU and we estimated a mass of $7.3 \pm 0.5$~\solmass . The new companion $\theta ^2$ Ori B$_2$ has a separation of $40 \pm 1$~AU and an approximate mass of $1.6 \pm 0.7$~\solmass . For $\theta ^2$ Ori C$_2$ we determined a separation of $15.7 \pm 0.2$~AU and estimated a mass of $1.7 \pm 0.2$. NU Ori$_4$ has a separation of $3.6 \pm 0.1$~AU and an estimated mass of $4 \pm 1$~\solmass . 

We confirmed companions for $\theta ^1$ Ori A, $\theta ^1$ Ori C, $\theta ^1$ Ori D, and $\theta ^2$ Ori A, all with substantially improved astrometry and photometric mass estimates. Additionally, we refined the orbit of the eccentric high-mass binary $\theta ^1$ Ori C and obtained a period $P = 11.4 \pm 0.2$~yr and a semi-major axis $a = 18.2 \pm 0.3$~AU. Furthermore, we derived a new orbit for $\theta ^1$ Ori D with a semi-major axis $a = 0.77 \pm 0.03$~AU and a period $P = 53.03 \pm 0.06$~d. The system mass is 21.7~\solmass , assuming a distance to the ONC of $414 \pm 7$~pc \citep{Menten2007}. We derived a multiplicity fraction of 0.69 and a companion fraction of 1.38 for our complete sample. Our observations are complete down to 3~\solmass . We illustrate the observed companion systems in \cref{fig:Orion_summary}.

The companion fraction rises with primary mass and extends from $\simsym 0.6$ for a mass range of $\leq 1$\textendash 2~\solmass\ to $2.3 \pm 0.3$ for objects with $> 16$~\solmass . The multiplicity fraction also increases with object mass. We obtain a multiplicity fraction of 0.5 for objects with $\leq 1 $\textendash  2 ~\solmass\ and it rises up to 100\% for stars with $> 16$~\solmass . 

The companion mass distribution of our sample resembles an IMF with $N \propto m ^{-2.3 \pm 0.3}$. We fit the distribution of the mass ratio $q$ with a power law $\propto q^{-1.7 \pm 0.1}$. The exponent of the power law is smaller compared to previous findings. We did not observe a preference for twin binaries in any mass range and more specifically, there is no tendency for high-mass stars to have companions of comparable masses in our sample. Additionally, we detected no correlation of the companion mass and the companion star separation. 

We observed a bimodal distribution of the mass ratio $q$ compared with the respective companion separation. This bimodal distribution resembles the distribution of A stars, even though our sample comprises mainly OB stars. We obtained a peak at separations $\simsym 0.5$~AU, followed by only few companions between 1\textendash 100~AU. This indicates a change of formation mechanism. Disk fragmentation becomes relevant at scales $\gtrsim 50$~AU. Hence, the observations indicate a transition to companions formed by disk fragmentation. We found a second peak for separations $\geq 100$~AU. Our sample covers separations up to $\simsym 600$~AU, thus we limit our conclusion to this separation range. 

Finally, we compared our observational results with the expected properties of star formation models. We found no clear tendency for either core accretion or competitive accretion. There are several aspects contradicting the predictions of core accretion, competitive accretion and stellar collisions. We excluded stellar collisions as the main formation mechanism.

We notice fundamental differences between our observations and previous observations of star forming regions. The main differences are that the companion mass distribution follows a Salpeter IMF, and that we find no tendency for high-mass binaries with equal mass companions. The differences in stellar mass distribution could result from the improved sensitivity with long baseline interferometry and GRAVITY. Another explanation could be the very young age of Orion, which is too short for a dynamical evolution of its binary systems. Close systems, as often observed for O-type stars, have not yet formed in the ONC. Further similar studies of other star forming regions are necessary.

\begin{acknowledgements}
This work is based on observations made with ESO Telescopes at the La Silla Paranal Observatory under programme ID 0100.D-0576, ID 098.D-0322, ID 60.A-9102 , and ID 0100.C-0608.
This research has made use of the Jean-Marie Mariotti Center \texttt{Aspro2} service \footnote{Available at http://www.jmmc.fr/aspro2} \citep{2013ascl.soft10005B} and \texttt{SearchCal} service \footnote{Available at http://www.jmmc.fr/searchcal}
co-developped by LAGRANGE and IPAG, of CDS Astronomical Databases SIMBAD and VIZIER \footnote{Available at http://cdsweb.u-strasbg.fr/} \citep{2016A&A...589A.112C}, of Astropy, a community-developed core Python package for Astronomy \citep{2013A&A...558A..33A, 2018arXiv180102634T}, of PyAstronomy\footnote{https://github.com/sczesla/PyAstronomy}, and of MiRA\footnote{https://github.com/emmt/MiRA}. R.G.L has received funding from the European Union’s Horizon 2020 research and innovation programme under the Marie Skłodowska-Curie Grant Agreement No. 706320. J.S.B. acknowledges the support from the ESO fellowship program. A.A., N.A., P.G. \&  P.G. acknowledge support from FCT-Portugal with reference UID\/FIS\/00099\/2013. This work has made use of data from the European Space Agency (ESA) mission
{\it Gaia} (\url{https://www.cosmos.esa.int/gaia}), processed by the {\it Gaia}
Data Processing and Analysis Consortium (DPAC,
\url{https://www.cosmos.esa.int/web/gaia/dpac/consortium}). Funding for the DPAC
has been provided by national institutions, in particular the institutions
participating in the {\it Gaia} Multilateral Agreement.
\end{acknowledgements}
\bibliographystyle{aa} % style aa.bst
\bibliography{library,library2} % your references Yourfile.bib

\begin{thebibliography}{148}
\expandafter\ifx\csname natexlab\endcsname\relax\def\natexlab#1{#1}\fi

\bibitem[{{Absil} {et~al.}(2011){Absil}, {Le Bouquin}, {Berger}, {Lagrange},
  {Chauvin}, {Lazareff}, {Zins}, {Haguenauer}, {Jocou}, {Kern}, {Millan-Gabet},
  {Rochat}, \& {Traub}}]{2011A&A...535A..68A}
{Absil}, O., {Le Bouquin}, J.-B., {Berger}, J.-P., {et~al.} 2011, \aap, 535, 10
  pp.

\bibitem[{Abt {et~al.}(1991)Abt, Wang, \& Cardona}]{Abt1991}
Abt, H.~A., Wang, R., \& Cardona, O. 1991, Astrophys. J., 367, 155

\bibitem[{{Aikman} \& {Goldberg}(1974)}]{1974JRASC..68..205A}
{Aikman}, G.~C.~L. \& {Goldberg}, B.~A. 1974, \jrasc, 68, 205

\bibitem[{Allen {et~al.}(2017)Allen, Costero, Ruelas-Mayorga, \&
  Sánchez}]{doi:10.1093/mnras/stx076}
Allen, C., Costero, R., Ruelas-Mayorga, A., \& Sánchez, L.~J. 2017, Monthly
  Notices of the Royal Astronomical Society, 466, 4937

\bibitem[{Allen \& Cox(2000)}]{allen2000allen}
Allen, C. \& Cox, A. 2000, Allen's Astrophysical Quantities, Allen's
  Astrophysical Quantities (Springer)

\bibitem[{{Astropy Collaboration} {et~al.}(2013){Astropy Collaboration},
  {Robitaille}, {Tollerud}, {Greenfield}, {Droettboom}, {Bray}, {Aldcroft},
  {Davis}, {Ginsburg}, {Price-Whelan}, {Kerzendorf}, {Conley}, {Crighton},
  {Barbary}, {Muna}, {Ferguson}, {Grollier}, {Parikh}, {Nair}, {Unther},
  {Deil}, {Woillez}, {Conseil}, {Kramer}, {Turner}, {Singer}, {Fox}, {Weaver},
  {Zabalza}, {Edwards}, {Azalee Bostroem}, {Burke}, {Casey}, {Crawford},
  {Dencheva}, {Ely}, {Jenness}, {Labrie}, {Lim}, {Pierfederici}, {Pontzen},
  {Ptak}, {Refsdal}, {Servillat}, \& {Streicher}}]{2013A&A...558A..33A}
{Astropy Collaboration}, {Robitaille}, T.~P., {Tollerud}, E.~J., {et~al.} 2013,
  \aap, 558, A33

\bibitem[{Balega {et~al.}(2004)Balega, Balega, Maksimov, Pluzhnik, Schertl,
  Shkhagosheva, \& Weigelt}]{Balega2004}
Balega, I., Balega, Y.~Y., Maksimov, A.~F., {et~al.} 2004, A{\&}A, 422, 627

\bibitem[{Balega {et~al.}(2007)Balega, Balega, Maksimov, Malogolovets,
  Rastegaev, Shkhagosheva, \& Weigelt}]{Balega2007}
Balega, I.~I., Balega, Y.~Y., Maksimov, A.~F., {et~al.} 2007, Astrophys. Bull.,
  62, 339

\bibitem[{{Balega} {et~al.}(2015){Balega}, {Chentsov}, {Rzaev}, \&
  {Weigelt}}]{2015ASPC..494...57B}
{Balega}, Y.~Y., {Chentsov}, E.~L., {Rzaev}, A.~K., \& {Weigelt}, G. 2015, in
  Astronomical Society of the Pacific Conference Series, Vol. 494, Physics and
  Evolution of Magnetic and Related Stars, ed. Y.~Y. {Balega}, I.~I.
  {Romanyuk}, \& D.~O. {Kudryavtsev}, 57

\bibitem[{Bally {et~al.}(2005)Bally, Moeckel, \& Throop}]{Bally2005}
Bally, J., Moeckel, N., \& Throop, H. 2005, in Chondrites Protoplanetary Disk,
  Vol. 341, 81--106

\bibitem[{{Bastian} {et~al.}(2010){Bastian}, {Covey}, \&
  {Meyer}}]{2010ARA&A..48..339B}
{Bastian}, N., {Covey}, K.~R., \& {Meyer}, M.~R. 2010, \araa, 48, 339

\bibitem[{{Bate} {et~al.}(2002){Bate}, {Bonnell}, \&
  {Bromm}}]{2002MNRAS.336..705B}
{Bate}, M.~R., {Bonnell}, I.~A., \& {Bromm}, V. 2002, \mnras, 336, 705

\bibitem[{{Bonnell}(2005{\natexlab{a}})}]{2005ASSL..327..425B}
{Bonnell}, I.~A. 2005{\natexlab{a}}, in Astrophysics and Space Science Library,
  Vol. 327, The Initial Mass Function 50 Years Later, ed. E.~{Corbelli},
  F.~{Palla}, \& H.~{Zinnecker}, 425

\bibitem[{{Bonnell}(2005{\natexlab{b}})}]{2005IAUS..227..266B}
{Bonnell}, I.~A. 2005{\natexlab{b}}, in IAU Symposium, Vol. 227, Massive Star
  Birth: A Crossroads of Astrophysics, ed. R.~{Cesaroni}, M.~{Felli},
  E.~{Churchwell}, \& M.~{Walmsley}, 266--275

\bibitem[{{Bonnell} \& {Bate}(2005)}]{2005MNRAS.362..915B}
{Bonnell}, I.~A. \& {Bate}, M.~R. 2005, \mnras, 362, 915

\bibitem[{{Bonnell} {et~al.}(1997){Bonnell}, {Bate}, {Clarke}, \&
  {Pringle}}]{1997MNRAS.285..201B}
{Bonnell}, I.~A., {Bate}, M.~R., {Clarke}, C.~J., \& {Pringle}, J.~E. 1997,
  \mnras, 285, 201

\bibitem[{{Bonnell} {et~al.}(2001){Bonnell}, {Bate}, {Clarke}, \&
  {Pringle}}]{2001MNRAS.323..785B}
{Bonnell}, I.~A., {Bate}, M.~R., {Clarke}, C.~J., \& {Pringle}, J.~E. 2001,
  \mnras, 323, 785

\bibitem[{{Bonnell} {et~al.}(2003){Bonnell}, {Bate}, \&
  {Vine}}]{2003MNRAS.343..413B}
{Bonnell}, I.~A., {Bate}, M.~R., \& {Vine}, S.~G. 2003, \mnras, 343, 413

\bibitem[{{Bossi} {et~al.}(1989){Bossi}, {Gaspani}, {Scardia}, \&
  {Tadini}}]{1989A&A...222..117B}
{Bossi}, M., {Gaspani}, A., {Scardia}, M., \& {Tadini}, M. 1989, \aap, 222, 117

\bibitem[{{Bourg{\`e}s} {et~al.}(2013){Bourg{\`e}s}, {Mella}, {Lafrasse}, \&
  {Duvert}}]{2013ascl.soft10005B}
{Bourg{\`e}s}, L., {Mella}, G., {Lafrasse}, S., \& {Duvert}, G. 2013, {ASPRO 2:
  Astronomical Software to PRepare Observations}, Astrophysics Source Code
  Library

\bibitem[{{Bragan{\c c}a} {et~al.}(2012){Bragan{\c c}a}, {Daflon}, {Cunha},
  {Bensby}, {Oey}, \& {Walth}}]{2012AJ....144..130B}
{Bragan{\c c}a}, G.~A., {Daflon}, S., {Cunha}, K., {et~al.} 2012, \aj, 144, 130

\bibitem[{Brice{\~{n}}o {et~al.}(2007)Brice{\~{n}}o, Preibisch, Sherry,
  Mamajek, Mathieu, Walter, \& Zinnecker}]{Briceno2007}
Brice{\~{n}}o, C., Preibisch, T., Sherry, W.~H., {et~al.} 2007, Protostars
  Planets V, 345

\bibitem[{{Chabrier}(2003)}]{2003PASP..115..763C}
{Chabrier}, G. 2003, \pasp, 115, 763

\bibitem[{{Chelli} {et~al.}(2016){Chelli}, {Duvert}, {Bourg{\`e}s}, {Mella},
  {Lafrasse}, {Bonneau}, \& {Chesneau}}]{2016A&A...589A.112C}
{Chelli}, A., {Duvert}, G., {Bourg{\`e}s}, L., {et~al.} 2016, \aap, 589, A112

\bibitem[{Chini {et~al.}(2012)Chini, Hoffmeister, Nasseri, Stahl, \&
  Zinnecker}]{Chini2012}
Chini, R., Hoffmeister, V.~H., Nasseri, A., Stahl, O., \& Zinnecker, H. 2012,
  Mon. Not. R. Astron. Soc., 424, 1925

\bibitem[{Chini {et~al.}(2011)Chini, Nasseri, Hoffmeister, Buda, \&
  Barr}]{Chini2011}
Chini, R., Nasseri, A., Hoffmeister, V.~H., Buda, L., \& Barr, A. 2011, in
  Evol. Compact Bin. ASP Conf. Ser., ed. L.~Schmidtobreick, M.~Schreiber, \&
  C.~Tappert, Vol. 447, 67

\bibitem[{{Clarke}(2001)}]{2001IAUS..200..346C}
{Clarke}, C.~J. 2001, in IAU Symposium, Vol. 200, The Formation of Binary
  Stars, ed. H.~{Zinnecker} \& R.~{Mathieu}, 346

\bibitem[{Close {et~al.}(2013)Close, Males, Morzinski, Kopon, Follette,
  Rodigas, Hinz, Wu, Puglisi, Esposito, Riccardi, Pinna, Xompero, Briguglio,
  Uomoto, \& Hare}]{Close2013}
Close, L.~M., Males, J.~R., Morzinski, K., {et~al.} 2013, Astrophys. J., 774,
  13pp

\bibitem[{Close {et~al.}(2012)Close, Puglisi, Males, Arcidiacono, Skemer,
  Guerra, Busoni, Brusa, Pinna, Miller, Riccardi, McCarthy, Xompero, Kulesa,
  Quiros-Pacheco, Argomedo, Brynnel, Esposito, Mannucci, Boutsia, Fini,
  Thompson, Hill, Woodward, Briguglio, Rodigas, Briguglio, Stefanini, Agapito,
  Hinz, Follette, \& Green}]{Close2012}
Close, L.~M., Puglisi, A., Males, J.~R., {et~al.} 2012, Astrophys. J., 749, 1

\bibitem[{Close {et~al.}(2003)Close, Wildi, Lloyd-Hart, Brusa, Fisher, Miller,
  Riccardi, Salinari, McCarthy, Angel, Allen, Martin, Sosa, Montoya,
  Rademacher, Rascon, Curly, Siegler, \& Duschl}]{Close2003}
Close, L.~M., Wildi, F., Lloyd-Hart, M., {et~al.} 2003, Astrophys. J., 599, 537

\bibitem[{Corporon \& Lagrange(1999)}]{Corporon1999}
Corporon, P. \& Lagrange, A.-M. 1999, Astron. Astrophys. Suppl. Ser., 136, 429

\bibitem[{Costero {et~al.}(2008)Costero, Allen, Echevarr, Georgiev, Poveda, \&
  Richer}]{Costero2008}
Costero, R., Allen, C., Echevarr, J., {et~al.} 2008, in RevMexAA (Seria Conf.,
  Vol.~34, 102--105

\bibitem[{{Costero} {et~al.}(2006){Costero}, {Echevarria}, {Richer}, {Poveda},
  \& {Li}}]{2006IAUC.8669....2C}
{Costero}, R., {Echevarria}, J., {Richer}, M.~G., {Poveda}, A., \& {Li}, W.
  2006, \iaucirc, 8669

\bibitem[{{Cutri} {et~al.}(2003){Cutri}, {Skrutskie}, {van Dyk}, {Beichman},
  {Carpenter}, {Chester}, {Cambresy}, {Evans}, {Fowler}, {Gizis}, {Howard},
  {Huchra}, {Jarrett}, {Kopan}, {Kirkpatrick}, {Light}, {Marsh}, {McCallon},
  {Schneider}, {Stiening}, {Sykes}, {Weinberg}, {Wheaton}, {Wheelock}, \&
  {Zacarias}}]{2003yCat.2246....0C}
{Cutri}, R.~M., {Skrutskie}, M.~F., {van Dyk}, S., {et~al.} 2003, VizieR Online
  Data Catalog, 2246

\bibitem[{{Da Rio} {et~al.}(2012){Da Rio}, M., {Hillenbrand}, {Henning}, \&
  {Stassun}}]{2012ApJ...748...14D}
{Da Rio}, N., M., R., {Hillenbrand}, L.~A., {Henning}, T., \& {Stassun}, K.~G.
  2012, \apj, 748, 14

\bibitem[{{Da Rio} {et~al.}(2010){Da Rio}, {Robberto}, {Soderblom}, {Panagia},
  {Hillenbrand}, {Palla}, \& {Stassun}}]{2010ApJ...722.1092D}
{Da Rio}, N., {Robberto}, M., {Soderblom}, D.~R., {et~al.} 2010, \apj, 722,
  1092

\bibitem[{{de Wit, W. J.} {et~al.}(2004){de Wit, W. J.}, {Testi, L.}, {Palla,
  F.}, {Vanzi, L.}, \& {Zinnecker, H.}}]{deWit2004}
{de Wit, W. J.}, {Testi, L.}, {Palla, F.}, {Vanzi, L.}, \& {Zinnecker, H.}
  2004, A\&A, 425, 937

\bibitem[{{de Wit, W. J.} {et~al.}(2005){de Wit, W. J.}, {Testi, L.}, {Palla,
  F.}, \& {Zinnecker, H.}}]{deWit2005}
{de Wit, W. J.}, {Testi, L.}, {Palla, F.}, \& {Zinnecker, H.} 2005, A\&A, 437,
  247

\bibitem[{{Donati} {et~al.}(2002){Donati}, {Babel}, {Harries}, {Howarth},
  {Petit}, \& {Semel}}]{2002MNRAS.333...55D}
{Donati}, J.-F., {Babel}, J., {Harries}, T.~J., {et~al.} 2002, \mnras, 333, 55

\bibitem[{{Drass} {et~al.}(2016){Drass}, {Haas}, {Chini}, {Bayo}, {Hackstein},
  {Hoffmeister}, {Godoy}, \& {Vogt}}]{2016MNRAS.461.1734D}
{Drass}, H., {Haas}, M., {Chini}, R., {et~al.} 2016, \mnras, 461, 1734

\bibitem[{{Ducati}(2002)}]{2002yCat.2237....0D}
{Ducati}, J.~R. 2002, VizieR Online Data Catalog, 2237

\bibitem[{Duch{\^{e}}ne \& Kraus(2013)}]{Duchene2013}
Duch{\^{e}}ne, G. \& Kraus, A. 2013, Annu. Rev. Astron. Astrophys, 51, 269

\bibitem[{{Feigelson} {et~al.}(2002){Feigelson}, {Broos}, {Gaffney}, {Garmire},
  {Hillenbrand}, {Pravdo}, {Townsley}, \& {Tsuboi}}]{2002ApJ...574..258F}
{Feigelson}, E.~D., {Broos}, P., {Gaffney}, III, J.~A., {et~al.} 2002, \apj,
  574, 258

\bibitem[{{Gaia Collaboration} {et~al.}(2018){Gaia Collaboration}, {Brown},
  {Vallenari}, {Prusti}, {de Bruijne}, {Babusiaux}, \&
  {Bailer-Jones}}]{2018arXiv180409365G}
{Gaia Collaboration}, {Brown}, A.~G.~A., {Vallenari}, A., {et~al.} 2018, ArXiv
  e-prints [\eprint[arXiv]{1804.09365}]

\bibitem[{{Gaia Collaboration} {et~al.}(2016){Gaia Collaboration}, {Prusti},
  {de Bruijne}, {Brown}, {Vallenari}, {Babusiaux}, {Bailer-Jones}, {Bastian},
  {Biermann}, {Evans}, \& et~al.}]{2016A&A...595A...1G}
{Gaia Collaboration}, {Prusti}, T., {de Bruijne}, J.~H.~J., {et~al.} 2016,
  \aap, 595, A1

\bibitem[{Gallenne {et~al.}(2015)Gallenne, M{\'{e}}rand, Kervella, Monnier,
  Schaefer, Baron, Breitfelder, {Le Bouquin}, Roettenbacher, Gieren,
  Pietrzy{\'{n}}ski, McAlister, {Ten Brummelaar}, Sturmann, Sturmann, Turner,
  Ridgway, \& Kraus}]{Gallenne2015}
Gallenne, A., M{\'{e}}rand, A., Kervella, P., {et~al.} 2015, Astron.
  Astrophys., 579, 1

\bibitem[{{Genzel} \& {Stutzki}(1989)}]{1989ARA&A..27...41G}
{Genzel}, R. \& {Stutzki}, J. 1989, \araa, 27, 41

\bibitem[{{Gravity Collaboration} {et~al.}(2017){Gravity Collaboration},
  {Abuter}, {Accardo}, {Amorim}, {Anugu}, {\'Avila}, {Azouaoui}, {Benisty},
  {Berger}, {Blind}, {Bonnet}, {Bourget}, {Brandner}, {Brast}, {Buron},
  {Burtscher}, {Cassaing}, {Chapron}, {Choquet}, {Cl\'enet}, {Collin}, {Coud\'e
  Du Foresto}, {de Wit}, {de Zeeuw}, {Deen}, {Delplancke-Str\"obele}, {Dembet},
  {Derie}, {Dexter}, {Duvert}, {Ebert}, {Eckart}, {Eisenhauer}, {Esselborn},
  {F\'edou}, {Finger}, {Garcia}, {Garcia Dabo}, {Garcia Lopez}, {Gendron},
  {Genzel}, {Gillessen}, {Gonte}, {Gordo}, {Grould}, {Gr\"ozinger}, {Guieu},
  {Haguenauer}, {Hans}, {Haubois}, {Haug}, {Haussmann}, {Henning}, {Hippler},
  {Horrobin}, {Huber}, {Hubert}, {Hubin}, {Hummel}, {Jakob}, {Janssen},
  {Jochum}, {Jocou}, {Kaufer}, {Kellner}, {Kendrew}, {Kern}, {Kervella},
  {Kiekebusch}, {Klein}, {Kok}, {Kolb}, {Kulas}, {Lacour}, {Lapeyr\`ere},
  {Lazareff}, {Le Bouquin}, {L\`ena}, {Lenzen}, {L\'ev\^eque}, {Lippa},
  {Magnard}, {Mehrgan}, {Mellein}, {M\'erand}, {Moreno-Ventas}, {Moulin},
  {M\"uller}, {M\"uller}, {Neumann}, {Oberti}, {Ott}, {Pallanca}, {Panduro},
  {Pasquini}, {Paumard}, {Percheron}, {Perraut}, {Perrin}, {Pfl\"uger},
  {Pfuhl}, {Phan Duc}, {Plewa}, {Popovic}, {Rabien}, {Ram\'{\i}rez}, {Ramos},
  {Rau}, {Riquelme}, {Rohloff}, {Rousset}, {Sanchez-Bermudez}, {Scheithauer},
  {Sch{\"o}ller}, {Schuhler}, {Spyromilio}, {Straubmeier}, {Sturm}, {Suarez},
  {Tristram}, {Ventura}, {Vincent}, {Waisberg}, {Wank}, {Weber}, {Wieprecht},
  {Wiest}, {Wiezorrek}, {Wittkowski}, {Woillez}, {Wolff}, {Yazici}, {Ziegler},
  \& {Zins}}]{GRAVITYCollaboration2017}
{Gravity Collaboration}, {Abuter}, R., {Accardo}, M., {et~al.} 2017, \aap, 602,
  A94

\bibitem[{Grellmann {et~al.}(2013)Grellmann, Preibisch, Ratzka, Kraus,
  Helminiak, \& Zinnecker}]{Grellmann2013}
Grellmann, R., Preibisch, T., Ratzka, T., {et~al.} 2013, A{\&}A, 550, 531

\bibitem[{Grunhut {et~al.}(2017)Grunhut, Wade, Neiner, Oksala, Petit, Alecian,
  Bohlender, Bouret, Henrichs, Hussain, Kochukhov, \&
  Collaboration}]{doi:10.1093/mnras/stw2743}
Grunhut, J.~H., Wade, G.~A., Neiner, C., {et~al.} 2017, Monthly Notices of the
  Royal Astronomical Society, 465, 2432

\bibitem[{{Habibi} {et~al.}(2017){Habibi}, {Gillessen}, {Martins},
  {Eisenhauer}, {Plewa}, {Pfuhl}, {George}, {Dexter}, {Waisberg}, {Ott}, {von
  Fellenberg}, {Baub{\"o}ck}, {Jimenez-Rosales}, \&
  {Genzel}}]{2017ApJ...847..120H}
{Habibi}, M., {Gillessen}, S., {Martins}, F., {et~al.} 2017, \apj, 847, 120

\bibitem[{{Hartwig}(1921)}]{1921AN....212..383H}
{Hartwig}, E. 1921, Astronomische Nachrichten, 212, 383

\bibitem[{Herbig(1950)}]{Herbig1950}
Herbig, G.~H. 1950, Astrophys. J., 11, 15

\bibitem[{Herbig \& Griffin(2006)}]{Herbig2006}
Herbig, G.~H. \& Griffin, R.~F. 2006, Astron. J., 132, 1763

\bibitem[{{Herrero} {et~al.}(1992){Herrero}, {Kudritzki}, {Vilchez}, {Kunze},
  {Butler}, \& {Haser}}]{1992A&A...261..209H}
{Herrero}, A., {Kudritzki}, R.~P., {Vilchez}, J.~M., {et~al.} 1992, \aap, 261,
  209

\bibitem[{Hillenbrand(1997)}]{Hillenbrand1997}
Hillenbrand, L.~A. 1997, Astron. J., 113, 1733

\bibitem[{Hillenbrand \& Hartmann(1998)}]{Hillenbrand1998a}
Hillenbrand, L.~A. \& Hartmann, L.~W. 1998, Astrophys. J., 492, 540

\bibitem[{{Houk} \& {Swift}(1999)}]{1999MSS...C05....0H}
{Houk}, N. \& {Swift}, C. 1999, in Michigan Spectral Survey, Ann Arbor, Dep.
  Astron., Univ. Michigan, Vol.~5

\bibitem[{Kennicutt(1998)}]{Kennicutt1998}
Kennicutt, R.~C. 1998, Annu. Rev. Astron. Astrophys., 36, 189

\bibitem[{K{\"{o}}hler {et~al.}(2006)K{\"{o}}hler, Petr-Gotzens, McCaughrean,
  Bouvier, Duch{\^{e}}ne, Quirrenbach, \& Zinnecker}]{Kohler2006}
K{\"{o}}hler, R., Petr-Gotzens, M.~G., McCaughrean, M.~J., {et~al.} 2006,
  Astron. Astrophys., 458, 461

\bibitem[{{Kratter} \&
  {Lodato}(2016)}]{doi:10.1146/annurev-astro-081915-023307}
{Kratter}, K. \& {Lodato}, G. 2016, Annual Review of Astronomy and
  Astrophysics, 54, 271

\bibitem[{{Kratter} \& {Matzner}(2006)}]{2006MNRAS.373.1563K}
{Kratter}, K.~M. \& {Matzner}, C.~D. 2006, \mnras, 373, 1563

\bibitem[{Kraus {et~al.}(2007)Kraus, Balega, Berger, Hofmann, Millan-Gabet,
  Monnier, Ohnaka, Pedretti, Preibisch, Schertl, Schloerb, Traub, \&
  Weigelt}]{Kraus2007}
Kraus, S., Balega, Y.~Y., Berger, J.-P., {et~al.} 2007, A{\&}A, 466, 649

\bibitem[{Kraus {et~al.}(2009)Kraus, Weigelt, Balega, Docobo, Hofmann,
  Preibisch, Schertl, Tamazian, Driebe, Ohnaka, Petrov, Sch{\"{o}}ller, \&
  Smith}]{Kraus2009}
Kraus, S., Weigelt, G., Balega, Y.~Y., {et~al.} 2009, a{\&}a, 497, 195

\bibitem[{{Kroupa}(2001)}]{2001MNRAS.322..231K}
{Kroupa}, P. 2001, \mnras, 322, 231

\bibitem[{Kroupa {et~al.}(2013)Kroupa, Weidner, Pflamm-Altenburg, Thies,
  Dabringhausen, Marks, \& Maschberger}]{Kroupa2013}
Kroupa, P., Weidner, C., Pflamm-Altenburg, J., {et~al.} 2013, The Stellar and
  Sub-Stellar Initial Mass Function of Simple and Composite Populations, ed.
  T.~D. Oswalt \& G.~Gilmore (Dordrecht: Springer Netherlands), 115--242

\bibitem[{{Krumholz}(2006)}]{2006ApJ...641L..45K}
{Krumholz}, M.~R. 2006, \apjl, 641, L45

\bibitem[{Krumholz(2015)}]{Krumholz2015}
Krumholz, M.~R. 2015, The Formation of Very Massive Stars, ed. J.~S. Vink
  (Cham: Springer International Publishing), 43--75

\bibitem[{Krumholz(2016)}]{Krumholz2016}
Krumholz, M.~R. 2016, {Notes on Star Formation}

\bibitem[{{Krumholz} {et~al.}(2012){Krumholz}, {Klein}, \&
  {McKee}}]{2012ApJ...754...71K}
{Krumholz}, M.~R., {Klein}, R.~I., \& {McKee}, C.~F. 2012, \apj, 754, 71

\bibitem[{Krumholz {et~al.}(2009)Krumholz, Klein, Mckee, Offner, \&
  Cunningham}]{Krumholz2009}
Krumholz, M.~R., Klein, R.~I., Mckee, C.~F., Offner, S. S.~R., \& Cunningham,
  A.~J. 2009, Science (80-. )., 323, 754

\bibitem[{{Ku} {et~al.}(1982){Ku}, {Righini-Cohen}, \&
  {Simon}}]{1982Sci...215...61K}
{Ku}, W.~H.-M., {Righini-Cohen}, G., \& {Simon}, M. 1982, Science, 215, 61

\bibitem[{Lada \& Lada(2003)}]{Lada2003}
Lada, C.~J. \& Lada, E.~A. 2003, Annu. Rev. Astron. Astrophys, 41, 57

\bibitem[{{Landstreet} {et~al.}(2017){Landstreet}, {Kochukhov}, {Alecian},
  {Bailey}, {Mathis}, {Neiner}, {Wade}, \& {BINaMIcS
  Collaboration}}]{2017A&A...601A.129L}
{Landstreet}, J.~D., {Kochukhov}, O., {Alecian}, E., {et~al.} 2017, \aap, 601,
  A129

\bibitem[{{Lapeyrere} {et~al.}(2014){Lapeyrere}, {Kervella}, {Lacour},
  {Azouaoui}, {Garcia-Dabo}, {Perrin}, {Eisenhauer}, {Perraut}, {Straubmeier},
  {Amorim}, \& {Brandner}}]{2014SPIE.9146E..2DL}
{Lapeyrere}, V., {Kervella}, P., {Lacour}, S., {et~al.} 2014, in \procspie,
  Vol. 9146, Optical and Infrared Interferometry IV, 91462D

\bibitem[{Lawson(2000)}]{Lawson2000}
Lawson, P.~R. 2000, in Princ. Long Baseline Stellar Interferom., ed. P.~R.
  Lawson

\bibitem[{Lehmann {et~al.}(2010)Lehmann, Vitrichenko, Bychkov, Bychkova, \&
  Klochkova}]{Lehmann2010}
Lehmann, H., Vitrichenko, {\'{E}}.~A., Bychkov, V., Bychkova, L., \& Klochkova,
  V. 2010, Astron. Astrophys., 514, A34

\bibitem[{Levato \& Abt(1976)}]{Levato1976}
Levato, H. \& Abt, H.~A. 1976, Publ. Astron. Soc. Pacific, 88, 712

\bibitem[{{Levenhagen} \& {Leister}(2006)}]{2006MNRAS.371..252L}
{Levenhagen}, R.~S. \& {Leister}, N.~V. 2006, \mnras, 371, 252

\bibitem[{{Levesque} {et~al.}(2005){Levesque}, {Massey}, {Olsen}, {Plez},
  {Josselin}, {Maeder}, \& {Meynet}}]{2005ApJ...628..973L}
{Levesque}, E.~M., {Massey}, P., {Olsen}, K.~A.~G., {et~al.} 2005, \apj, 628,
  973

\bibitem[{{Lohsen}(1975)}]{1975IBVS..988....1L}
{Lohsen}, E. 1975, Information Bulletin on Variable Stars, 988

\bibitem[{{Luri} {et~al.}(2018){Luri}, {Brown}, {Sarro}, {Arenou},
  {Bailer-Jones}, {Castro-Ginard}, {de Bruijne}, {Prusti}, {Babusiaux}, \&
  {Delgado}}]{2018arXiv180409376L}
{Luri}, X., {Brown}, A.~G.~A., {Sarro}, L.~M., {et~al.} 2018, ArXiv e-prints
  [\eprint[arXiv]{1804.09376}]

\bibitem[{Martins {et~al.}(2005)Martins, Schaerer, \& Hillier}]{Martins2005}
Martins, F., Schaerer, D., \& Hillier, D.~J. 2005, Astron. Astrophys., 436,
  1049

\bibitem[{{Mason} {et~al.}(1998){Mason}, {Gies}, {Hartkopf}, {Bagnuolo}, {ten
  Brummelaar}, \& {McAlister}}]{1998AJ....115..821M}
{Mason}, B.~D., {Gies}, D.~R., {Hartkopf}, W.~I., {et~al.} 1998, \aj, 115, 821

\bibitem[{{Mattei} \& {Baldwin}(1976)}]{1976IAUC.3004....1M}
{Mattei}, J. \& {Baldwin}, M. 1976, \iaucirc, 3004

\bibitem[{{McKee} \& {Ostriker}(2007)}]{2007ARA&A..45..565M}
{McKee}, C.~F. \& {Ostriker}, E.~C. 2007, \araa, 45, 565

\bibitem[{{McKee} \& {Tan}(2002)}]{2002Natur.416...59M}
{McKee}, C.~F. \& {Tan}, J.~C. 2002, \nat, 416, 59

\bibitem[{Menten {et~al.}(2007)Menten, Reid, Forbrich, \&
  Brunthaler}]{Menten2007}
Menten, K.~M., Reid, M.~J., Forbrich, J., \& Brunthaler, A. 2007, a{\&}a, 474,
  515

\bibitem[{{Moe} \& {Di Stefano}(2017)}]{2017ApJS..230...15M}
{Moe}, M. \& {Di Stefano}, R. 2017, \apjs, 230, 15

\bibitem[{{Moe} \& {Kratter}(2018)}]{2018ApJ...854...44M}
{Moe}, M. \& {Kratter}, K.~M. 2018, \apj, 854, 44

\bibitem[{{Moeckel} \& {Clarke}(2011)}]{2011MNRAS.410.2799M}
{Moeckel}, N. \& {Clarke}, C.~J. 2011, \mnras, 410, 2799

\bibitem[{Morales-Calder{\'{o}}n {et~al.}(2012)Morales-Calder{\'{o}}n,
  Stauffer, Stassun, Vrba, Prato, Hillenbrand, Terebey, Covey, Rebull,
  Terndrup, Gutermuth, Song, Plavchan, Carpenter, Marchis, Garc{\'{i}}a,
  Margheim, Luhman, Angione, \& Irwin}]{Morales-Calderon2012}
Morales-Calder{\'{o}}n, M., Stauffer, J.~R., Stassun, K.~G., {et~al.} 2012,
  Astrophys. J., 753, (17pp)

\bibitem[{Morrell \& Levato(1991)}]{Morrell1991}
Morrell, N. \& Levato, H. 1991, Astrophys. J. Suppl. Ser., 75, 965

\bibitem[{Motte {et~al.}(2018)Motte, Bontemps, \& Louvet}]{Motte2018}
Motte, F., Bontemps, S., \& Louvet, F. 2018, Annu. Rev. Astron. Astrophys., 56,
  1.1

\bibitem[{Muench {et~al.}(2008)Muench, Getman, Hillenbrand, \&
  Preibisch}]{Muench2008}
Muench, A., Getman, K., Hillenbrand, L., \& Preibisch, T. 2008, in Handb. Star
  Form. Reg., Vol.~I, 483

\bibitem[{Muench {et~al.}(2002)Muench, Lada, Lada, \& Alves}]{Muench2002}
Muench, A.~A., Lada, E.~A., Lada, C.~J., \& Alves, J. 2002, Astrophys. J., 573,
  366

\bibitem[{{Mu\v{z}i\'c} {et~al.}(2017){Mu\v{z}i\'c}, {Sch\"odel}, {Scholz},
  {Geers}, {Jayawardhana}, {Ascenso}, \& {Cieza}}]{doi:10.1093/mnras/stx1906}
{Mu\v{z}i\'c}, K., {Sch\"odel}, R., {Scholz}, A., {et~al.} 2017, Monthly
  Notices of the Royal Astronomical Society, 471, 3699

\bibitem[{Newville {et~al.}(2014)Newville, Stensitzki, Allen, \&
  Ingargiola}]{newville_2014_11813}
Newville, M., Stensitzki, T., Allen, D.~B., \& Ingargiola, A. 2014, {LMFIT:
  Non-Linear Least-Square Minimization and Curve-Fitting for Python}

\bibitem[{Nieva \& Przybilla(2014)}]{Nieva2014}
Nieva, M.-F. \& Przybilla, N. 2014, Astron. Astrophys., 566, A7

\bibitem[{{O'Dell} \& {Wong}(1996)}]{1996AJ....111..846O}
{O'Dell}, C.~R. \& {Wong}, K. 1996, \aj, 111, 846

\bibitem[{{Oskinova} {et~al.}(2013){Oskinova}, {Steinke}, {Hamann}, {Sander},
  {Todt}, \& {Liermann}}]{2013MNRAS.436.3357O}
{Oskinova}, L.~M., {Steinke}, M., {Hamann}, W.-R., {et~al.} 2013, \mnras, 436,
  3357

\bibitem[{{Patience} {et~al.}(2008){Patience}, {Zavala}, {Prato}, {Franz},
  {Wasserman}, {Tycner}, {Hutter}, \& {Hummel}}]{2008ApJ...674L..97P}
{Patience}, J., {Zavala}, R.~T., {Prato}, L., {et~al.} 2008, \apjl, 674, L97

\bibitem[{{Peter} {et~al.}(2012){Peter}, {Feldt}, {Henning}, \&
  {Hormuth}}]{2012A&A...538A..74P}
{Peter}, D., {Feldt}, M., {Henning}, T., \& {Hormuth}, F. 2012, \aap, 538, A74

\bibitem[{Petr {et~al.}(1998)Petr, {Coude du Foresto}, Beckwith, Richichi, \&
  McCaughrean}]{Petr1998}
Petr, M.~G., {Coude du Foresto}, V., Beckwith, S. V.~W., Richichi, A., \&
  McCaughrean, M.~J. 1998, Astrophys. J., 500, 825

\bibitem[{Popper \& Plavec(1976)}]{Popper1976}
Popper, D.~M. \& Plavec, M. 1976, Astrophys. J., 205, 462

\bibitem[{{Portegies Zwart} {et~al.}(2010){Portegies Zwart}, {McMillan}, \&
  {Gieles}}]{2010ARA&A..48..431P}
{Portegies Zwart}, S.~F., {McMillan}, S.~L.~W., \& {Gieles}, M. 2010, \araa,
  48, 431

\bibitem[{Preibisch {et~al.}(1999)Preibisch, Balega, Hofmann, Weigelt, \&
  Zinnecker}]{Preibisch1999}
Preibisch, T., Balega, Y., Hofmann, K.~H., Weigelt, G., \& Zinnecker, H. 1999,
  New Astron., 4, 531

\bibitem[{{Reid} {et~al.}(2014){Reid}, {Menten}, {Brunthaler}, {Zheng}, {Dame},
  {Xu}, {Wu}, {Zhang}, {Sanna}, {Sato}, {Hachisuka}, {Choi}, {Immer},
  {Moscadelli}, {Rygl}, \& {Bartkiewicz}}]{2014ApJ...783..130R}
{Reid}, M.~J., {Menten}, K.~M., {Brunthaler}, A., {et~al.} 2014, \apj, 783, 130

\bibitem[{Reiter {et~al.}(2018)Reiter, Calvet, Thanathibodee, Kraus, Cauley,
  Monnier, Rubinstein, Aarnio, \& Harries}]{0004-637X-852-1-5}
Reiter, M., Calvet, N., Thanathibodee, T., {et~al.} 2018, The Astrophysical
  Journal, 852, 5

\bibitem[{Rio {et~al.}(2016)Rio, Tan, Covey, Cottaar, Foster, Cullen, Tobin,
  Kim, Meyer, Nidever, Stassun, Chojnowski, Flaherty, Majewski, Skrutskie,
  Zasowski, \& Pan}]{0004-637X-818-1-59}
Rio, N.~D., Tan, J.~C., Covey, K.~R., {et~al.} 2016, The Astrophysical Journal,
  818, 59

\bibitem[{{R{\"o}ser} {et~al.}(1994){R{\"o}ser}, {Bastian}, \&
  {Kuzmin}}]{1994A&AS..105..301R}
{R{\"o}ser}, S., {Bastian}, U., \& {Kuzmin}, A. 1994, \aaps, 105, 301

\bibitem[{{Roy}(2005)}]{2005ormo.book.....R}
{Roy}, A.~E. 2005, {Orbital motion}, 4th edn. (Institute of Physics Publishing)

\bibitem[{{Salaris} \& {Cassisi}(2005)}]{2005essp.book.....S}
{Salaris}, M. \& {Cassisi}, S. 2005, {Evolution of Stars and Stellar
  Populations}, 400

\bibitem[{{Salpeter}(1955)}]{1955ApJ...121..161S}
{Salpeter}, E.~E. 1955, \apj, 121, 161

\bibitem[{{Samus'} {et~al.}(2017){Samus'}, {Kazarovets}, {Durlevich},
  {Kireeva}, \& {Pastukhova}}]{2017ARep...61...80S}
{Samus'}, N.~N., {Kazarovets}, E.~V., {Durlevich}, O.~V., {Kireeva}, N.~N., \&
  {Pastukhova}, E.~N. 2017, Astronomy Reports, 61, 80

\bibitem[{Sana {et~al.}(2012)Sana, de~Mink, de~Koter, Langer, Evans, Gieles,
  Gosset, Izzard, {Le Bouquin}, \& Schneider}]{Sana2012}
Sana, H., de~Mink, S.~E., de~Koter, A., {et~al.} 2012, Science (80-. )., 337,
  444

\bibitem[{{Sana} {et~al.}(2014){Sana}, {Le Bouquin}, {Lacour}, {Berger},
  {Duvert}, {Gauchet}, {Norris}, {Olofsson}, {Pickel}, {Zins}, {Absil}, {de
  Koter}, {Kratter}, {Schnurr}, \& {Zinnecker}}]{2014ApJS..215...15S}
{Sana}, H., {Le Bouquin}, J.-B., {Lacour}, S., {et~al.} 2014, \apjs, 215, 15

\bibitem[{{Sana} {et~al.}(2017){Sana}, {Ram{\'{\i}}rez-Tannus}, {de Koter},
  {Kaper}, {Tramper}, \& {Bik}}]{2017A&A...599L...9S}
{Sana}, H., {Ram{\'{\i}}rez-Tannus}, M.~C., {de Koter}, A., {et~al.} 2017,
  \aap, 599, L9

\bibitem[{{Sanchez-Bermudez} {et~al.}(2017){Sanchez-Bermudez}, {Alberdi},
  {Barb{\'a}}, {Bestenlehner}, {Cantalloube}, {Brandner}, {Henning}, {Hummel},
  {Ma{\'\i}z Apell{\'a}niz}, {Pott}, {Sch{\"o}del}, \& {van
  Boekel}}]{2017ApJ...845...57S}
{Sanchez-Bermudez}, J., {Alberdi}, A., {Barb{\'a}}, R., {et~al.} 2017, \apj,
  845, 57

\bibitem[{{Scalo}(1986)}]{1986FCPh...11....1S}
{Scalo}, J.~M. 1986, Fundamentals of Cosmic Physics, 11, 1

\bibitem[{Schertl {et~al.}(2003)Schertl, Balega, Preibisch, \&
  Weigelt}]{Schertl2003}
Schertl, D., Balega, Y.~Y., Preibisch, T., \& Weigelt, G. 2003, Astron.
  Astrophys., 402, 267

\bibitem[{Schneider {et~al.}(2018)Schneider, Sana, Evans, Bestenlehner, Castro,
  Fossati, Gr{\"a}fener, Langer, Ram{\'\i}rez-Agudelo,
  Sab{\'\i}n-Sanjuli{\'a}n, Sim{\'o}n-D{\'\i}az, Tramper, Crowther, de~Koter,
  de~Mink, Dufton, Garcia, Gieles, H{\'e}nault-Brunet, Herrero, Izzard, Kalari,
  Lennon, Ma{\'\i}z~Apell{\'a}niz, Markova, Najarro, Podsiadlowski, Puls,
  Taylor, van Loon, Vink, \& Norman}]{Schneider69}
Schneider, F. R.~N., Sana, H., Evans, C.~J., {et~al.} 2018, Science, 359, 69

\bibitem[{{Schneller}(1948)}]{1948AN....276..144S}
{Schneller}, H. 1948, Astronomische Nachrichten, 276, 144

\bibitem[{{Shu} {et~al.}(1987){Shu}, {Adams}, \&
  {Lizano}}]{1987ARA&A..25...23S}
{Shu}, F.~H., {Adams}, F.~C., \& {Lizano}, S. 1987, \araa, 25, 23

\bibitem[{Simon {et~al.}(1999)Simon, Close, \& Beck}]{Simon1999}
Simon, M., Close, L.~M., \& Beck, T.~L. 1999, Astron. J., 117, 1375

\bibitem[{Sim{\'{o}}n-D{\'{i}}az {et~al.}(2006)Sim{\'{o}}n-D{\'{i}}az, Herrero,
  Esteban, \& Najarro}]{Simon-Diaz2006}
Sim{\'{o}}n-D{\'{i}}az, S., Herrero, A., Esteban, C., \& Najarro, F. 2006,
  Astron. Astrophys., 448, 351

\bibitem[{{Sota} {et~al.}(2011){Sota}, {Ma{\'{\i}}z Apell{\'{a}}niz},
  {Walborn}, {Alfaro}, {Barb{\'{a}}}, {Morrell}, {Gamen}, \&
  {Arias}}]{2011ApJS..193...24S}
{Sota}, A., {Ma{\'{\i}}z Apell{\'{a}}niz}, J., {Walborn}, N.~R., {et~al.} 2011,
  \apjs, 193, 24

\bibitem[{Stahl {et~al.}(1996)Stahl, Kaufer, Rivinius, Szeifert, Wolff,
  G{\"{a}}ng, Gummersbach, Jankovics, Kovacs, Mandel, Pakull, \&
  Peitz}]{Stahl1996}
Stahl, O., Kaufer, A., Rivinius, T., {et~al.} 1996, Astron. Astrophys., 312,
  539

\bibitem[{Stahl {et~al.}(2008)Stahl, Wade, Petit, Stober, \&
  Schanne}]{Stahl2008}
Stahl, O., Wade, G., Petit, V., Stober, B., \& Schanne, L. 2008, Astron.
  Astrophys., 487, 323

\bibitem[{{Stahl} {et~al.}(1993){Stahl}, {Wolf}, {Gang}, {Gummersbach},
  {Kaufer}, {Kovacs}, {Mandel}, \& {Szeifert}}]{1993A&A...274L..29S}
{Stahl}, O., {Wolf}, B., {Gang}, T., {et~al.} 1993, \aap, 274, L29

\bibitem[{{Stelzer} {et~al.}(2005){Stelzer}, {Flaccomio}, {Montmerle},
  {Micela}, {Sciortino}, {Favata}, {Preibisch}, \&
  {Feigelson}}]{2005ApJS..160..557S}
{Stelzer}, B., {Flaccomio}, E., {Montmerle}, T., {et~al.} 2005, \apjs, 160, 557

\bibitem[{Tan {et~al.}(2014)Tan, Beltran, Caselli, Fontani, Fuente, Krumholz,
  McKee, \& Stolte}]{Tan2014}
Tan, J.~C., Beltran, M.~T., Caselli, P., {et~al.} 2014, Protostars and Planets,
  VI, 149

\bibitem[{{Tatulli} {et~al.}(2007){Tatulli}, {Millour}, {Chelli}, {Duvert},
  {Acke}, {Hernandez Utrera}, {Hofmann}, {Kraus}, {Malbet}, {M{\`e}ge},
  {Petrov}, {Vannier}, {Zins}, {Antonelli}, {Beckmann}, {Bresson}, {Dugu{\'e}},
  {Gennari}, {Gl{\"u}ck}, {Kern}, {Lagarde}, {Le Coarer}, {Lisi}, {Perraut},
  {Puget}, {Rantakyr{\"o}}, {Robbe-Dubois}, {Roussel}, {Weigelt}, {Accardo},
  {Agabi}, {Altariba}, {Arezki}, {Aristidi}, {Baffa}, {Behrend}, {Bl{\"o}cker},
  {Bonhomme}, {Busoni}, {Cassaing}, {Clausse}, {Colin}, {Connot},
  {Delboulb{\'e}}, {Domiciano de Souza}, {Driebe}, {Feautrier}, {Ferruzzi},
  {Forveille}, {Fossat}, {Foy}, {Fraix-Burnet}, {Gallardo}, {Giani}, {Gil},
  {Glentzlin}, {Heiden}, {Heininger}, {Kamm}, {Kiekebusch}, {Le Contel}, {Le
  Contel}, {Lesourd}, {Lopez}, {Lopez}, {Magnard}, {Marconi}, {Mars},
  {Martinot-Lagarde}, {Mathias}, {Monin}, {Mouillet}, {Mourard}, {Nussbaum},
  {Ohnaka}, {Pacheco}, {Perrier}, {Rabbia}, {Rebattu}, {Reynaud}, {Richichi},
  {Robini}, {Sacchettini}, {Schertl}, {Sch{\"o}ller}, {Solscheid}, {Spang},
  {Stee}, {Stefanini}, {Tallon}, {Tallon-Bosc}, {Tasso}, {Testi}, {Vakili},
  {von der L{\"u}he}, {Valtier}, \& {Ventura}}]{2007A&A...464...29T}
{Tatulli}, E., {Millour}, F., {Chelli}, A., {et~al.} 2007, \aap, 464, 29

\bibitem[{{The Astropy Collaboration} {et~al.}(2018){The Astropy
  Collaboration}, {Price-Whelan}, {Sip{\H o}cz}, {G{\"u}nther}, {Lim},
  {Crawford}, {Conseil}, {Shupe}, {Craig}, {Dencheva}, {Ginsburg},
  {VanderPlas}, {Bradley}, {P{\'e}rez-Su{\'a}rez}, {de Val-Borro}, {Aldcroft},
  {Cruz}, {Robitaille}, {Tollerud}, {Ardelean}, {Babej}, {Bachetti}, {Bakanov},
  {Bamford}, {Barentsen}, {Barmby}, {Baumbach}, {Berry}, {Biscani}, {Boquien},
  {Bostroem}, {Bouma}, {Brammer}, {Bray}, {Breytenbach}, {Buddelmeijer},
  {Burke}, {Calderone}, {Cano Rodr{\'{\i}}guez}, {Cara}, {Cardoso},
  {Cheedella}, {Copin}, {Crichton}, {D{\'A}vella}, {Deil}, {Depagne},
  {Dietrich}, {Donath}, {Droettboom}, {Earl}, {Erben}, {Fabbro}, {Ferreira},
  {Finethy}, {Fox}, {Garrison}, {Gibbons}, {Goldstein}, {Gommers}, {Greco},
  {Greenfield}, {Groener}, {Grollier}, {Hagen}, {Hirst}, {Homeier}, {Horton},
  {Hosseinzadeh}, {Hu}, {Hunkeler}, {Ivezi{\'c}}, {Jain}, {Jenness}, {Kanarek},
  {Kendrew}, {Kern}, {Kerzendorf}, {Khvalko}, {King}, {Kirkby}, {Kulkarni},
  {Kumar}, {Lee}, {Lenz}, {Littlefair}, {Ma}, {Macleod}, {Mastropietro},
  {McCully}, {Montagnac}, {Morris}, {Mueller}, {Mumford}, {Muna}, {Murphy},
  {Nelson}, {Nguyen}, {Ninan}, {N{\"o}the}, {Ogaz}, {Oh}, {Parejko}, {Parley},
  {Pascual}, {Patil}, {Patil}, {Plunkett}, {Prochaska}, {Rastogi}, {Reddy
  Janga}, {Sabater}, {Sakurikar}, {Seifert}, {Sherbert}, {Sherwood-Taylor},
  {Shih}, {Sick}, {Silbiger}, {Singanamalla}, {Singer}, {Sladen}, {Sooley},
  {Sornarajah}, {Streicher}, {Teuben}, {Thomas}, {Tremblay}, {Turner},
  {Terr{\'o}n}, {van Kerkwijk}, {de la Vega}, {Watkins}, {Weaver}, {Whitmore},
  {Woillez}, \& {Zabalza}}]{2018arXiv180102634T}
{The Astropy Collaboration}, {Price-Whelan}, A.~M., {Sip{\H o}cz}, B.~M.,
  {et~al.} 2018, ArXiv e-prints [\eprint[arXiv]{1801.02634}]

\bibitem[{Vasileiskii \& Vitrichenko(2000)}]{Vasileiskii2000}
Vasileiskii, A.~S. \& Vitrichenko, {\'{E}}.~A. 2000, Astron. Lett., 26, 529

\bibitem[{Vitrichenko(2002{\natexlab{a}})}]{Vitrichenko2002b}
Vitrichenko, {\'{E}}.~A. 2002{\natexlab{a}}, Astron. Lett., 12, 843

\bibitem[{Vitrichenko(2002{\natexlab{b}})}]{Vitrichenko2002a}
Vitrichenko, {\'{E}}.~A. 2002{\natexlab{b}}, Astron. Lett., 28, 324

\bibitem[{Vitrichenko {et~al.}(1998)Vitrichenko, Klochkova, \&
  Plachinda}]{Vitrichenko1998}
Vitrichenko, {\'{E}}.~A., Klochkova, V.~G., \& Plachinda, S.~I. 1998, Astron.
  Lett., 24, 296

\bibitem[{{Vitrichenko} {et~al.}(2006){Vitrichenko}, {Klochkova}, \&
  {Tsymbal}}]{2006Ap.....49...96V}
{Vitrichenko}, {\'E}.~A., {Klochkova}, V.~G., \& {Tsymbal}, V.~V. 2006,
  Astrophysics, 49, 96

\bibitem[{Vitrichenko \& Plachinda(2001)}]{Vitrichenko2001}
Vitrichenko, {\'{E}}.~A. \& Plachinda, S.~I. 2001, Astron. Lett., 27, 581

\bibitem[{{Voss} {et~al.}(2010){Voss}, {Diehl}, {Vink}, \&
  {Hartmann}}]{2010A&A...520A..51V}
{Voss}, R., {Diehl}, R., {Vink}, J.~S., \& {Hartmann}, D.~H. 2010, \aap, 520,
  A51

\bibitem[{Wade {et~al.}(2006)Wade, Fullerton, Donati, Landstreet, Petit, \&
  Strasser}]{Wade2006}
Wade, G.~A., Fullerton, A.~W., Donati, J.-F., {et~al.} 2006, Astron.
  Astrophys., 451, 195

\bibitem[{Wales \& Doye(1997)}]{Wales1997}
Wales, D.~J. \& Doye, J. P.~K. 1997, J. Phys. Chem., 101, 5111

\bibitem[{{Wang} {et~al.}(2010){Wang}, {Li}, {Abel}, \&
  {Nakamura}}]{2010ApJ...709...27W}
{Wang}, P., {Li}, Z.-Y., {Abel}, T., \& {Nakamura}, F. 2010, \apj, 709, 27

\bibitem[{Weigelt {et~al.}(1999)Weigelt, Balega, Preibisch, Schertl, Schoeller,
  \& Zinnecker}]{Weigelt1999}
Weigelt, G., Balega, Y., Preibisch, T., {et~al.} 1999, Astron. Astrophys., 347,
  L15

\bibitem[{Wolff {et~al.}(2004)Wolff, Strom, \&
  Hillenbrand}]{0004-637X-601-2-979}
Wolff, S.~C., Strom, S.~E., \& Hillenbrand, L.~A. 2004, The Astrophysical
  Journal, 601, 979

\bibitem[{{Yorke} \& {Sonnhalter}(2002)}]{2002ApJ...569..846Y}
{Yorke}, H.~W. \& {Sonnhalter}, C. 2002, \apj, 569, 846

\bibitem[{Zinnecker \& Yorke(2007)}]{Zinnecker2007}
Zinnecker, H. \& Yorke, H.~W. 2007, Annu. Rev. Astron. Astrophys., 45, 481

\end{thebibliography}
\begin{appendix}
\section{Data Results}
%TODO tabelle mit den fit results fuer alle sterne
We present our resulting separations and flux ratios for object. We list the values for $\theta ^1$ Ori A and $\theta ^1$ Ori C that we took from the literature as well, but in separate tables. 

\subsection{$\theta ^1$ Ori A}

% Table generated by Excel2LaTeX from sheet 'Tabelle1'
\begin{table*}[tbh]
  \centering
    \begin{tabular}{llllllll}
    \toprule
   Date &{MJD} & {x [mas]} & {$\Delta$ x} & {y [mas]} & {$\Delta$y} & {$f$} & {$\Delta f$}   \\
    \midrule
2016.900	&57717.34 & 37.84 & 0.06  & 177.55 & 0.09 & 0.240  & 0.006  \\
2017.779	&58038.38 &40.6 & 0.1  & 175.1 & 0.3 & 0.34 & 0.05 \\
2017.779	&58038.39  & 40.69 & 0.09 & 175.1 & 0.2 & 0.36 & 0.05\\
2018.031	&58130.16  &41.55 & 0.04 & 174.49 & 0.07 & 0.230  & 0.008\\
2018.031	&58130.16 & 41.70 & 0.04 & 174.25 & 0.06 & 0.209 & 0.007 \\
  \bottomrule
    \end{tabular}%
      \caption[Fit results from GRAVITY data of $\theta ^1$ Ori A]{Results of the binary fit for GRAVITY data of $\theta ^1$ Ori A. Relative positions x and y in respect to the primary star, with x pointing towards the east and y towards the north. The uncertainties are $\Delta$x and $\Delta$y. $f$ is the resulting flux ratio, $\Delta f$ the uncertainty. The first column lists the Date and the second column the MJD of the observation.}%
  \label{tab:t1A_new}%
\end{table*}%

% Table generated by Excel2LaTeX from sheet 'Tabelle1'
\begin{table*}[tbh]
  \centering
    \begin{tabular}{lllllll}
    \toprule
    {Date} & MJD & {PA [$^{\circ}$]} & {$\Delta$PA [$^{\circ}$]} & {Sep[mas]} & {$\Delta$ Sep [mas]} & {Reference} \\
    \midrule
    1995.775 &50001.319 &350.6 & 2     & 227   & 5     & [1] \\
    1996.247 &50173.717 &352.8 & 2     & 227   & 4     &  [1]\\
    1996.746 &50355.977 &352.7 & 2     & 223   & 4     &  [1]\\
    1997.788 &50736.567 &353   & 2     & 224   & 2     &  [1]\\
    1998.838 &51120.080 &353.8 & 2     & 221   & 5     &  [1]\\
    1998.841 &51121.175 &353.8 & 2     & 221.5 & 5     & [2] \\
    1999.715 &51440.404 &355.4 & 2     & 219   & 3     & [1] \\
    1999.737 &51448.439 &354.8 & 2     & 215   & 3     & [1] \\
    1999.819 &51478.390 &355.1 & 0.5   & 212   & 2.5   & [3] \\
    2000.765 &51823.916 &356.2 & 2     & 215   & 4     & [1] \\
    2000.781 &51829.760 &356.1 & 2     & 216   & 4     &  [1]\\
    2000.781 &51829.760 &356   & 2     & 215   & 3     &  [1]\\
    2001.186 &51977.687 &356   & 2     & 215   & 3     &  [1]\\
    2001.718 &52172.000 &356.9 & 1     & 205.1 & 3     & [4] \\
    2003.701 &52896.290 &3.9   & 1     & 210   & 5     & [5] \\
    2003.945 &52985.411 &3.9   & 1     & 209   & 5     & [5] \\
    2004.816 &53303.544 &0.3   & 1.6   & 203   & 2     & [6] \\
    2004.822 &53305.736 &0.9   & 0.8   & 205   & 3     & [6] \\
    2004.945 &53350.661 &4.6   & 1     & 207   & 5     & [5] \\
    2005.06 &53392.665 &5.3   & 1     & 208   & 5     &  [5]\\
    2005.94 &53714.085 &5.9   & 1     & 204   & 5     &  [5]\\
    2007.704 &54358.386 &6.1   & 1     & 202   & 5     &  [5]\\
    2009.019 &54838.690 &7.5   & 1     & 199   & 5     &  [5]\\
    2009.885 &55154.996 &8.2   & 1     & 197   & 5     & [5] \\
    2009.899 &55160.110 &8.5   & 1     & 198   & 5     &  [5]\\
    2010.26 &55291.965 &9.4   & 1     & 197   & 5     &  [5]\\
    2010.877 &55517.324 &6.5   & 0.3   & 193.1 & 0.5   &[4] \\
    2010.953 &55545.083 &6.2   & 2     & 193   & 1     & [5]\\
    2011.827 &55864.312 &7.3   & 2     & 193.2 & 1     &  [5]\\
    \bottomrule
    \end{tabular}%
      \caption[Positions of $\theta ^1$ Ori A$_2$ from the literature]{Positions of $\theta ^1$ Ori A$_2$, in position angle (PA) and separation (Sep) with respect to the primary star. The observation time is listed in the first column. The corresponding uncertainties are denoted as $\Delta$PA and $\Delta$Sep  Positions are taken from indicated references: [1] \citet{Schertl2003}, [2] \citet{Weigelt1999}, [3] \citet{Balega2004}, [4] \citet{Close2012}, [5] \citet{Grellmann2013}, [6] \citet{Balega2007}.}%
  \label{tab:t1A_old}%
\end{table*}%

\FloatBarrier

\subsection{$\theta ^1$ Ori B}

\begin{table*}[tbh]
  \centering
    \begin{tabular}{llllllll}
    \toprule
    Date & {MJD} & {x} & {$\Delta$} & {y} & {$\Delta$y} & {$f$} & {$\Delta f$} \\
    \midrule
   2017.028 &57764.11 & -7.10 & 0.02  & -4.68 & 0.02 & 0.224 & 0.003 \\
   2017.138 &57804.11 & -8.64 & 0.03 & -4.5 & 0.1  & 0.30 & 0.01 \\
   2017.209 &57830.03 & -9.656 & 0.002 & -4.304 & 0.002 & 0.325 & 0.001 \\
   2017.217 &57833.03 & -9.897 & 0.002 & -4.196 & 0.002 & 0.330  & 0.005 \\
   2017.777 &58037.38 & -15.841 & 0.003 & -2.081 & 0.007 & 0.442 & 0.002 \\
   2017.779 &58038.35 & -15.763 & 0.004 & -2.136 & 0.008 & 0.296 & 0.001 \\
   2017.782 &58039.38 & -15.841 & 0.003 & -2.081 & 0.007 & 0.442 & 0.002 \\
   2017.782 &58039.4 & -15.821 & 0.004 & -2.104 & 0.007 & 0.453 & 0.003 \\
   2018.028 &58129.12 & -17.212 & 0.003 & -0.987 & 0.005 & 0.303 & 0.001 \\
    \bottomrule
    \end{tabular}%
      \caption[Fit results from GRAVITY data of $\theta ^1$ Ori B]{Results of the binary fit for GRAVITY data of $\theta ^1$ Ori B. Relative positions x and y in respect to the primary star, with x pointing towards the east and y towards the north. The uncertainties are $\Delta$x and $\Delta$y. $f$ is the resulting flux ratio, $\Delta f$ the uncertainty. The first column lists the observation time and the second column the MJD of the observation.}%
  \label{tab:t1B}%
\end{table*}%

\FloatBarrier
\subsection{$\theta ^1$ Ori C}

\begin{table*}[tbh]
  \centering
    \begin{tabular}{llllllll}
    \toprule
   Date & {MJD} & {x} & {$\Delta$x} & {y} & {$\Delta$y} & {$f$} & {$\Delta f$} \\
    \midrule
   2015.854 & 57335.34 & 6.53 & 0.03 & 26.23 & 0.030 & 0.266 & 0.002 \\
 2016.021   & 57396.23 & 5.572 & 0.003 & 25.217 & 0.006 & 0.296 & 0.001 \\
 2016.758   & 57665.36 & 0.368 & 0.003 & 19.243 & 0.008 & 0.285 & 0.002 \\
 2016.900   & 57717.21 & -0.616 & 0.007 & 17.841 & 0.007 & 0.27  & 0.002 \\
  2017.209  & 57830.05 & -2.751 & 0.009 & 14.726 & 0.007 & 0.309 & 0.002 \\
 2017.776   & 58037.34 & -6.85 & 0.03 & 8.67  & 0.05 & 0.29 & 0.04 \\
  2017.779  & 58038.32 & -6.92 & 0.02 & 8.57 & 0.02 & 0.41 & 0.03 \\
 2018.031   & 58130.2 & -8.346 & 0.007 & 5.97  & 0.01 & 0.307 & 0.002 \\
    \bottomrule
    \end{tabular}%
     \caption[Fit results from GRAVITY data of $\theta ^1$ Ori C]{Results of the binary fit for GRAVITY data of $\theta ^1$ Ori C. Relative positions x and y in respect to the primary star, with x pointing towards the east and y towards the north. The uncertainties are $\Delta$x and $\Delta$y. $f$ is the resulting flux ratio, $\Delta f$ the uncertainty. The first column lists the observation time and the second column the MJD of the observation.}%
  \label{tab:t1C_new}%
\end{table*}%

\begin{table*}[tbh]
  \centering
    \begin{tabular}{lllllll}
    \toprule
    {Date} & MJD & {PA [deg]} & {$\Delta$ PA [deg]} & {Sep [mas]} & {$\Delta$ Sep [mas]} &Reference\\
\midrule
    1997.784 &50735.106 & 226   & 3     & 33    & 2 &[1] \\
    1998.383 &50953.891 & 222   & 5     & 37    & 4 &[1]\\
    1999.737 &51448.439 & 214   & 2     & 43    & 1 &[2]\\
    1999.819 &51478.353 & 213.5 & 2     & 42    & 1 &[3]\\
    2000.873 &51863.509 & 210   & 2     & 40    & 1 &[3]\\
    2001.184 &51976.956 & 208   & 2     & 38    & 1 &[2]\\
    2003.8 &   52932.450& 19.3  & 2     & 29    & 2 &[3]\\
    2003.925 & 52978.252& 19    & 2     & 29    & 2 &[3]\\
    2003.928 &52979.202 & 19.1  & 2     & 29    & 2 &[3]\\
    2004.822 &53305.589 & 10.5  & 4     & 24    & 4 &[3]\\
    2005.921 &53706.981 & 342.74 & 2     & 13.55 & 0.5 &[3]\\
    2006.149 &53790.276 & 332.3 & 3.5   & 11.8  & 1.11 &[4]\\
    2007.019 &54108.263 & 274.9 & 1     & 11.04 & 0.5 &[5]\\
    2007.143 &54153.298 & 268.1 & 5.2   & 11.94 & 0.31 &[4] \\
    2007.151 &54156.293 & 272.9 & 8.8   & 12.13 & 1.58 &[4]\\
    2007.175 &54165.278 & 266.6 & 2.1   & 12.17 & 0.37 &[4]\\
    2007.206 &54176.309 & 265.6 & 1.9   & 12.28 & 0.41 &[4]\\
    2007.214 &54179.304 & 263   & 2.3   & 12.14 & 0.43 &[4]\\
    2007.901 &54430.486 & 238   & 2     & 19.8  & 2 &[5]\\
    2007.923 &54438.485 & 241.2 & 1     & 19.07 & 0.5 &[5]\\
    2008.027 &54476.508 & 237   & 3     & 19.7  & 3 &[5]\\
    2008.027 &54476.508 & 236.5 & 3     & 19.6  & 3 &[5]\\
    2008.071 &54492.506 & 236.2 & 2     & 20.1  & 2 &[5]\\
    2008.148 &54520.520 & 234.6 & 1     & 21.17 & 0.5 &[5]\\
    2008.173 &54529.542 & 236.4 & 1     & 21.27 & 0.5 &[5]\\
    2010.762 &55475.321 & 216.3 & 2     & 42.6  & 1 &[6]\\
    2010.986 &55557.137 & 215.7 & 2     & 43.4  & 1 &[6]\\
    2010.989 &55558.232 & 215   & 2     & 43.1  & 1 &[6]\\
    \bottomrule
    \end{tabular}%
      \caption[Positions of $\theta ^1$ Ori C$_2$ from the literature]{Positions of $\theta ^1$ Ori C$_2$, in position angle (PA) and separation (Sep) with respect to the primary star. The first column lists the observation time and the second column the MJD of the observation. The corresponding uncertainties are denoted as $\Delta$PA and $\Delta$Sep  Positions are taken from indicated references: [1] \citet{Weigelt1999}, [2] \citet{Schertl2003}, [3] \citet{Kraus2007}, [4] \citet{2008ApJ...674L..97P}, [5] \citet{Kraus2009}, [6] \citet{Grellmann2013}.}%
  \label{tab:t1C_old}%
\end{table*}%

\FloatBarrier

\subsection{$\theta ^1$ Ori D}
\begin{table*}[tbh]
  \centering
    \begin{tabular}{llllllll}
    \toprule
    Date & {MJD} & {x} & {$\Delta$x} & {y} & {$\Delta$y} & {$f$} & {$\Delta f$} \\
    \midrule
 2016.903   & 57718.315 & -0.967 & 0.01  & 2.378 & 0.02  & 0.422 & 0.012 \\
  2017.212  & 57831.052 & -1.571 & 0.004 & 1.4   & 0.005 & 0.365 & 0.002 \\
 2017.217   & 57833.005 & -1.576 & 0.02  & 0.98 & 0.02 & 0.331 & 0.01 \\
 2017.217   & 57833.005 & -1.614 & 0.007 & 1.01 & 0.01  & 0.349 & 0.003 \\
  2017.782  & 58039.332 & -1.27 & 0.01 & 1.95 & 0.02 & 0.295 & 0.006 \\
  2018.025  & 58128.128 & 0.881 & 0.004 & 2.315 & 0.008 & 0.338 & 0.002 \\
  2018.030  & 58130.094 & 0.662 & 0.002 & 2.477 & 0.004 & 0.335 & 0.001 \\
    \bottomrule
    \end{tabular}%
         \caption[Fit results from GRAVITY data of $\theta ^1$ Ori D]{Results of the binary fit for GRAVITY data of $\theta ^1$ Ori D. Relative positions x and y in respect to the primary star, with x pointing towards the east and y towards the north. The uncertainties are $\Delta$x and $\Delta$y. $f$ is the resulting flux ratio, $\Delta f$ the uncertainty. The first column lists the observation time and the second column the MJD of the observation.}%
  \label{tab:t1D_dat}%
\end{table*}%

\FloatBarrier

\subsection{$\theta ^2$ Ori A}

% Table generated by Excel2LaTeX from sheet 'Tabelle1'
\begin{table*}[tbh]
  \centering
    \begin{tabular}{llllllll}
    \toprule
    Date & {MJD} & {x} & {$\Delta$x} & {y} & {$\Delta$y} & {$f$} & {$\Delta f$} \\
    \midrule
   2016.903 &57718.263 & 0.36 & 0.01  & -1.21 & 0.02 & 0.53 & 0.06 \\
  2018.030  &58130.111 & -0.062 & 0.002 & 0.993 & 0.008 & 0.52 & 0.02 \\
  2018.031  &58130.157   & -0.291 & 0.002 & 0.946 & 0.005 & 0.72  & 0.04 \\
  
    \bottomrule
    \end{tabular}%
     \caption[Fit results from GRAVITY data of $\theta ^2$ Ori A]{Results of the binary fit for GRAVITY data of $\theta ^2$ Ori A. Relative positions x and y in respect to the primary star, with x pointing towards the east and y towards the north. The uncertainties are $\Delta$x and $\Delta$y. $f$ is the resulting flux ratio, $\Delta f$ the uncertainty. The first column lists the MJD of the observation.}%
  \label{tab:t2A}%
\end{table*}%
\FloatBarrier

\subsection{$\theta ^2$ Ori B}
We observed $\theta ^2$ Ori B on January 10th 2018 with: x$~= -66.93 \pm 0.04$, y$~= 68.52 \pm 0.07$, $f = 0.022 \pm 0.001$.

\subsection{$\theta ^2$ Ori C}
We observed $\theta ^2$ Ori C on January 12th 2018 with: x$~= -36.74 \pm 0.02	$, y$~= 10.21 \pm 0.03$, $	f = 0.114 \pm 0.002$.

\subsection{NU Ori}

% Table generated by Excel2LaTeX from sheet 'Tabelle1'
\begin{table*}[htbp]
  \centering
    \begin{tabular}{llllllll}
    \toprule
    Date & {MJD} & {x} & {$\Delta$x} & {y} & {$\Delta$y} & {$f$} & {$\Delta f$} \\
    \midrule
  2017.782  &58039.3443 & -8.480 & 0.007 & 1.07 & 0.01 & 0.179 & 0.001 \\
   2018.025 &58128.0987 & -2.695 & 0.005 & 3.46 & 0.01 & 0.189 & 0.001 \\
    \bottomrule
    \end{tabular}%
      \caption[Fit results from GRAVITY data of NU Ori.]{Results of the binary fit for GRAVITY data of NU Ori. Relative positions x and y in respect to the primary star, with x pointing towards the east and y towards the north. The uncertainties are $\Delta$x and $\Delta$y. $f$ is the resulting flux ratio, $\Delta f$ the uncertainty. The first column lists the date and the second column the MJD of the observation.}%
  \label{tab:NUO}%
\end{table*}%

\FloatBarrier
\end{appendix}
\end{document}